\documentclass{aa}  

\usepackage{graphicx}

\usepackage{amsmath,amssymb}
\usepackage{relsize}
\usepackage[flushleft]{threeparttable}
\usepackage{array,booktabs,makecell}
\usepackage{enumitem}
\usepackage{multirow}

\usepackage{color}
\usepackage{xcolor}

\newcommand{\boldvec}[1]{\vec{\mbox{\boldmath{$#1$}}}}

\usepackage{txfonts}
\usepackage[]{hyperref}
\begin{document} 
   
   \title{A three-dimensional reconstruction of the interstellar magnetic field toward a star-forming region}
   
   \titlerunning{A 3D reconstruction of the interstellar magnetic field}

   \author{K. Ferrière\inst{1} \and
          L. Montier\inst{1} \and
          J.-S. Carrière \inst{1}
          }

   \institute{IRAP, Université de Toulouse, CNRS, 9 avenue du Colonel Roche, BP 44346, 31028 Toulouse Cedex 4, France\\
              \email{katia.ferriere@irap.omp.eu}
             }

   %\date{Received September 15, 1996; accepted March 16, 1997}

% \abstract{}{}{}{}{} 
% 5 {} token are mandatory
 
  \abstract
  % context heading (optional), leave it empty if necessary  
{The polarized thermal emission from interstellar dust offers a valuable tool for probing both the dust and the magnetic field in the interstellar medium (ISM).
However, existing observations only yield the total amount of dust emission along the line of sight (LoS), with no information on its LoS distribution.}
{We present a new method designed to give access to the LoS distribution of the dust emission, both in terms of intensity and polarization.}
{We relied on three kinematic gas tracers (H{\sc i}, $^{12}$CO, and $^{13}$CO emission lines) to identify the different clouds present along the LoS. We decomposed the measured intensity of the dust emission, $I_{\rm d}$, into separate contributions from these clouds.
We performed a similar decomposition of the measured Stokes parameters for linear polarization, $Q_{\rm d}$ and $U_{\rm d}$, to derive the polarization parameters of the different clouds, and from this we inferred the clouds' magnetic field orientations.
}
{We applied our method to a $3~{\rm deg}^2$ region of the sky, centered on $(l,b) = (139^{\circ}30',-3^{\circ}16')$ and exhibiting signs of star formation activity.
We found this region to be dominated by an extended and bright cloud with nearly horizontal magnetic field, as expected from the nearly vertical polarization angles measured by {\it Planck}.
More importantly, we detected the presence of two smaller, depolarizing molecular clouds with very different magnetic field orientations in the plane of the sky ($\simeq 65^\circ$ and $\simeq 45^\circ$ from the horizontal).
This is a novel and viable result, which cannot be directly read off the {\it Planck} polarization maps.
}
{The application of our method to the G139 region convincingly demonstrates the need to complement 2D polarization maps with 3D kinematic information when looking for reliable estimates of magnetic field orientations.
}


   \maketitle
%
%-------------------------------------------------------------------

\section{Introduction}
\label{sec:introduction}

The sky offers a 2D view of the cosmos and one of the main challenges of observational astrophysics is to gain access to the line-of-sight (LoS) dimension.
The latter is crucial to uncover the true physical properties and understand the exact inner workings of our cosmic environment.
For instance, in the case of our Galaxy, access to the LoS dimension makes it possible to retrieve the 3D distribution of interstellar matter \citep{lallement_etal_19, lallement&vbc_22, green_etal_19, chen_etal_19, leike&e_19, leike&ge_21, hottier&ba_20, zucker_etal_21, oneill_etal_2024}; in turn, this paves the way to studying the dynamics of the interstellar medium (ISM), the formation and evolution of interstellar structures, and, ultimately, the whole cycle of matter between stars and the ISM.
It is evident that 2D images alone are insufficient to reach that goal, as they afford only a partial, and possibly misleading, perspective on the observed medium.
In general, integration along the LoS results in a loss of information on the spatial variations of physical properties such as temperature, volume density, emissivity, and so on.
The loss of information is more severe in the case of vectorial quantities, such as polarization vectors, which can add up either constructively or destructively.
As a result, LoS integration can cause depolarization of an intrinsically polarized signal; for instance, when the magnetic field orientation varies along the LoS \citep{Planck_XIX_2015, Planck_L_2017, clark_18, pelgrims_etal_21} or in the presence of Faraday rotation \citep{sokoloff_etal_98, beck_01}.
Another issue is that projecting onto the plane of the sky (PoS) can distort our perception of certain classes of objects (cores, clumps, filaments, etc.) as well as our estimation of their geometric characteristics (size, aspect ratio, relative orientation angle, etc.) and, accordingly, bias the related statistical analyses \citep[e.g.,][]{Planck_XXXII_2016, padoan_etal_23}.

There are a range of tools available to help probe the LoS dimension.
For instance, turning to spectroscopy, we can infer, from the profile of a spectral line, the brightness distribution as a function of radial velocity (RV) and then we can use the RV as a proxy for LoS coordinate.
The direct output is  a 3D map of the line brightness as a function of position in the sky and  RV, known as a spectral cube.
To convert from RV to LoS coordinate (and, thus, obtain a physical cube), we can rely on objects whose distances can be measured directly (e.g., via the parallax) or indirectly (e.g., via bracketing or standard candles).

Observations of the Galactic magnetic field are no exception, as they are also plagued by the lack of information along the LoS.
The classical observational methods based on Faraday rotation, synchrotron emission, and dust polarization provide only LoS-integrated quantities, with no information on how the integrant (Faraday rotation rate, synchrotron emissivity, or dust emissivity) varies along the LoS.
Several avenues have been proposed to (partially) overcome this problem and, thus, gain (partial) access to the 3D structure of the Galactic magnetic field.
Here, we note three possible approaches, based on polarization of starlight, Faraday rotation of pulsar signals, and Faraday tomography, respectively.

The polarization of starlight presumably results from its interaction with interstellar dust grains that are aligned by the interstellar magnetic field.
The measured polarization orientation directly gives the orientation of $\boldvec{B}_\perp$,\footnote{
In this paper, the magnetic field vector is denoted by $\boldvec{B}$, its component in the plane of the sky (PoS) by $\boldvec{B}_\perp$, and its component along the line of sight (LoS) by $B_\parallel$.
}
while the measured polarization fraction can, under certain conditions, provide an estimate of the inclination of $\boldvec{B}$ to the PoS.
If $\boldvec{B}$ varies along the LoS, the inferred orientation and inclination angles are dust-weighted averages between the considered star and the observer.
By considering a large number of stars distributed in space and combining their measured polarization parameters with their measured distances (e.g., from {\it Gaia}), we can reconstruct the orientation of $\boldvec{B}_\perp$ (or possibly $\boldvec{B}$) in 3D.
This type of tomographic mapping of $\boldvec{B}_\perp$ was performed
toward the Perseus molecular cloud \citep{doi_etal_2021}, toward the Southern Coalsack dark nebula \citep{versteeg_etal_2024} and toward a portion of the Sagittarius spiral arm \citep{doi_etal_2024}.
For the larger datasets from the optical polarization survey PASIPHAE, an automated, Bayesian LoS-inversion method was developed by \cite{pelgrims_etal_23}
and later applied by \cite{pelgrims_etal_24} to a $4~{\rm deg}^2$ region of the sky, centered on $(l,b) = (103.3^{\circ},22.3^{\circ})$.

Faraday rotation of the linearly polarized radio waves emitted by Galactic pulsars occurs in ionized regions of the ISM, i.e., mainly in the warm ionized medium (WIM), where radio waves interact with the thermal electrons of the medium.
The angle by which the polarization orientation rotates is equal to the observing wavelength squared, $\lambda^2$, times the rotation measure, ${\rm RM} \propto  \int _0 ^L n_{\rm e} \ B_\parallel \ ds$, where $n_{\rm e}$ is the thermal-electron density and $L$ the path length from the pulsar to the observer.
Aside from the RM of a pulsar, we can also measure its dispersion measure, ${\rm DM} =  \int _0 ^L n_{\rm e} \ ds$.
The ratio of RM to DM directly yields the $n_{\rm e}$-weighted average value of $B_\parallel$ between the pulsar and the observer.
By combining the average values of $B_\parallel$ toward all pulsars with measured RM and DM (roughly 1\,500 at the present time) with their measured distances, we can map out the large-scale 3D distribution of $B_\parallel$ \citep[e.g.,][]{han&mvd_18, sobey_etal_19}. 

Faraday tomography also relies on Faraday rotation, but instead of considering the Faraday rotation of the linearly polarized radiation from a background radio source, we can exploit the $\lambda^2$-dependent Faraday rotation of the synchrotron radiation from the Galaxy itself \citep{burn_66,brentjens&d_05}.
More precisely, we can measure the Galactic polarized intensity at many different radio wavelengths and then convert its variation with $\lambda^2$ into a variation with LoS coordinate.
The standard output of Faraday tomography is a so-called Faraday cube, namely, a 3D map of the synchrotron polarized emission as a function of position in the sky and Faraday depth, which is the equivalent of physical depth measured in terms of Faraday rotation.
In practice, Faraday tomography can be used to separate synchrotron-emitting regions located at different Faraday depths and to estimate their respective synchrotron polarized intensities, which, in turn, can lead to the strength and the orientation of their $\boldvec{B}_\perp$.
Faraday tomography can also be used to uncover intervening Faraday screens and to estimate their Faraday thicknesses, which, in turn, can lead to their $B_\parallel$.
The method is particularly interesting when the uncovered Faraday screens can be identified with known gaseous structures because it then offers a new way of probing their magnetic fields.
Faraday tomography was successfully applied to several regions of the sky, including small fields centered on the nearby galaxy IC\,342 \citep{vaneck_etal_2017} and the extra-galactic point source 3C\,196 \citep{turic_etal_2021}, as well as a much larger area toward the high-latitude outer Galaxy \citep{erceg_etal_2022,erceg_etal_2024}.

Aside from these three classical methods, \cite{hu&l_2023MNRAS.524.2379H} proposed a more indirect approach to map out the 3D distribution of the orientation and strength of $\boldvec{B}_\perp$, based on the application of the velocity gradient and two Mach number techniques to H{\sc i} spectroscopic observations.

In this paper, we propose a new 3D polarimetric approach, which combines two of the well-proven observational tools described above, namely, 3D spectral cubes and dust polarization.
Compared to the starlight polarimetric approach, which utilizes the polarization of stars with measured distances, we rely on the polarization of the thermal emission from interstellar dust, which we connect to kinematic gas tracers with measured spectral cubes.
Thus, our method combines two complementary datasets: spectral cubes of atomic (H{\sc i}) and molecular ($^{12}$CO and $^{13}$CO) gas tracers and polarization maps of the dust emission.
The former contain the kinematic information needed to locate the dust-emitting structures along the LoS, albeit in terms of RV rather than physical distance, and the latter provide the polarization information needed to reconstruct the magnetic field orientation. 

In Sect.~\ref{sec:method}, we lay out the general method.
In Sect.~\ref{sec:application}, we apply our method to a $104' \times 104'$ area of the sky containing a star-forming region.
In Sect.~\ref{sec:conclusion}, we discuss our results and present our conclusions.

\section{General method}
\label{sec:method}

In this section, we introduce the general equations needed in our study and we explain how we can identify the different dust-emitting clouds along the LoS and estimate their magnetic field orientations.
In Sect.~\ref{sec:method_dust_polar}, we present the basic equations describing the polarized dust emission.
In Sect.~\ref{sec:method_gas_tracers}, we introduce our atomic (H{\sc i}) and molecular ($^{12}$CO and $^{13}$CO) kinematic gas tracers, combine the latter into a single molecular (CO) tracer, and connect the dust emission to the H{\sc i} and CO tracers via conversion factors.
In Sect.~\ref{sec:method_clouds}, we explain how the measured intensity of the dust emission can be decomposed along the LoS into the contributions from different clouds identified with the help of the kinematic gas tracers.
In Sect.~\ref{sec:method_Bfield}, we show how the polarization parameters of the different clouds can be fitted to the observed polarization maps of the dust emission and used to infer the orientations of their internal magnetic fields, which we assume to be uniform.
Along the way, we make it clear that our LoS decomposition has degeneracies, with implications for the validity of our derived magnetic field orientations. We also propose a convergence test to identify the magnetic field orientations that are truly reliable.

\subsection{Polarized dust emission}
\label{sec:method_dust_polar}

The polarized thermal emission from interstellar dust, integrated along the LoS, can be described by three quantities: the intensity,
\begin{equation}
I_{\rm d} = \int_0^\infty {\cal E}_{\rm d}(r) \ dr \ ,
\label{eq:totintensity}
\end{equation}
and the two Stokes parameters for linear polarization,
\begin{equation}
Q_{\rm d} = \int_0^\infty p_{\rm d}^{\rm loc}(r) \ {\cal E}_{\rm d}(r) \ \cos{\big( 2 \psi_{\rm d}^{\rm loc}(r) \big)} \ dr
\label{eq:stokesQ}
\end{equation}
and
\begin{equation}
U_{\rm d} = \int_0^\infty p_{\rm d}^{\rm loc}(r) \ {\cal E}_{\rm d}(r) \ \sin{\big( 2 \psi_{\rm d}^{\rm loc}(r) \big)} \ dr \,, 
\label{eq:stokesU}
\end{equation}
where $r$ denotes the LoS distance from the observer, subscript ${\rm d}$ stands for dust, ${\cal E}_{\rm d}(r)$ is the local emissivity of the dust thermal emission, $p_{\rm d}^{\rm loc}(r)$ the local polarization fraction, and $\psi_{\rm d}^{\rm loc}(r)$ the local polarization angle (increasing counterclockwise from Galactic north).
Equations~(\ref{eq:stokesQ}) and (\ref{eq:stokesU}) can be combined into a single equation for the complex polarized intensity,
\begin{equation}
{\cal P}_{\rm d} \ \equiv \ Q_{\rm d} + i \, U_{\rm d}
\ = \ \int_0^\infty p_{\rm d}^{\rm loc}(r) \ {\cal E}_{\rm d}(r) \ e^{2 i \psi_{\rm d}^{\rm loc}(r)} \ dr \ \cdot
\label{eq:polintensity_complex}
\end{equation}
The norm of the complex polarized intensity is generally referred to as the (real) polarized intensity,
\begin{equation}
P_{\rm d} = |{\cal P}_{\rm d}| = \sqrt{Q_{\rm d}^2 + U_{\rm d}^2} \ \cdot
\label{eq:polintensity_real}
\end{equation}

The local polarization fraction can be written as
\begin{equation}
p_{\rm d}^{\rm loc} = (p_{\rm d}^{\rm loc})_{\rm max} \ \cos^2{\gamma_B} \ ,
\label{eq:polfrac_local}
\end{equation}
where $\gamma_B$ is the inclination angle of the local magnetic field to the PoS,
$(p_{\rm d}^{\rm loc})_{\rm max} = (p_{\rm intr} \ R)$ is the theoretical maximum polarization fraction that can be achieved locally,
$p_{\rm intr}$ is the intrinsic polarization fraction, which depends on the shape, elongation, and material of dust grains,
and $R$ is the Rayleigh reduction factor, which accounts for the imperfect alignment of dust grains\footnote{
There is no consensus in the community on the exact definition of the intrinsic polarization fraction, $p_{\rm intr}$.
For some authors, $p_{\rm intr}$ is the local polarization fraction achieved when $\boldvec{B}$ lies in the PoS (the quantity that we denote $(p_{\rm d}^{\rm loc})_{\rm max}$).
For others, $p_{\rm intr}$ is the local polarization fraction that would be achieved if, in addition to $\boldvec{B}$ lying in the PoS, dust grains were perfectly aligned (the definition adopted here).
}
\citep[see, e.g.,][and references therein]{Planck_XX_2015}.
The local polarization angle is related to the orientation angle of the local magnetic field in the PoS, $\psi_{\rm B}$, via
\begin{equation}
\psi_{\rm d}^{\rm loc} = \psi_B \pm 90^\circ. \ 
\label{eq:polangle_local}
\end{equation}
The $\pm 90^\circ$ term in Eq.~(\ref{eq:polangle_local}) arises because both orientation angles are defined within a $180^\circ$ range, for instance, in the range $[-90^{\circ}, +90^{\circ}]$.

The LoS-averaged polarization fraction is given by
\begin{equation}
p_{\rm d}^{\rm los} = \frac{P_{\rm d}}{I_{\rm d}} = \frac{\sqrt{Q_{\rm d}^2 + U_{\rm d}^2}}{I_{\rm d}} \ ,
\label{eq:polfrac}
\end{equation}
and the LoS-averaged polarization angle by
\begin{equation}
\psi_{\rm d}^{\rm los} = \frac{1}{2} \ \arctan \left( \frac{U_{\rm d}}{Q_{\rm d}} \right) \,
\label{eq:polangle}
,\end{equation}
with $\arctan$ the two-argument arctangent function defined from $-180^\circ$ to $+180^\circ$.
Equations~(\ref{eq:polfrac})  and (\ref{eq:polangle}) are equivalent to the pair of equations
\begin{equation}
Q_{\rm d} = p_{\rm d}^{\rm los} \ I_{\rm d} \ \cos{\big( 2 \psi_{\rm d}^{\rm los} \big)}
\label{eq:stokesQ_obs}
\end{equation}
and
\begin{equation}
U_{\rm d} = p_{\rm d}^{\rm los} \ I_{\rm d} \ \sin{\big( 2 \psi_{\rm d}^{\rm los} \big)} \ ,
\label{eq:stokesU_obs}
\end{equation}
which can be rewritten in complex form as
\begin{equation}
{\cal P}_{\rm d} \ \equiv \ Q_{\rm d} + i \, U_{\rm d}
\ = \ p_{\rm d}^{\rm los} \ I_{\rm d} \ e^{2 i \psi_{\rm d}^{\rm los}} \ \cdot
\label{eq:polintensity_complex_obs}
\end{equation}

The link between the LoS-averaged polarization fraction and angle and their local counterparts is easily obtained by equating Eq.~(\ref{eq:polintensity_complex_obs}) to Eq.~(\ref{eq:polintensity_complex}), while keeping Eq.~(\ref{eq:totintensity}) in mind,
\begin{equation}
p_{\rm d}^{\rm los} \ = \
\frac{\displaystyle
\Bigg\vert \int_0^\infty p_{\rm d}^{\rm loc}(r) \ {\cal E}_{\rm d}(r) \ e^{2 i \psi_{\rm d}^{\rm loc}(r)} \ dr \ \Bigg\vert}
{\displaystyle
\int_0^\infty {\cal E}_{\rm d}(r) \ dr
}
\label{eq:polfrac_link}
%%\nonumber
\end{equation}
and
\begin{equation}
e^{2 i \psi_{\rm d}^{\rm los}} \ = \
\frac{\displaystyle
\int_0^\infty p_{\rm d}^{\rm loc}(r) \ {\cal E}_{\rm d}(r) \ e^{2 i \psi_{\rm d}^{\rm loc}(r)} \ dr
}
{\displaystyle
\Bigg\vert \int_0^\infty p_{\rm d}^{\rm loc}(r) \ {\cal E}_{\rm d}(r) \ e^{2 i \psi_{\rm d}^{\rm loc}(r)} \ dr \ \Bigg\vert
} \ \cdot
\label{eq:polangle_link} 
%%\nonumber
\end{equation}
The physical meaning of Eqs.~(\ref{eq:polfrac_link}) and (\ref{eq:polangle_link}) is pretty straightforward: 
the LoS-averaged polarization fraction, $p_{\rm d}^{\rm los}$, is a dust-weighted LoS average of the local polarization fraction, $p_{\rm d}^{\rm loc}(r)$, reduced by a LoS depolarization factor, $e^{2 i \psi_{\rm d}^{\rm loc}(r)}$, due to fluctuations in the local polarization angle; 
the LoS-averaged polarization orientation, defined by the LoS-averaged polarization angle, $\psi_{\rm d}^{\rm los}$, is a dust-weighted LoS average of the local polarization orientation, defined by the local polarization angle, $\psi_{\rm d}^{\rm loc}(r)$.
In reality, observations do not strictly capture a single LoS, but rather a whole telescope beam. 
Therefore, in practice, the LoS integrals in Eqs.~(\ref{eq:polfrac_link}) and (\ref{eq:polangle_link}) are actually integrals over a telescope beam, and the associated averaging and depolarization actually occur both along the LoS and across the telescope beam. Observational values of $p_{\rm d}^{\rm los}$ are discussed in Appendix~\ref{sec:polfrac}.

\subsection{Kinematic gas tracers and conversion factors}
\label{sec:method_gas_tracers}

\subsubsection{Kinematic gas tracers}

It is often implicitly assumed that the local polarization fraction and angle are uniform along the LoS through (most of) the dust-emitting region.
This assumption makes it possible to infer their values from the observed LoS-averaged polarization fraction and angle, respectively, $p_{\rm d}^{\rm loc}(r) = p_{\rm d}^{\rm los}$ and $\psi_{\rm d}^{\rm loc}(r) = \psi_{\rm d}^{\rm los}$.
In reality, however, the LoS is likely to intersect several clouds with different polarization properties.
In that case, the polarization fraction and angle of the different clouds cannot be immediately inferred from the observed dust emission.
Other tracers, such as kinematic gas tracers, are needed to estimate them separately.

Here, we rely on spectral cubes, namely, 3D maps in $(l,b,v)$ space,\footnote{Throughout the paper, $l$ = Galactic longitude; $b$ = Galactic latitude; $v$ = radial velocity (RV); $d$ = distance.} 
of the brightness temperature, $T_{\rm b}$,
of the H{\sc i} 21\,cm, $^{12}$CO ($1-0$) 2.6\,mm, and $^{13}$CO ($1-0$) 2.7\,mm emission lines, 
based on the notion that the H{\sc i} line traces atomic gas, while the $^{12}$CO and $^{13}$CO lines together trace molecular gas.
In principle, we should also include a spectral cube of the ionized gas, but we assume that the contribution from ionized gas is negligible in the region of interest.

The brightness temperature, $T_{\rm b}$, is affected by optical depth effects, undergoing gradual saturation with increasing opacity. 
Correcting for opacity saturation is a difficult and subtle task, which we address in Appendix~\ref{sec:opacity}.
There, we define an opacity-corrected brightness temperature, $T_{\rm b\star}$, and we derive the equation relating $T_{\rm b\star}$ to $T_{\rm b}$ (Eq.~(\ref{eq:Tb*_Tb_T})).
Since this equation involves the excitation temperature, $T$, we discuss the choice of the value of $T$.
In brief, we argue in favor of choosing $T = 80~{\rm K}$ for H{\sc i} (see paragraph preceding Eq.~(\ref{eq:sigma2_Tb*})) and $T = (T_{\rm b}^{^{12}{\rm CO}})_{\rm max}$ for CO (see paragraph preceding Eq.~(\ref{eq:sigma2_Tb*_1})).
Once the values of $T$ are set, we can convert the spectral cubes of $T_{\rm b}^{{\rm H}\textsc{i}}$, $T_{\rm b}^{^{12}{\rm CO}}$, and $T_{\rm b}^{^{13}{\rm CO}}$ into spectral cubes of $T_{\rm b\star}^{{\rm H}\textsc{i}}$, $(T_{\rm b\star}^{^{12}{\rm CO}})_1$, and $(T_{\rm b\star}^{^{12}{\rm CO}})_2$, respectively, where $(T_{\rm b\star}^{^{12}{\rm CO}})_1$ and $(T_{\rm b\star}^{^{12}{\rm CO}})_2$ are two complementary estimates of $T_{\rm b\star}^{^{12}{\rm CO}}$.
We can then combine the $(T_{\rm b\star}^{^{12}{\rm CO}})_1$ and $(T_{\rm b\star}^{^{12}{\rm CO}})_2$ spectral cubes into the spectral cube of a best-estimate $T_{\rm b\star}^{^{12}{\rm CO}}$ (Eq.~(\ref{eq:Tb*_best})).
From now on, superscript $12$ in $^{12}$CO is dropped for notational simplicity.

The spectral cubes of $T_{\rm b\star}^{{\rm H}\textsc{i}}$ and $T_{\rm b\star}^{\rm CO}$ generally have different angular and spectral resolutions, so they first need to be brought to a common angular resolution, $\delta \theta$, a common spectral resolution, $\delta v$, and a common $(l,b,v)$ grid with, say, $n_1 \times n_2$ pixels.
The maps of $I_{\rm d}$, $Q_{\rm d}$, and $U_{\rm d}$ (see Sect.~\ref{sec:method_dust_polar}) also need to be brought to the angular resolution $\delta \theta$ and the $n_1 \times n_2$ pixel grid.

We assume that our $T_{\rm b\star}^{{\rm H}\textsc{i}}$ and $T_{\rm b\star}^{\rm CO}$ cubes are complementary, in the sense that together they account for all the gas along the LoS, with no omission and no overlap.
It then follows that the hydrogen column density, $N_{\rm H}$, can be decomposed into the contributions from the gas components probed by 
%%our three tracers (denoted with superscripts):
H{\sc i} and CO,
\begin{equation}
N_{\rm H} = N_{\rm H}^{{\rm H}\textsc{i}} 
%%+ N_{\rm H}^{^{12}{\rm CO}} + N_{\rm H}^{^{13}{\rm CO}} 
+ N_{\rm H}^{\rm CO} \ \cdot
\label{eq:decomposition_coldensity}
\end{equation}
The intensity of the dust emission can be decomposed in a similar way,
\begin{equation}
I_{\rm d} = I_{\rm d}^{{\rm H}\textsc{i}} 
%%+ I_{\rm d}^{^{12}{\rm CO}} + I_{\rm d}^{^{13}{\rm CO}} 
+ I_{\rm d}^{\rm CO} \ \cdot
\label{eq:decomposition_totintensity}
\end{equation}

We make the standard assumption that for each gas tracer ${\rm A}$ (${\rm A} =$ H{\sc i} or CO), the intensity of the dust emission associated with that tracer, $I_{\rm d}^{\rm A}$, is simply proportional to the hydrogen column density probed by that tracer, $N_{\rm H}^{\rm A}$ \citep[e.g.,][]{hildebrand_83}. 
If we denote the conversion factor from hydrogen column density to dust intensity
by $X^{\rm A}_{I_{\rm d}/N_{\rm H}}$, we can then write
\begin{equation}
I_{\rm d}^{\rm A} = X^{\rm A}_{I_{\rm d}/N_{\rm H}} \ N_{\rm H}^{\rm A} \ \cdot
\label{eq:conversion_coldensity_totintensity}
\end{equation}
We note that H{\sc i} and CO are expected to have slightly different conversion factors, mostly because dust grains have different properties (composition, size, emissivity) in the atomic and molecular media.
We further assume that for each tracer ${\rm A}$, the hydrogen column density, $N_{\rm H}^{\rm A}$, is proportional to the opacity-corrected brightness temperature, $T_{\rm b\star}^{\rm A}(v)$, integrated over the RV (see Eq.~(\ref{eq:coldensity}) in Appendix~\ref{sec:opacity}),
\begin{equation}
N_{\rm H}^{\rm A} = X^{\rm A}_{N_{\rm H}/T_{\rm b}} \ \int T_{\rm b\star}^{\rm A}(v) \ dv \ ,
\label{eq:conversion_brighttemp_coldensity}
\end{equation}
where $X^{\rm A}_{N_{\rm H}/T_{\rm b}}$ is the conversion factor from velocity-integrated opacity-corrected brightness temperature to hydrogen column density.
Combining Eqs.~(\ref{eq:conversion_coldensity_totintensity}) and (\ref{eq:conversion_brighttemp_coldensity}) leads to
\begin{equation}
I_{\rm d}^{\rm A} = X^{\rm A}_{I_{\rm d}/T_{\rm b}} \ \int T_{\rm b\star}^{\rm A}(v) \ dv \ ,
\label{eq:conversion_brighttemp_totintensity}
\end{equation}
with $X^{\rm A}_{I_{\rm d}/T_{\rm b}} = X^{\rm A}_{I_{\rm d}/N_{\rm H}} \ X^{\rm A}_{N_{\rm H}/T_{\rm b}}$.
If we now insert Eq.~(\ref{eq:conversion_brighttemp_totintensity}) into Eq.~(\ref{eq:decomposition_totintensity}), we finally obtain for the dust intensity
\begin{equation}
I_{\rm d} \ = \ 
X^{{\rm H}\textsc{i}}_{I_{\rm d}/T_{\rm b}} \ \int T_{\rm b\star}^{{\rm H}\textsc{i}} \ dv
\ + \ X^{\rm CO}_{I_{\rm d}/T_{\rm b}} \ \int T_{\rm b\star}^{\rm CO} \ dv \ \cdot
\label{eq:finalexpr_totintensity}
\end{equation}

\subsubsection{Conversion factors}

The observables in Eq.~(\ref{eq:finalexpr_totintensity}) are the dust intensity, $I_{\rm d}$, and the opacity-corrected brightness temperatures, $T_{\rm b\star}^{{\rm H}\textsc{i}}$ and $T_{\rm b\star}^{\rm CO}$.
The conversion factors, $X^{{\rm H}\textsc{i}}_{I_{\rm d}/T_{\rm b}}$ and $X^{\rm CO}_{I_{\rm d}/T_{\rm b}}$, are not directly observable, and we did not find any values for them in the literature.
We did find separate estimates for some of the intermediate conversion factors, $X_{I_{\rm d}/N_{\rm H}}$ and $X_{N_{\rm H}/T_{\rm b}}$ 
(see Sect.~\ref{sec:application_conversion_factors}), but they were obtained for restricted regions of the sky and they tend to have wide scatter, so they cannot be directly applied to our region of interest.

Here, we treat the conversion factors, $X^{{\rm H}\textsc{i}}_{I_{\rm d}/T_{\rm b}}$ and $X^{\rm CO}_{I_{\rm d}/T_{\rm b}}$, as free parameters, and we derive their best-fit values in the considered region by minimizing the reduced $\chi^2$, defined by
\begin{equation}
\label{eq:chi_reduced}
\chi_{\rm r}^2 
= \frac{1}{\left( n_{\rm dat} - n_{\rm par} \right)} \ \mathlarger{\sum}_{n_{\rm pix}}
\frac{\left[ I_{\rm d}^{\rm obs} - \left( I_{\rm d}^{{\rm H}\textsc{i}} + I_{\rm d}^{\rm CO} \right) \right]^2}
{\sigma^2_I} \ ,
\end{equation}
where $I_{\rm d}^{\rm obs}$ is the observed dust intensity,
$I_{\rm d}^{{\rm H}\textsc{i}}$ and $I_{\rm d}^{\rm CO}$ are the dust intensities associated with H{\sc i} and CO, respectively (Eq.~(\ref{eq:conversion_brighttemp_totintensity}) with ${\rm A} =$ H{\sc i} and CO), 
$\sigma_I$ is the total observational uncertainty,
$n_{\rm dat} = n_{\rm pix} = n_1 \times n_2$ is the total number of data points,
$n_{\rm par} = 2$ is the number of free parameters, 
and the sum runs over the $n_{\rm pix}$ pixels.
The total observational uncertainty is the quadratic sum of the measurement errors in $I_{\rm d}^{\rm obs}$, $I_{\rm d}^{{\rm H}\textsc{i}}$, and $I_{\rm d}^{\rm CO}$,
\begin{equation}
\sigma^2_I = \sigma^2(I_{\rm d}^{\rm obs})
+ \sigma^2(I_{\rm d}^{{\rm H}\textsc{i}}) 
+ \sigma^2(I_{\rm d}^{\rm CO}) \ ,
\label{eq:sigma2_Id_tot}
\end{equation}
where
\begin{equation}
\sigma^2(I_{\rm d}^{\rm A}) 
= \left( X^{\rm A}_{I_{\rm d}/T_{\rm b}} \right)^2 \ 
\left( \sum_v \sigma^2(T_{\rm b\star}^{\rm A}) \right) \ 
\delta v^2
\label{eq:sigma2_Id_A}
\end{equation}
for ${\rm A} =$ H{\sc i} and CO,
$\sigma^2(T_{\rm b\star}^{{\rm H}\textsc{i}})$ and $\sigma^2(T_{\rm b\star}^{\rm CO})$ are given by Eqs.~(\ref{eq:sigma2_Tb*}) and (\ref{eq:sigma2_Tb*_best}), respectively,
$\delta v$ is the spectral resolution, 
and the sum runs over all the velocity bins.

The minimization of $\chi_{\rm r}^2$ is performed through Markov chain Monte Carlo (MCMC) simulations.
Since MCMC simulations alone consider only measurement errors, which in the case at hand are potentially dominated by modeling errors (namely, errors in Eq.~(\ref{eq:finalexpr_totintensity}) with constant conversion factors), they are likely to underestimate the uncertainties in the best-fit parameters.
To obtain meaningful uncertainties, $\sigma(X^{{\rm H}\textsc{i}}_{I_{\rm d}/T_{\rm b}})$ and $\sigma(X^{\rm CO}_{I_{\rm d}/T_{\rm b}})$, we resort to parametric bootstrap sampling, with each bootstrap sample involving MCMC simulations.

\subsection{LoS decomposition into clouds}
\label{sec:method_clouds}

Consider a given LoS and assume that the dust emission measured on that LoS arises from $n_{\rm cl}$ distinct clouds.
Here, the term "cloud" is to be understood in a broad sense, which may include intercloud regions.
The intensity of the dust emission can then be modeled as the sum of the contributions from the $n_{\rm cl}$ clouds,
\begin{equation}
I_{\rm d}^{\rm mod} = \sum_{i=1}^{n_{\rm cl}} \ I_{{\rm d},i} \ ,
\label{eq:totintensity_sum}
\end{equation}
and similarly for the two Stokes parameters for linear polarization,
\begin{equation}
Q_{\rm d}^{\rm mod} \ = \ \sum_{i=1}^{n_{\rm cl}} \ Q_{{\rm d},i}
\label{eq:stokesQ_sum}
,\end{equation}
and
\begin{equation}
U_{\rm d}^{\rm mod} \ = \ \sum_{i=1}^{n_{\rm cl}} \ U_{{\rm d},i} \ ,
\label{eq:stokesU_sum}
\end{equation}
where the subscript $i$ refers to cloud $i$.
The contributions $I_{{\rm d},i}$, $Q_{{\rm d},i}$, and $U_{{\rm d},i}$ from cloud $i$ have the same expressions as the total $I_{\rm d}$ (Eq.~(\ref{eq:totintensity})), $Q_{\rm d}$ (Eq.~(\ref{eq:stokesQ})), and $U_{\rm d}$ (Eq.~(\ref{eq:stokesU})), respectively, with the LoS integral over an infinite path length replaced by a LoS integral over the path length through cloud $i$, $L_i$.

Next, we use the spectral cubes of the H{\sc i} and CO opacity-corrected brightness temperatures, $T_{\rm b\star}^{{\rm H}\textsc{i}}$ and $T_{\rm b\star}^{\rm CO}$, introduced in Sect.~\ref{sec:method_gas_tracers} to identify the different clouds along the LoS.

\subsubsection{{\tt ROHSA} decomposition}

In a first step, we apply the algorithm {\tt ROHSA} (Regularized Optimization for Hyper-Spectral Analysis) developed by \cite{Marchal_etal_2019} to our $T_{\rm b\star}^{{\rm H}\textsc{i}}$ and $T_{\rm b\star}^{\rm CO}$ spectral cubes separately.
This algorithm is designed to decompose a spectral data cube, say, a cube of $T_{\rm b\star}(l,b,v)$, into several spatially coherent Gaussian kinematic components.
The Gaussian decomposition is optimized over the entire data cube at once, with the requirement that the solution must be spatially smooth.

In mathematical terms, the observed $T_{\rm b\star}(l,b,v)$ is approximated by a modeled $\tilde{T}_{\rm b\star}(l,b,v)$, equal to the sum of $n_{\rm G}$ Gaussian components,
\begin{equation}
\tilde{T}_{\rm b\star}(l,b,v) = \sum_{j=1}^{n_{\rm G}} \ \tilde{T}_{{\rm b\star},j}(l,b,v) \ ,
\label{eq:brighttemp_sum}
\end{equation}
where each component $j$ is described by an amplitude $A_j(l,b)$, a mean RV, $\overline{v}_j(l,b)$, and a standard deviation, $\sigma_j(l,b)$,
\begin{equation}
\tilde{T}_{{\rm b\star},j}(l,b,v) \ = \ 
A_j(l,b) \ \exp \left[ -
\frac{\left( v - \overline{v}_j(l,b) \right)^2}
{2 \, \sigma^2_j(l,b)}
\right] \ \cdot
\label{eq:brighttemp_compon}
\end{equation}
The best-fit values of the $3 {n_{\rm G}}$ Gaussian parameters, $A_j$, $\overline{v}_j$, and $\sigma_j$, are derived through minimization of a cost function that includes the standard $\chi^2$ term plus a regularization term meant to ensure a spatially smooth solution. 

{\tt ROHSA} has several free parameters (six in the initial version described in \cite{Marchal_etal_2019}, and more in the present online version).
We keep the default values of these parameters, except for the three hyper-parameters entering the regularization term, $\lambda_A$, $\lambda_{\overline{v}}$, and $\lambda_\sigma$.
Since our ultimate purpose is to construct smooth, coherent clouds, we need to impose strong constraints of spatial coherence on $A_j$, $\overline{v}_j$, and $\sigma_j$, which is done by choosing large values for $\lambda_A$, $\lambda_{\overline{v}}$, and $\lambda_\sigma$.
After verifying that the exact values are not critical, we adopt $\lambda_A = \lambda_\sigma = 1000$ and $\lambda_{\overline{v}} = 100$ for our application in Sect.~\ref{sec:application}.
Furthermore, since we want to retain only truly physical components, we discard the extracted components whose velocity-integrated %%amplitudes 
$\tilde{T}_{{\rm b\star},j}$ fall below twice the noise level at every pixel.

\subsubsection{Cloud reconstruction}

The second step is to reconstruct the different clouds along the LoS.
To that end, we collect all the Gaussian components from both H{\sc i} and CO, and we group together the components that have similar velocity profiles.
Since the very definition and the exact outline of an interstellar cloud are moot points, the criterion we choose to rely on is necessarily subjective.

To start with, there is a total of $n_{\rm G}^{\rm tot} = n_{\rm G}^{{\rm H}\textsc{i}} + n_{\rm G}^{\rm CO}$ Gaussian components, which we consider two by two.
For every pair of components $j$ and $j'$ ($j \not= j'$), we compute a velocity correlation coefficient at each pixel, $(l,b)$,
\begin{equation}
\mathcal{C}_{jj'}(l,b) = 
\frac{\displaystyle \sum_v \tilde{T}_{{\rm b\star},j} \ \tilde{T}_{{\rm b\star},j'}}
{\sqrt{\displaystyle \sum_v \tilde{T}_{{\rm b\star},j}^2} \ \sqrt{\displaystyle \sum_v \tilde{T}_{{\rm b\star},j'}^2}} \ ,
\label{eq:correlation}
\end{equation}
where the sums run over all the velocity bins.
Clearly, $\mathcal{C}_{jj'}$ depends only on the velocity profiles (mean velocities, $\overline{v}_j$ and $\overline{v}_{j'}$, and standard deviations, $\sigma_j$ and $\sigma_{j'}$) of both components, not on their amplitudes ($A_j$ and $A_{j'}$).
We then compute the average value of $\mathcal{C}_{jj'}(l,b)$ over all the pixels, weighted by the sum of the dust intensities of components $j$ and $j'$,
\begin{equation}
\langle \mathcal{C}_{jj'} \rangle = 
\frac{\displaystyle \sum_{n_{\rm pix}} (\tilde{I}_{{\rm d},j} + \tilde{I}_{{\rm d},j'}) \ \mathcal{C}_{jj'}}
{\displaystyle \sum_{n_{\rm pix}} (\tilde{I}_{{\rm d},j} + \tilde{I}_{{\rm d},j'})} \ ,
\label{eq:correlation_average}
\end{equation}
where $\tilde{I}_{{\rm d},j}$ is the dust intensity of component $j$ (defined by Eq.~(\ref{eq:totintensity_compon}) below) and the sum runs over the $n_{\rm pix}$ pixels. 

We consider that components $j$ and $j'$ are correlated, and thus parts of a same cloud, if $\langle \mathcal{C}_{jj'} \rangle$ lies above a certain threshold, $\mathcal{C}_{\rm th}$:
\begin{equation}
\langle \mathcal{C}_{jj'} \rangle \geq \mathcal{C}_{\rm th} \ \cdot
\label{eq:threshold_correlation}
\end{equation}
It is important to realize that this is a purely kinematic criterion, independent of any possible spatial correlation.
This choice is motivated by the fact that while two components of a given cloud are expected to have similar velocity profiles, they do not need to be correlated in space; for instance, they could very well be co-spatial (e.g., a region containing both H{\sc i} and CO) or adjacent (e.g., an H{\sc i} envelope around a CO core).

A critical issue concerns the choice of the value of $\mathcal{C}_{\rm th}$.
For our application in Sect.~\ref{sec:application}, we tested all the integer values of $\mathcal{C}_{\rm th}$ %%in the range $[0,100\,\%]$
between 0 and 100\,\%.
Based on the results (summarized in Table~\ref{tab:G139_dependence_Cth}), we decided to present a detailed analysis in the case $\mathcal{C}_{\rm th} = 50\,\%$ (Sects.~\ref{sec:application_4clouds} -- \ref{sec:application_Bfield}),
considering that this value strikes a good balance between preserving physical clouds and separating distinct clouds along the LoS.
The results obtained for other values of $\mathcal{C}_{\rm th}$ as well as their sensitivity to the exact value of $\mathcal{C}_{\rm th}$ are discussed in Sect.~\ref{sec:application_threshold}.

Pairs satisfying Eq.~(\ref{eq:threshold_correlation}) are themselves grouped together into one multicomponent cloud if they possess a component in common.
For instance, if pairs $jk$ and $kl$ satisfy $\langle \mathcal{C}_{jk} \rangle \geq \mathcal{C}_{\rm th}$ and $\langle \mathcal{C}_{kl} \rangle \geq \mathcal{C}_{\rm th}$, they are grouped into one cloud enclosing components $j$, $k$, and $l$, even if $\langle \mathcal{C}_{jl} \rangle < \mathcal{C}_{\rm th}$.

All the remaining Gaussian components, i.e., the components that are not correlated with any other component (in the sense of satisfying Eq.~(\ref{eq:threshold_correlation})), are considered to each form a separate cloud.

Altogether, we end up with $n_{\rm cl}$ distinct clouds made up of one, two, or more Gaussian components.
The dust intensity of cloud $i$, $I_{{\rm d},i}$ can be written as a sum over the $n_{{\rm G},i}$ Gaussian components of cloud $i$:
\begin{equation}
I_{{\rm d},i} = \sum_j \ \tilde{I}_{{\rm d},j} \ ,
\label{eq:totintensity_cloud}
\end{equation}
where the dust intensity of component $j$, $\tilde{I}_{{\rm d},j}$, is related to its opacity-corrected brightness temperature, $\tilde{T}_{{\rm b\star},j}$ (Eq.~(\ref{eq:brighttemp_compon})), through an equation similar to Eq.~(\ref{eq:conversion_brighttemp_totintensity}):
\begin{equation}
\tilde{I}_{{\rm d},j} = X^{\rm A}_{I_{\rm d}/T_{\rm b}} \ \int \tilde{T}_{{\rm b\star},j}(v) \ dv \ \cdot
\label{eq:totintensity_compon}
\end{equation}
Superscript ${\rm A}$ in the conversion factor $X^{\rm A}_{I_{\rm d}/T_{\rm b}}$ refers to the gas tracer (H{\sc i} or CO) associated with component $j$.
As a reminder, the best-fit value of $X^{\rm A}_{I_{\rm d}/T_{\rm b}}$ and the attendant uncertainty, $\sigma(X^{\rm A}_{I_{\rm d}/T_{\rm b}})$, are determined as explained in Sect.~\ref{sec:method_gas_tracers}.

\subsection{Derivation of the magnetic field orientation in each cloud}
\label{sec:method_Bfield}

At this point, we have identified $n_{\rm cl}$ clouds along the LoS, and we have derived their contributions $I_{{\rm d},i}$ (Eqs.~(\ref{eq:totintensity_cloud}) and (\ref{eq:totintensity_compon})) to the dust intensity, $I_{\rm d}^{\rm mod}$ (Eq.~(\ref{eq:totintensity_sum})).
We now turn to their contributions $Q_{{\rm d},i}$ and $U_{{\rm d},i}$ to the two Stokes parameters for linear polarization, $Q_{\rm d}^{\rm mod}$ and $U_{\rm d}^{\rm mod}$ (Eqs.~(\ref{eq:stokesQ_sum}) and (\ref{eq:stokesU_sum})).
As mentioned below Eq.~(\ref{eq:stokesU_sum}), $I_{{\rm d},i}$, $Q_{{\rm d},i}$, and $U_{{\rm d},i}$ have the same expressions as $I_{\rm d}$, $Q_{\rm d}$, and $U_{\rm d}$ (Eqs.~(\ref{eq:totintensity}), (\ref{eq:stokesQ}), and (\ref{eq:stokesU})), respectively, with the LoS integral reduced to the path length through cloud $i$, $L_i$.
By analogy with Eqs.~(\ref{eq:stokesQ_obs}) -- (\ref{eq:polangle_link}), we can then directly write
\begin{equation}
Q_{{\rm d},i} = p_{{\rm d},i} \ I_{{\rm d},i} \ \cos{\big( 2 \psi_{{\rm d},i} \big)}
\label{eq:stokesQ_cloud}
\end{equation}
and
\begin{equation}
U_{{\rm d},i} = p_{{\rm d},i} \ I_{{\rm d},i} \ \sin{\big( 2 \psi_{{\rm d},i} \big)} \ ,
\label{eq:stokesU_cloud}
\end{equation}
with
\begin{equation}
p_{{\rm d},i} \ = \
\frac{\displaystyle
\Bigg\vert \int_{L_i} p_{\rm d}^{\rm loc}(r) \ {\cal E}_{\rm d}(r) \ e^{2 i \psi_{\rm d}^{\rm loc}(r)} \ dr \ \Bigg\vert}
{\displaystyle
\int_{L_i} {\cal E}_{\rm d}(r) \ dr
}
\label{eq:polfrac_cloud}
\end{equation}
and
\begin{equation}
e^{2 i \psi_{{\rm d},i}} \ = \
\frac{\displaystyle
\int_{L_i} p_{\rm d}^{\rm loc}(r) \ {\cal E}_{\rm d}(r) \ e^{2 i \psi_{\rm d}^{\rm loc}(r)} \ dr
}
{\displaystyle
\Bigg\vert \int_{L_i} p_{\rm d}^{\rm loc}(r) \ {\cal E}_{\rm d}(r) \ e^{2 i \psi_{\rm d}^{\rm loc}(r)} \ dr \ \Bigg\vert
} \ \cdot
\label{eq:polangle_cloud}
\end{equation}
The physical meaning of $p_{{\rm d},i}$ and $\psi_{{\rm d},i}$ for cloud $i$ is similar to that of $p_{\rm d}^{\rm los}$ and $\psi_{\rm d}^{\rm los}$ given below Eq.~(\ref{eq:polangle_link}) for the entire LoS.

To proceed, we consider that $p_{{\rm d},i}$ and $\psi_{{\rm d},i}$ are uniform across the PoS surface of cloud $i$, in contrast to $I_{{\rm d},i}$, which generally varies.
Strictly speaking, this is unlikely to be correct, as magnetic field lines have probably been distorted by internal motions. 
However, we are {\it a priori} entitled to take this approach if we are only interested in dust-weighted average values of $p_{{\rm d},i}$ and $\psi_{{\rm d},i}$ at the cloud scale, which we denote with an overscore.\footnote{The exact expressions of the dust-weighted average values of $p_{{\rm d},i}$ and $\psi_{{\rm d},i}$ obtained through minimization of $\chi_{\rm r}^2$ (Eq.~(\ref{eq:chi_reduced_QU})) are derived in Appendix~\ref{sec:bestfit_polar_parameters}.}
Thus, the pair of equations we work here with is obtained by inserting Eqs.~(\ref{eq:stokesQ_cloud}) and (\ref{eq:stokesU_cloud}), with  $p_{{\rm d},i}$ and $\psi_{{\rm d},i}$ overscored, into Eqs.~(\ref{eq:stokesQ_sum}) and (\ref{eq:stokesU_sum}):
\begin{equation}
Q_{\rm d}^{\rm mod} \ = \ \sum_{i=1}^{n_{\rm cl}} \ \overline{p}_{{\rm d},i} \ I_{{\rm d},i} \ \cos{\big( 2 \overline{\psi}_{{\rm d},i} \big)}
\label{eq:stokesQ_sum_cloud}
\end{equation}
and
\begin{equation}
U_{\rm d}^{\rm mod} \ = \ \sum_{i=1}^{n_{\rm cl}} \ \overline{p}_{{\rm d},i} \ I_{{\rm d},i} \ \sin{\big( 2 \overline{\psi}_{{\rm d},i} \big)} \ \cdot
\label{eq:stokesU_sum_cloud}
\end{equation}
In the above equations, the dust intensity of cloud $i$, $I_{{\rm d},i}$, was derived in Sect.~\ref{sec:method_clouds} (Eqs.~(\ref{eq:totintensity_cloud}) and (\ref{eq:totintensity_compon})), while its average polarization fraction and angle, $\overline{p}_{{\rm d},i}$ and $\overline{\psi}_{{\rm d},i}$, are treated as free parameters.
This leaves us with two free parameters per cloud and hence a total of $2 n_{\rm cl}$ free parameters.

The best-fit values of $\overline{p}_{{\rm d},i}$ and $\overline{\psi}_{{\rm d},i}$ are the values that minimize 
\begin{equation}
\label{eq:chi_reduced_QU}
\chi_{\rm r}^2
= \frac{1}{\left( n_{\rm dat} - n_{\rm par} \right)} \ \mathlarger{\sum}_{n_{\rm pix}} \left[
\frac{\left( Q_{\rm d}^{\rm obs} - Q_{\rm d}^{\rm mod} \right)^2}{\sigma^2_Q}
+
\frac{\left( U_{\rm d}^{\rm obs} - U_{\rm d}^{\rm mod} \right)^2}{\sigma^2_U}
\right] \ ,
\end{equation}
where $Q_{\rm d}^{\rm obs}$ and $U_{\rm d}^{\rm obs}$ are the observed Stokes parameters, 
$Q_{\rm d}^{\rm mod}$ and $U_{\rm d}^{\rm mod}$ are the modeled Stokes parameters given by Eqs.~(\ref{eq:stokesQ_sum_cloud}) and (\ref{eq:stokesU_sum_cloud}), respectively,
$\sigma_Q$ and $\sigma_U$ are the total "observational uncertainties" associated with $Q_{\rm d}$ and $U_{\rm d}$, respectively,
$n_{\rm dat} = 2\,n_{\rm pix}$ is the total number of data points, 
$n_{\rm par} = 2\,n_{\rm cl}$ is the number of free parameters, 
and the sum runs over the $n_{\rm pix}$ pixels.
The total "observational uncertainties" are the quadratic sums of the measurement errors in the observed Stokes parameters, $\sigma(Q_{\rm d}^{\rm obs})$ and $\sigma(U_{\rm d}^{\rm obs})$, and the "decomposition errors" in the modeled Stokes parameters arising from decomposition errors in the modeled cloud intensities, $I_{{\rm d},i}$.
The contributions from the different clouds cannot be derived separately, but we may reasonably consider that the "decomposition errors" in $Q_{\rm d}^{\rm mod}$ and $U_{\rm d}^{\rm mod}$ are both equal to the error in the modeled intensity, $I_{\rm d}^{\rm mod}$, times the observed LoS-averaged polarization fraction, $p_{\rm d}^{\rm los,obs}$.
We may further approximate the error in $I_{\rm d}^{\rm mod}$ by the residual $\Delta(I_{\rm d}) = I_{\rm d}^{\rm obs} - I_{\rm d}^{\rm mod}$.
Altogether, we have
\begin{equation}
\sigma^2_Q 
= \sigma^2(Q_{\rm d}^{\rm obs}) 
+ (p_{\rm d}^{\rm los,obs})^2 \ \Delta^2(I_{\rm d})
\label{eq:sigma2_Qd_tot}
\end{equation}
and 
\begin{equation}
\sigma^2_U 
= \sigma^2(U_{\rm d}^{\rm obs}) 
+ (p_{\rm d}^{\rm los,obs})^2 \ \Delta^2(I_{\rm d}) \ \cdot
\label{eq:sigma2_Ud_tot}
\end{equation}
Minimization of $\chi_{\rm r}^2$ is performed through MCMC simulations,
leading to the best-fit values of $\overline{p}_{{\rm d},i}$ and $\overline{\psi}_{{\rm d},i}$, together with the attendant uncertainties, $\sigma(\overline{p}_{{\rm d},i})$ and $\sigma(\overline{\psi}_{{\rm d},i})$.

Once the polarization fraction, $\overline{p}_{{\rm d},i}$, and the polarization angle, $\overline{\psi}_{{\rm d},i}$, of cloud $i$ have been determined, it is possible to estimate the orientation of its internal magnetic field, $\boldvec{B}_i$. 
Here, too, we are referring to dust-weighted average values, denoted with an overscore.
The orientation angle of $\overline{\boldvec{B}}_i$ in the PoS, $\overline{\psi}_{B,i}$, can be directly inferred from the polarization angle, $\overline{\psi}_{{\rm d},i}$, with the help of Eq.~(\ref{eq:polangle_local}) applied to cloud $i$:
\begin{equation}
\overline{\psi}_{B,i} = \overline{\psi}_{{\rm d},i} \pm 90^\circ \ \cdot
\label{eq:polangle_cloud_inverted}
\end{equation}
The inclination angle of $\overline{\boldvec{B}}_i$ to the PoS, $\overline{\gamma}_{B,i}$, can in principle be inferred from the polarization fraction, $\overline{p}_{{\rm d},i}$, using Eq.~(\ref{eq:polfrac_local}) applied to cloud $i$,
\begin{equation}
\cos^2{\overline{\gamma}_{B,i}} = \frac{\overline{p}_{{\rm d},i}}{(\overline{p}_{{\rm d},i})_{\rm max}} \ ,
\label{eq:polfrac_cloud_inverted}
\end{equation}
together with an adopted value of the maximum polarization fraction averaged at the cloud scale, $(\overline{p}_{{\rm d},i})_{\rm max}$. 
Following \cite{Planck_XX_2015}, we could, for instance, take $(\overline{p}_{{\rm d},i})_{\rm max} = 23\,\%$ (see Appendix~\ref{sec:polfrac}).
Because of the uncertainty in $(\overline{p}_{{\rm d},i})_{\rm max}$, the derived value of $\overline{\gamma}_{B,i}$ is much less reliable than the derived value of $\overline{\psi}_{B,i}$.
Moreover, the existence of two opposite-signed solutions for $\overline{\gamma}_{B,i}$ implies that the orientation of $\overline{\boldvec{B}}_i$ can only be determined with a mirror ambiguity with respect to the PoS.
Zeeman observations, which are sensitive to $B_\parallel$, would be needed to set the sign of $\overline{\gamma}_{B,i}$ for the dominant cloud.

\section{Application to the G139 region}
\label{sec:application}

To illustrate the method presented in Sect.~\ref{sec:method}, we now apply it to a small region of the sky encompassing the {\it Herschel} G139 field, which is one of the 116 Galactic fields observed with {\it Herschel} as part of the {\it Herschel} Galactic cold core (GCC) key-program \citep[][]{Juvela_GCCI_2010, Juvela_GCCIII_2012}.
This small region, which we refer to as the G139 region, is a $104' \times 104'$ square centered on $(l,b) = (139^{\circ}30',-3^{\circ}16')$. 
The {\it Herschel} maps reveal a long filamentary structure with signs of star formation activity, including a number of embedded cores and young stellar object candidates \citep{montillaud_etal_2015}. This filamentary structure is surrounded by a more diffuse and extended emission.

In Sect.~\ref{sec:application_input}, we present the relevant available data.
In Sect.~\ref{sec:application_conversion_factors}, we derive the conversion factors from gas tracers to dust emission.
In Sect.~\ref{sec:application_4clouds}, we decompose the measured intensity of the dust emission into the contributions from seven separate clouds along the LoS.
In Sect.~\ref{sec:application_Bfield}, we derive the magnetic field orientation in each cloud.
In Sect.~\ref{sec:application_threshold}, we examine alternative configurations of clouds, involving an increasing number of clouds. The input maps of the G139 region used in our study are listed in Table~\ref{tab:G139_inputmaps}, along with their angular resolution, $\delta \theta$, and (when relevant) their spectral resolution, $\delta v$, and their spectral extent, $\Delta v$.

\begin{table*}
\caption{
   Characteristics of the input maps of the G139 region displayed in the top rows of Figs.~\ref{fig:G139_planckmaps} and \ref{fig:G139_gasmaps}.
}
\begin{threeparttable}
\begin{tabular}{lccccc}
\toprule\toprule
Input map & $\delta \theta$\tablefootmark{a} & $\delta v \ [{\rm km\,s^{-1}}]$\tablefootmark{b} & $\Delta v \ [{\rm km\,s^{-1}}]$\tablefootmark{c} & Survey/instrument & Reference \\
\midrule
Dust 353\,GHz & $7'$ & & & {\it Planck} & 1 \\
H{\sc i} 21\,cm & $10.8'$ & $1.44$ & $[-600,+600]$ & EBHIS & 2 \\
$^{12}$CO 2.6\,mm & $50"$ & $0.16$ & $[-95,+25]$ & MWISP & 3 \\
$^{13}$CO 2.7\,mm & $50"$ & $0.17$ & $[-95,+25]$ & MWISP & 4 \\
\bottomrule\bottomrule
\end{tabular}
\end{threeparttable}
\tablefoot{
\tablefoottext{a}{Angular resolution.}
\tablefoottext{b}{Spectral resolution.}
\tablefoottext{c}{Spectral extent.}
}
\tablebib{
(1)~\citet{Planck_III_2020}; (2) \citet{winkel_etal_2016}; (3) \citet{Su_etal_2019,yuan_etal_2021}; (4) \citet{yuan_etal_2022}.
}
\label{tab:G139_inputmaps}
\end{table*}

\subsection{Available data for G139}
\label{sec:application_input}

\begin{figure*}
%%\centering
\includegraphics[width=\textwidth]{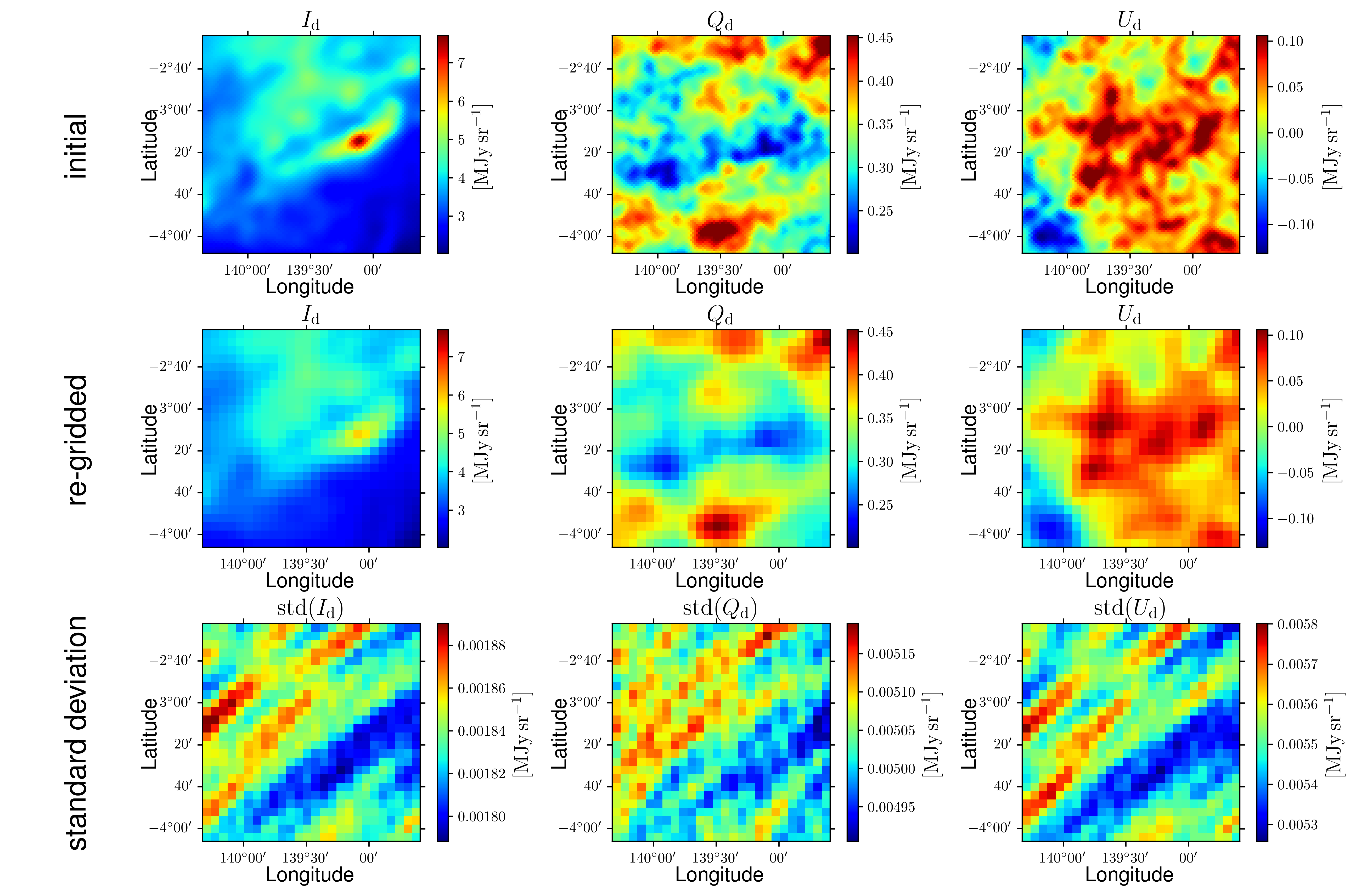}
\caption{
    Maps of the intensity, $I_{\rm d}$ ({\it left}), and of the two Stokes parameters for linear polarization, $Q_{\rm d}$ ({\it middle}) and $U_{\rm d}$ ({\it right}), of the polarized dust emission at 353\,GHz toward the G139 region defined in the first paragraph of Sect.~\ref{sec:application}.
    {\it Top row}: Observational maps from {\it Planck} \citep{Planck_III_2020}.
    {\it Middle row}: Same maps resampled to the common $26 \times 26$ pixel grid of the gas tracers (pixel size = 4').
    {\it Bottom row}: Statistical uncertainties in the {\it Planck} maps resampled to the $26 \times 26$ pixel grid.
    The total uncertainties are equal to the quadratic sums of the statistical uncertainties and the photometric calibration uncertainties, which in turn are given by %%$0.0078 \, I_{\rm d}^{\it Planck}$ 
    $0.0078 \, I_{\rm d}$ for $I_{\rm d}$\citep{Planck_VIII_2016,Planck_III_2020} and  %%$0.015 \, P_{\rm d}^{\it Planck}$ 
    $0.015 \, P_{\rm d}$ for $Q_{\rm d}$ and $U_{\rm d}$ \citep{Planck_III_2020,Planck_XI_2020}.
}
\label{fig:G139_planckmaps}
\end{figure*}

The polarization information needed to infer the magnetic field orientations toward G139 can be extracted from the all-sky maps of the polarized dust emission measured by {\it Planck} at 353\,GHz \citep{Planck_XIX_2015,Planck_III_2020}.
The three panels in the top row of Fig.~\ref{fig:G139_planckmaps} show the maps of the intensity, $I_{\rm d}$ (Eq.~(\ref{eq:totintensity})), and of the two Stokes parameters for linear polarization, $Q_{\rm d}$ (Eq.~(\ref{eq:stokesQ})) and $U_{\rm d}$ (Eq.~(\ref{eq:stokesU})), of the 353\,GHz dust emission toward G139.

\begin{figure*}
\centering
\includegraphics[width=\textwidth]{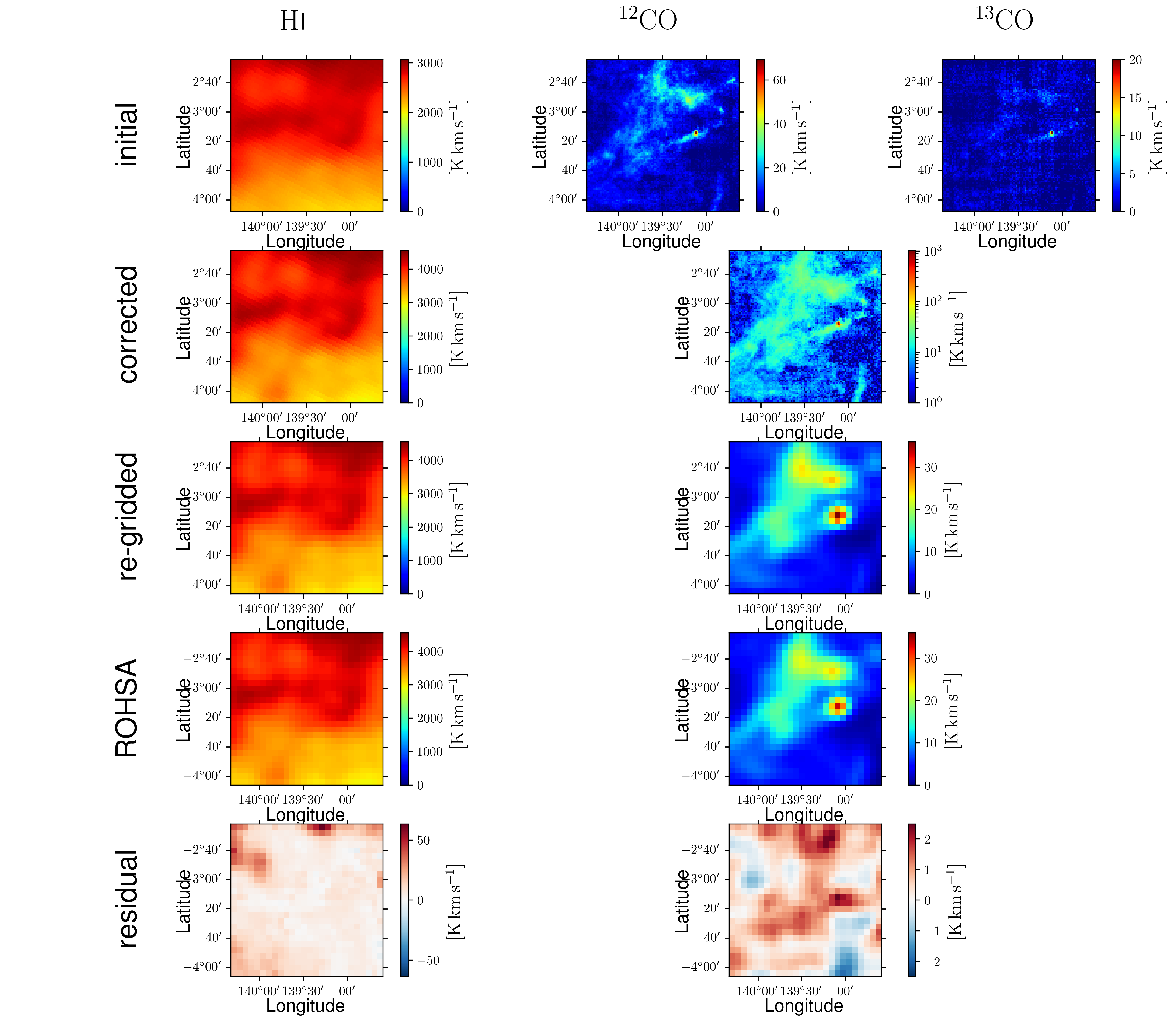}
\caption{
    {\it Top row}: Observational maps of the velocity-integrated brightness temperatures, $T_{\rm b}$,
    of the H{\sc i} 21\,cm ({\it left}), $^{12}$CO 2.6\,mm ({\it middle}), and $^{13}$CO 2.7\,mm ({\it right}) emission lines toward the G139 region;
    these maps are from \cite{winkel_etal_2016}, \cite{yuan_etal_2021}, and \cite{yuan_etal_2022}, respectively.
    {\it Second row}: Maps of the velocity-integrated opacity-corrected brightness temperatures, $T_{\rm b\star}$, of the H{\sc i} ({\it left}) and $^{12}$CO ({\it right}) lines, 
    where $T_{\rm b\star}^{^{12}{\rm CO}}$ is derived based on the combined $^{12}$CO and $^{13}$CO data.
    {\it Third row}: Same maps resampled to the common $26 \times 26$ pixel grid (pixel size = 4').
    {\it Fourth row}: Reconstructed maps after application of the Gaussian decomposition algorithm {\tt ROHSA}.
    {\it Bottom row}: Maps of the residuals obtained by subtracting the reconstructed {\tt ROHSA} maps from the resampled input maps.
    The color bars for H{\sc i} and CO (superscript $12$ dropped) can be rescaled to dust intensity at 353\,GHz using the best-fit values of the conversion factors derived in Sect.~\ref{sec:application_conversion_factors} (see Fig.~\ref{fig:G139_cornerplot_conversionfactors}).
    The resulting ranges in the third and fourth rows are $\simeq [0, 3.5\,{\rm MJy\,sr^{-1}}]$ for H{\sc i} and $\simeq [0, 2.5\,{\rm MJy\,sr^{-1}}]$ for CO.
}
\label{fig:G139_gasmaps}
\end{figure*}

The kinematic information needed to separate the different clouds along the LoS is provided by the spectral cubes of the brightness temperature, $T_{\rm b}$, of the three gas tracers introduced in Sect.~\ref{sec:method_gas_tracers}, namely, the H{\sc i} 21\,cm, $^{12}$CO 2.6\,mm, and $^{13}$CO 2.7\,mm emission lines.
Here, we resort to the following cubes: 
the H{\sc i} cube from the Effelsberg-Bonn H{\sc i} Survey (EBHIS) of the whole northern sky, with $\delta \ge -5^\circ$ \citep{winkel_etal_2016}, and the $^{12}$CO and $^{13}$CO cubes from the Milky Way Imaging Scroll Painting (MWISP) survey of the Galactic plane, with $-10^\circ \le l \le +250^\circ$ and $|b| \lesssim 5.2^\circ$ \citep{Su_etal_2019, yuan_etal_2021, yuan_etal_2022}.
The corresponding maps of the velocity-integrated $T_{\rm b}$ toward G139 are displayed in the three panels of the top row of Fig.~\ref{fig:G139_gasmaps}.
We see that H{\sc i} (left panel) is quite uniformly distributed, with a weak northward gradient, while $^{12}$CO (middle panel) and $^{13}$CO (right panel) have more structured distributions, with a prominent peak at the position of the bright core in the dust intensity map (top-left panel of Fig.~\ref{fig:G139_planckmaps}).
Interestingly, the $^{12}$CO peak does not stand out as prominently as the $^{13}$CO peak, which is most likely because $^{12}$CO emission from the underlying emitting region is optically very thick.

Following the procedure explained in detail in Appendix~\ref{sec:opacity}, we correct $T_{\rm b}$ for opacity saturation, thereby obtaining an opacity-corrected brightness temperature, $T_{\rm b\star}$.
For H{\sc i}, we derive $T_{\rm b\star}^{{\rm H}\textsc{i}}$ using Eq.~(\ref{eq:Tb*_Tb_T}) with $T = 80~{\rm K}$.
For CO, we derive two complementary estimates of $T_{\rm b\star}^{^{12}{\rm CO}}$ assuming $T = (T_{\rm b}^{^{12}{\rm CO}})_{\rm max}$:
$(T_{\rm b\star}^{^{12}{\rm CO}})_1$ based on the $^{12}$CO cube (Eq.~(\ref{eq:Tb*_Tb_T_1})) and $(T_{\rm b\star}^{^{12}{\rm CO}})_2$ based on the rescaled $^{13}$CO cube (Eq.~(\ref{eq:Tb*_Tb_T_2}));
we then combine $(T_{\rm b\star}^{^{12}{\rm CO}})_1$ and $(T_{\rm b\star}^{^{12}{\rm CO}})_2$ to obtain a best-estimate $T_{\rm b\star}^{^{12}{\rm CO}}$ (Eq.~(\ref{eq:Tb*_best})), from now on referred to as $T_{\rm b\star}^{\rm CO}$.
We find that $T_{\rm b\star}^{\rm CO}$ is nearly equal to the estimate with the smaller uncertainty, which is $(T_{\rm b\star}^{^{12}{\rm CO}})_1$ away from the CO peak and $(T_{\rm b\star}^{^{12}{\rm CO}})_2$ toward the CO peak.
This dichotomy results from the huge jump in the opacity correction to $T_{\rm b}^{^{12}{\rm CO}}$ toward the CO peak, which renders $(T_{\rm b\star}^{^{12}{\rm CO}})_1$ %%completely unreliable.
extremely uncertain.

The maps of the velocity-integrated $T_{\rm b\star}^{{\rm H}\textsc{i}}$ and $T_{\rm b\star}^{\rm CO}$ are displayed in the second row of Fig.~\ref{fig:G139_gasmaps}.
The opacity-corrected H{\sc i} map (left panel) is similar to, but more contrasted than, its observed counterpart (left panel in the first row).
It also differs by the emergence of a slight over-intensity along the lower-left boundary, which probably indicates the presence of a small, H{\sc i}-bright cloud.
The opacity-corrected $^{12}$CO and $^{13}$CO maps (not shown) are quite similar, except toward the CO peak, where only $T_{\rm b\star}^{^{13}{\rm CO}}$ is reliable.
Altogether, the combined opacity-corrected CO map (right panel in the second row) is very close to the opacity-corrected $^{12}$CO map, with only the CO peak taken from the rescaled opacity-corrected $^{13}$CO map.

Before jointly exploiting the $T_{\rm b\star}^{{\rm H}\textsc{i}}$ and $T_{\rm b\star}^{\rm CO}$ spectral cubes, we bring them to common angular and spectral resolutions and to a common $(l,b,v)$ grid.
We also bring the {\it Planck} maps of $I_{\rm d}$, $Q_{\rm d}$, and $U_{\rm d}$ to the common angular resolution and common $(l,b)$ grid.
Since the H{\sc i} cube has the lowest angular resolution (see  Table~\ref{tab:G139_inputmaps}), we adopt its angular resolution, $\delta \theta = 10.8'$, together with a pixel size at the (rounded) Nyquist limit, $\delta \theta _{\rm pix} = 4'$.
Accordingly, the $(l,b)$ grid needed to cover our $104' \times 104'$ G139 region possesses $26 \times 26$ pixels.
Clearly, this grid undersamples the CO and dust maps.
Along the $v$-axis, we retain a total velocity range $[-120,+20]\,{\rm km\,s^{-1}}$, which is broad enough to encompass all the true emission from H{\sc i} and CO, and we adopt a spectral resolution $\delta v = 0.3\,{\rm km\,s^{-1}}$, which is much finer than the spectral resolution of the H{\sc i} cube and roughly twice the spectral resolution of the CO cube.
This spectral oversampling of $T_{\rm b\star}^{{\rm H}\textsc{i}}$  makes it easier to extract meaningful Gaussian kinematic components with {\tt ROHSA}. The re-gridded maps of $I_{\rm d}$, $Q_{\rm d}$, and $U_{\rm d}$ and those of the velocity-integrated $T_{\rm b\star}^{{\rm H}\textsc{i}}$ and $T_{\rm b\star}^{\rm CO}$ are displayed in the second row of Fig.~\ref{fig:G139_planckmaps} and the third row of Fig.~\ref{fig:G139_gasmaps}, respectively.

\subsection{Derivation of the conversion factors}
\label{sec:application_conversion_factors}

For each gas tracer ${\rm A}$ (${\rm A} =$ H{\sc i} or CO), we need to determine the conversion factor from velocity-integrated opacity-corrected brightness temperature to dust intensity at 353\,GHz, $X^{\rm A}_{I_{\rm d}/T_{\rm b}}$ (see Eq.~(\ref{eq:conversion_brighttemp_totintensity})).
As explained in Sect.~\ref{sec:method_gas_tracers}, the best-fit values of $X^{{\rm H}\textsc{i}}_{I_{\rm d}/T_{\rm b}}$ and $X^{\rm CO}_{I_{\rm d}/T_{\rm b}}$, together with their uncertainties, are obtained by minimizing $\chi_{\rm r}^2$ in Eq.~(\ref{eq:chi_reduced}) through bootstrap sampling + MCMC simulations.

\begin{figure*}
\sidecaption
\centering
\includegraphics[width=12cm]{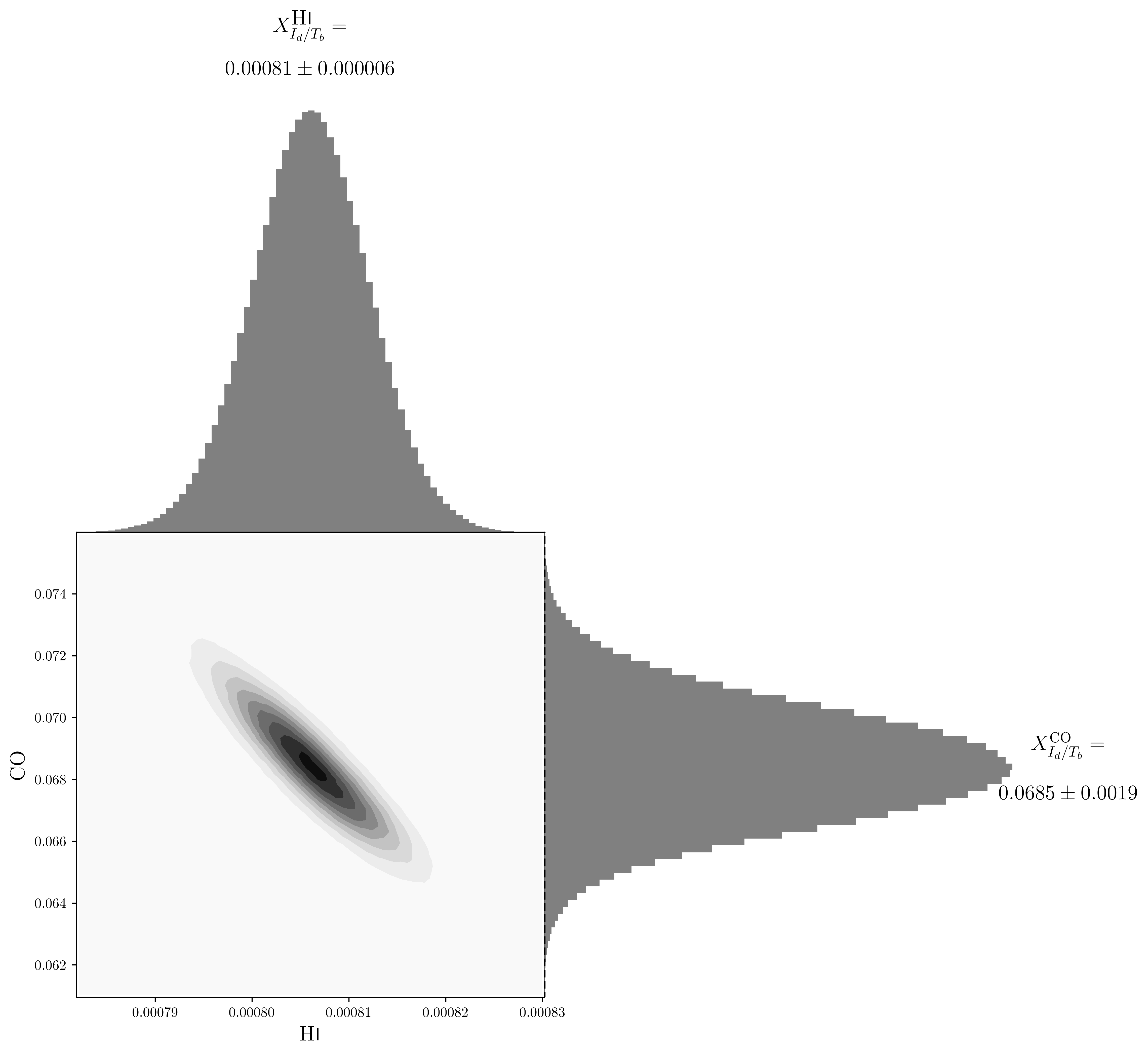}
\caption{
    Corner plot of the marginal (1D) and joint (2D) probability density functions of the conversion factors from velocity-integrated opacity-corrected brightness temperature to dust intensity at 353\,GHz, $X^{{\rm H}\textsc{i}}_{I_{\rm d}/T_{\rm b}}$ and $X^{\rm CO}_{I_{\rm d}/T_{\rm b}}$ [in ${\rm (MJy\,sr^{-1})\,(K\,km\,s^{-1})^{-1}}$], for the two gas tracers, H{\sc i} and CO, in the G139 region.
}
\label{fig:G139_cornerplot_conversionfactors}
\end{figure*}

We find that the best fit has $\chi_{\rm r} = 1.63$.
This small value of $\chi_{\rm r}$ indicates that our model is globally satisfactory, in the sense that the errors in the reconstruction of the dust intensity with Eq.~(\ref{eq:finalexpr_totintensity}) are only slightly larger than the total observational uncertainty, $\sigma_I$ (Eq.~(\ref{eq:sigma2_Id_tot})). 
The latter is dominated by $\sigma(I_{\rm d}^{{\rm H}\textsc{i}})$, which generally exceeds $\sigma(I_{\rm d}^{\rm obs})$ and $\sigma(I_{\rm d}^{\rm CO})$ by a factor $\sim 2 - 10$, except toward the CO peak, where $\sigma(I_{\rm d}^{\rm CO}) \sim 2 \ \sigma(I_{\rm d}^{{\rm H}\textsc{i}}) \sim 3.5 \ \sigma(I_{\rm d}^{\rm obs})$.
The dominant contribution to $\sigma(I_{\rm d}^{{\rm H}\textsc{i}})$, in turn, comes by far from the uncertainty in the spin temperature of H{\sc i} (see second term in the r.h.s. of Eq.~(\ref{eq:sigma2_Tb*})), 
while the dominant contribution to $\sigma(I_{\rm d}^{\rm CO})$ comes from measurement errors (see first terms in the r.h.s. of Eqs.~(\ref{eq:sigma2_Tb*_1}) and (\ref{eq:sigma2_Tb*_2})).

Displayed in Fig.~\ref{fig:G139_cornerplot_conversionfactors} is a corner plot of the marginal and joint probability density functions of $X^{{\rm H}\textsc{i}}_{I_{\rm d}/T_{\rm b}}$ and $X^{\rm CO}_{I_{\rm d}/T_{\rm b}}$.
Their best-fit values and standard deviations, written above the corresponding 1D histograms (upper-left and lower-right panels), are $X^{{\rm H}\textsc{i}}_{I_{\rm d}/T_{\rm b}} = (0.00081 \pm 0.000006)\,{\rm (MJy\,sr^{-1})\,(K\,km\,s^{-1})^{-1}}$ and $X^{\rm CO}_{I_{\rm d}/T_{\rm b}} = (0.0685 \pm 0.0019)\,{\rm (MJy\,sr^{-1})\,(K\,km\,s^{-1})^{-1}}$, respectively.
The shape of the joint distribution (lower-left panel) indicates that $X^{{\rm H}\textsc{i}}_{I_{\rm d}/T_{\rm b}}$ and $X^{\rm CO}_{I_{\rm d}/T_{\rm b}}$ are strongly anticorrelated.
This result can be explained by the fact that H{\sc i} and CO both have widespread spatial distributions, with significant overlap (see third row of Fig.~\ref{fig:G139_gasmaps}), such that an increase in the dust emission associated with one tracer must be accompanied by a decrease in the dust emission associated with the other tracer.

\begin{figure*}
\centering
\includegraphics[width=\textwidth]{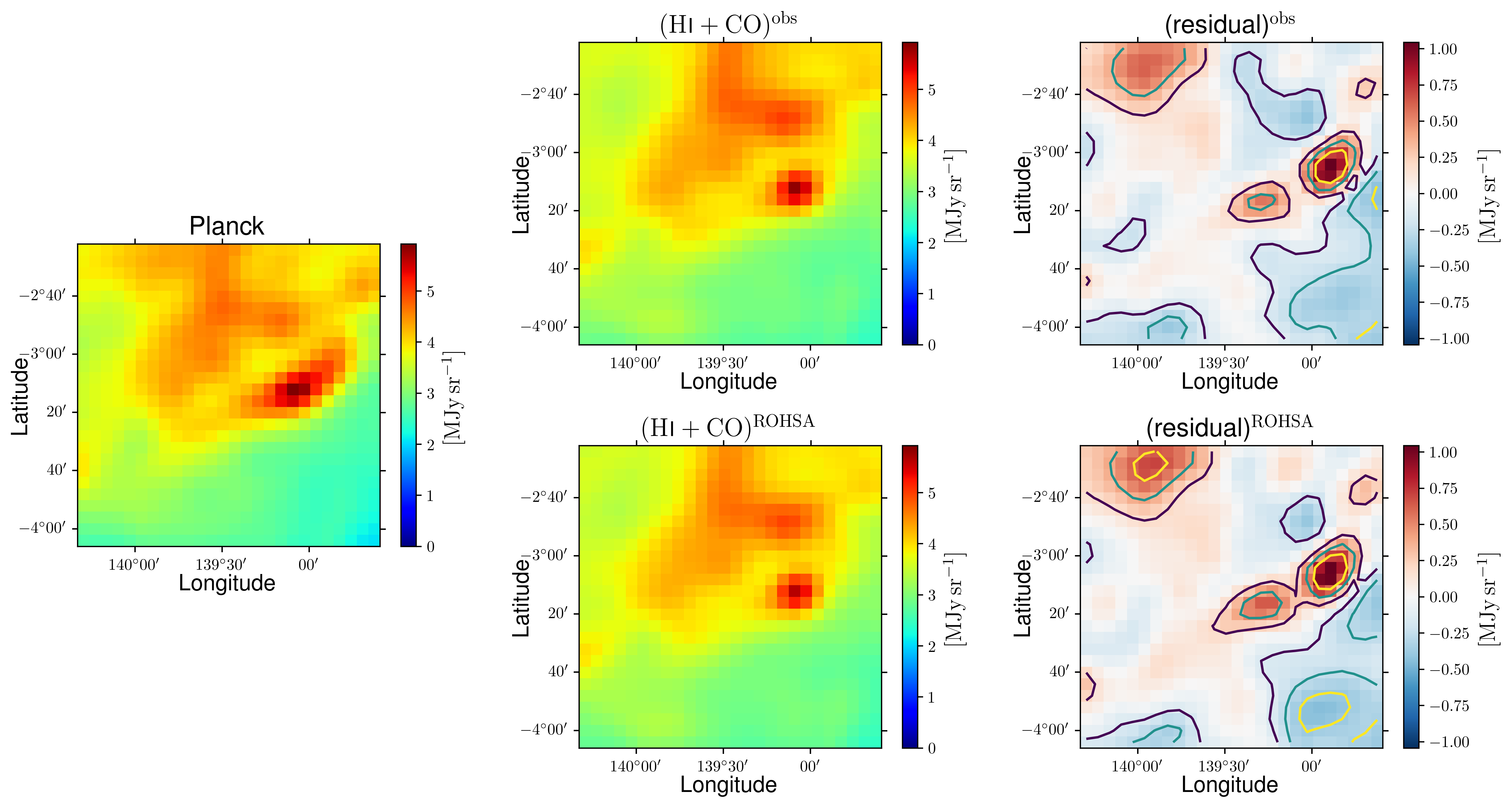}
\caption{
    Maps of the intensity, $I_{\rm d}$, of the dust emission at 353\,GHz toward the G139 region.
    {\it Left}: Observational map from {\it Planck}. 
    {\it Middle}: Best-fit maps reconstructed with the two gas tracers, H{\sc i} and CO, before ({\it top}) and after ({\it bottom}) application of the Gaussian decomposition algorithm {\tt ROHSA} to each gas tracer.
    {\it Right}: Maps of the residuals obtained by subtracting the respective reconstructed maps from the observational {\it Planck} map.
    The color code refers to the absolute residuals, whereas the contour lines follow the relative residuals.
}
\label{fig:G139_dustmaps_tracers}
\end{figure*}

The best-fit values of the conversion factors enable us to rescale the H{\sc i} and CO maps in the third row of Fig.~\ref{fig:G139_gasmaps} to dust intensity at 353\,GHz and thus obtain the best-fit maps of $I_{\rm d}^{{\rm H}\textsc{i}}$ and $I_{\rm d}^{\rm CO}$ as well as the best-fit map of the reconstructed dust intensity, $I_{\rm d}^{{\rm H}\textsc{i}} + I_{\rm d}^{\rm CO}$.
The latter is displayed in the top-middle panel of Fig.~\ref{fig:G139_dustmaps_tracers}, where it can be compared to the observational {\it Planck} map, %%$I_{\rm d}^{\it Planck}$,
$I_{\rm d}^{\rm obs}$, in the left panel.
Both maps look similar, and the bright regions in the {\it Planck} map are well reproduced.
The H{\sc i} medium provides the general dust-emission background, including its northward gradient, while the CO medium is responsible for the bright core to the right and for the weaker dust enhancement that runs obliquely north of the southeast-northwest diagonal.
Globally, the H{\sc i} and CO media account for $84.3\,\%$ and $15.7\,\%$, respectively, of the dust emission from the G139 region.

The map of the residuals, $\left[ I_{\rm d}^{\rm obs} - \left( I_{\rm d}^{{\rm H}\textsc{i}} + I_{\rm d}^{\rm CO} \right) \right]$,
is shown in the top-right panel of Fig.~\ref{fig:G139_dustmaps_tracers}.
The largest residuals are positive and arise on the northwest and southeast sides of the bright core (which itself is almost residual-free). 
These positive residuals result from the bright core being intrinsically more extended in the observational {\it Planck} map than in the opacity-corrected CO map.
They possibly reveal the presence of so-called dark gas, namely, gas that is undetected in H{\sc i} and CO \citep[e.g.,][]{grenier&ct_2005, Planck_XIX_2011}.
Large positive and negative residuals also arise in the upper-left and lower-right corners, respectively, i.e., in two regions with little CO emission and where the northward gradient of the opacity-corrected H{\sc i} emission remains too shallow to reproduce the observed dust-emission gradient.
The negative residual along the lower-left boundary coincides with 
the slight over-intensity appearing in the H{\sc i} map after opacity correction (see Sect.~\ref{sec:application_input}),
which suggests that the H{\sc i} emission could have been over-corrected.
Other residuals could also result from imperfect opacity corrections, following a poor estimation of the excitation temperature.
Alternatively, residuals could indicate that dust emission does not exactly follow the gas distribution -- in other words, that the conversion factors are not perfectly uniform.

It would be interesting to compare our best-fit values of the conversion factors $X^{{\rm H}\textsc{i}}_{I_{\rm d}/T_{\rm b}}$ and $X^{\rm CO}_{I_{\rm d}/T_{\rm b}}$ to previous estimates, but we did not find such estimates in the literature.
However, we found a number of estimates for the intermediate conversion factors $X^{{\rm H}\textsc{i}}_{I_{\rm d}/N_{\rm H}}$, $X^{{\rm H}\textsc{i}}_{N_{\rm H}/T_{\rm b}}$, $X^{^{12}{\rm CO}}_{I_{\rm d}/N_{\rm H}}$, and $X^{^{12}{\rm CO}}_{N_{\rm H}/T_{\rm b}}$ defined in Sect.~\ref{sec:method_gas_tracers}. 

Regarding H{\sc i}, the value of the conversion factor from velocity-integrated brightness temperature to hydrogen column density is generally taken as $X^{{\rm H}\textsc{i}}_{N_{\rm H}/T_{\rm b}} = 0.0182\,{\rm (10^{20}\,cm^{-2})\,(K\,km\,s^{-1})^{-1}}$ (from \cite{Wilson&rh_2013}), which is strictly valid in the optically thin case.
This value was used by \cite{Bailin_etal_2016} and by \cite{Planck_XIX_2011,Planck_XXIV_2011}, with, in the latter study, an opacity correction equivalent to that in our Eq.~(\ref{eq:Tb*_Tb_T}) with $T = 80\,{\rm K}$.
The two {\it Planck} papers also discussed the value of the conversion factor from hydrogen column density to dust intensity at 353\,GHz.
\cite{Planck_XIX_2011} obtained 
$X^{{\rm H}\textsc{i}}_{I_{\rm d}/N_{\rm H}} = (0.0484 \pm 0.0010)\,{\rm (MJy\,sr^{-1})\,(10^{20}\,cm^{-2})^{-1}}$ in their reference region defined by $|b| > 20^\circ$ and $N_{{\rm H}\textsc{i}} < 1.2 \times 10^{21}\,{\rm cm^{-2}}$, 
while \cite{Planck_XXIV_2011} derived values in the range $[0.021,0.071]\,{\rm (MJy\,sr^{-1})\,(10^{20}\,cm^{-2})^{-1}}$ for %%the local diffuse ISM 
local (low-velocity) clouds toward 14 high-latitude fields covering $\simeq 825\,{\rm deg}^2$ on the sky.
Combining the above values of $X^{{\rm H}\textsc{i}}_{N_{\rm H}/T_{\rm b}}$ and $X^{{\rm H}\textsc{i}}_{I_{\rm d}/N_{\rm H}}$ gives for the conversion factor from velocity-integrated (opacity-corrected in the second case) brightness temperature to dust intensity at 353\,GHz 
$X^{{\rm H}\textsc{i}}_{I_{\rm d}/T_{\rm b}} = 
X^{{\rm H}\textsc{i}}_{I_{\rm d}/N_{\rm H}} \ X^{{\rm H}\textsc{i}}_{N_{\rm H}/T_{\rm b}}
= (0.00088 \pm 0.000018)\,{\rm (MJy\,sr^{-1})\,(K\,km\,s^{-1})^{-1}}$ %%\citep{Planck_XIX_2011,Bailin_etal_2016} 
and $[0.00038,0.00129]\,{\rm (MJy\,sr^{-1})\,(K\,km\,s^{-1})^{-1}}$, %%\citep{Planck_XXIV_2011,Bailin_etal_2016}, 
respectively, where the uncertainties include only uncertainties in $X^{{\rm H}\textsc{i}}_{I_{\rm d}/N_{\rm H}}$, not uncertainties in $X^{{\rm H}\textsc{i}}_{N_{\rm H}/T_{\rm b}}$ associated with, for instance, opacity saturation.
Both estimates are consistent.
Our best-fit value, $X^{{\rm H}\textsc{i}}_{I_{\rm d}/T_{\rm b}} = (0.00081 \pm 0.000006)\,{\rm (MJy\,sr^{-1})\,(K\,km\,s^{-1})^{-1}}$, is slightly smaller than the value from \cite{Planck_XIX_2011}, and it falls right within the range from \cite{Planck_XXIV_2011}.

Regarding CO, \cite{remy&gmc_2017} derived values of $X_{\rm CO}$ and $({\tau_{353}/N_{\rm H}})^{\rm CO}$ for six nearby molecular clouds between longitude $l = 139^\circ$ and $191^\circ$ and between latitude $b = -3^\circ$ and $-56^\circ$.
Relying on 353\,GHz dust emission data and on $[0.4,100]\,{\rm GeV}$ $\gamma$-ray emission data, they obtained values of $X_{\rm CO}$ in the range $[0.86,1.73]\,{\rm (10^{20}\,cm^{-2})\,(K\,km\,s^{-1})^{-1}}$ and $[0.36,1.14]\,{\rm (10^{20}\,cm^{-2})\,(K\,km\,s^{-1})^{-1}}$, respectively.
They further obtained values of $({\tau_{353}/N_{\rm H}})^{\rm CO}$ in the range $[12.6,44]\,{\rm (10^{27}\,cm^{-2})^{-1}}$.
Noting that $X^{\rm CO}_{N_{\rm H}/T_{\rm b}} = 2 \, X_{\rm CO}$ and $I_{\rm d} = {\rm (4.76 \times 10^4\,MJy\,sr^{-1})} \ \tau_{353}$ for a dust temperature $T_{\rm d} = 19.7\,{\rm K}$ \citep{Planck_XI_2014}, we find that the corresponding ranges of $X^{\rm CO}_{I_{\rm d}/T_{\rm b}} = X^{\rm CO}_{I_{\rm d}/N_{\rm H}} \ X^{\rm CO}_{N_{\rm H}/T_{\rm b}}$
are $[0.10,0.73]\,{\rm (MJy\,sr^{-1})\,(K\,km\,s^{-1})^{-1}}$ and $[0.043,0.48]\,{\rm (MJy\,sr^{-1})\,(K\,km\,s^{-1})^{-1}}$, respectively.
These ranges include no opacity correction for the CO line, which might explain why they are so broad.
Our best-fit value, $X^{\rm CO}_{I_{\rm d}/T_{\rm b}} = (0.0685 \pm 0.0019)\,{\rm (MJy\,sr^{-1})\,(K\,km\,s^{-1})^{-1}}$ falls somewhat below the former range and within, though close to the lower end, of the latter range.
Finding a lower value of $X^{\rm CO}_{I_{\rm d}/T_{\rm b}}$ here is not surprising given that our opacity correction increases the brightness temperatures, and this increase must be offset by a decrease in the conversion factor.

\subsection{LoS decomposition into clouds}
\label{sec:application_4clouds}

\begin{figure*}
\centering
\includegraphics[width=\textwidth]{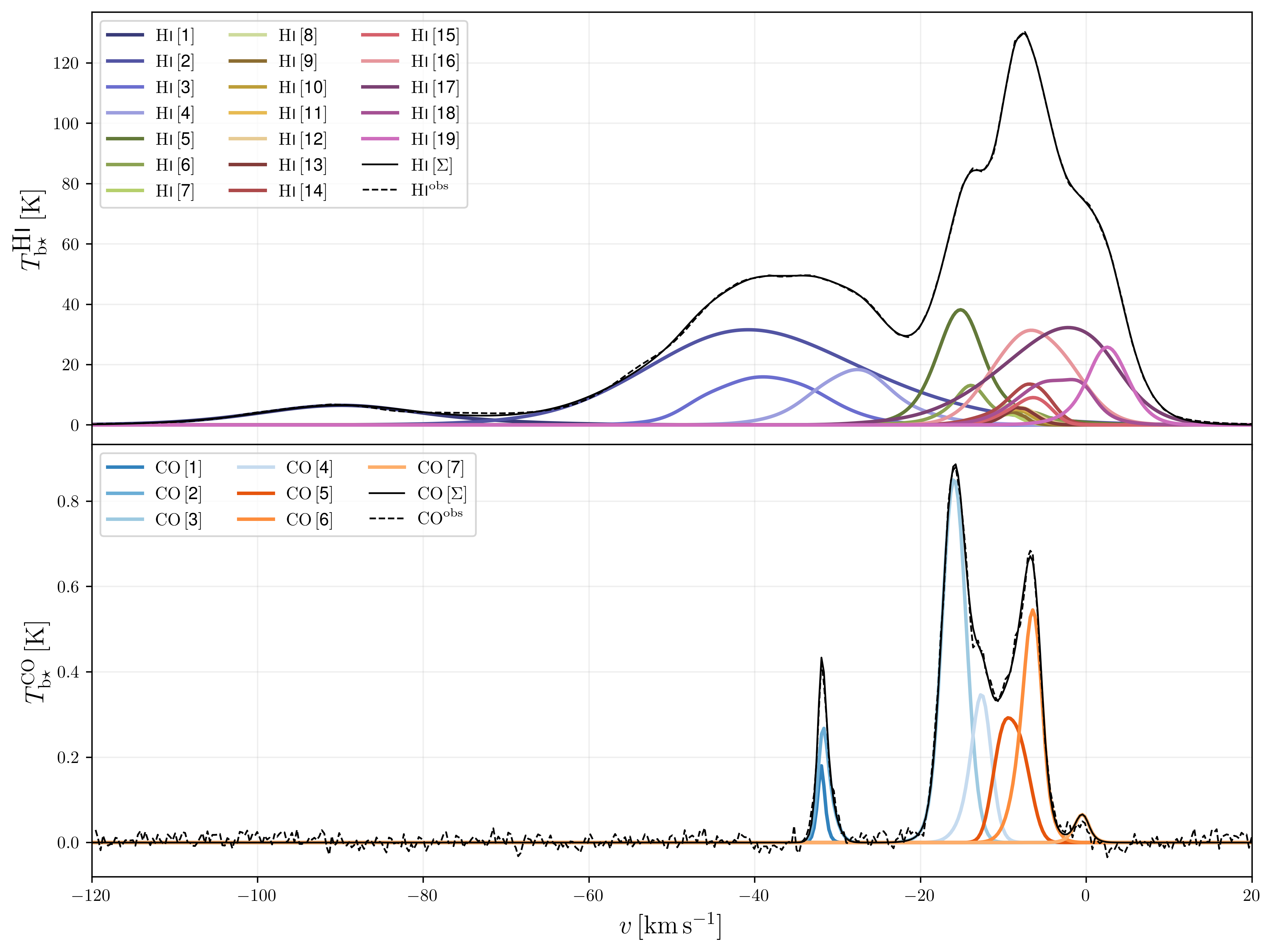}
\caption{
    Spectra of the opacity-corrected brightness temperatures, $T_{\rm b\star}^{{\rm H}\textsc{i}}$ ({\it top}) and $T_{\rm b\star}^{\rm CO}$ ({\it bottom}), averaged over the $26 \times 26$ pixels of the common $(l,b)$ grid, toward the G139 region.
    The observed spectra are plotted in black dashed line,
    the reconstructed spectra obtained after application of the Gaussian decomposition algorithm {\tt ROHSA} are plotted in black solid line,
    and the spectra of the individual Gaussian kinematic components, which are ordered (for each tracer) by increasing mean velocity, are plotted in color.
    The $T_{\rm b\star}^{{\rm H}\textsc{i}}$ and $T_{\rm b\star}^{\rm CO}$ spectra can be rescaled to dust intensity per unit velocity at 353\,GHz using the best-fit values of the conversion factors derived in Sect.~\ref{sec:application_conversion_factors} (see Fig.~\ref{fig:G139_cornerplot_conversionfactors}).
}
\label{fig:G139_tracers_components_spectra}
\end{figure*}

\subsubsection{{\tt ROHSA} decomposition}

Next, we apply the Gaussian decomposition algorithm {\tt ROHSA} (presented in Sect.~\ref{sec:method_clouds}) to the %%observational H{\sc i}, $^{12}$CO, and $^{13}$CO spectral cubes
observed spectral cubes of the opacity-corrected brightness temperatures, $T_{\rm b\star}$, of our two gas tracers, H{\sc i} and CO, separately.
The procedure yields 19 and 7 Gaussian kinematic components, respectively.
The spectra of these components, averaged over the $26 \times 26$ pixels of the common $(l,b)$ grid, are plotted in the top and bottom panels, respectively, of Fig.~\ref{fig:G139_tracers_components_spectra}.
For simplicity, the components of each tracer are ordered by increasing mean velocity.
Also plotted in Fig.~\ref{fig:G139_tracers_components_spectra} are the total reconstructed $T_{\rm b\star}^{{\rm H}\textsc{i}}$ and $T_{\rm b\star}^{\rm CO}$ spectra, obtained by summing the averaged spectra of their respective kinematic components (black solid lines), as well as the corresponding observed spectra (black dashed lines).

Comparing the reconstructed and observed spectra indicates that {\tt ROHSA} does on average a very good job at reconstructing the observed spectra.
Since we discarded the extracted components falling everywhere below twice the noise level, the reconstructed spectra are automatically free of the measurement noise apparent in the observational $^{12}$CO and $^{13}$CO cubes.

The H{\sc i} Gaussian decomposition (top panel of Fig.~\ref{fig:G139_tracers_components_spectra}) results in 19 components, which peak at various velocities between $\simeq -90$ and $+2\,{\rm km\,s^{-1}}$ and which, together, cover almost the entire velocity range $[-120,+20]\,{\rm km\,s^{-1}}$.
The H{\sc i} spectrum is dominated by a cluster of 15 components (H{\sc i}\,[5] - H{\sc i}\,[19]) peaking between $\simeq -15$ and $+2\,{\rm km\,s^{-1}}$.
Also prominent in the H{\sc i} spectrum is a strong and broad component (H{\sc i}\,[2]) centered at $v \simeq -40\,{\rm km\,s^{-1}}$.

The CO decomposition (bottom panel of Fig.~\ref{fig:G139_tracers_components_spectra}) leads to 7 components peaking between $\simeq -32$ and $0\,{\rm km\,s^{-1}}$.
The four dominant components (CO\,[3] - CO\,[6]) are clustered around $v \sim -10\,{\rm km\,s^{-1}}$, with peak velocities between $\simeq -16$ and $0\,{\rm km\,s^{-1}}$;
each of these components could easily be related to one of the 15 clustered H{\sc i} components.
The CO spectrum also contains a very weak component (CO\,[7]) centered at $v \simeq 0\,{\rm km\,s^{-1}}$ as well as two very close components (CO\,[1] and CO\,[2])
centered at $v \simeq -32\,{\rm km\,s^{-1}}$.
The latter probably form a single physical entity, which  was artificially split by the sharp transition between our two estimates of $T_{\rm b\star}^{^{12}{\rm CO}}$; they will naturally be recombined at the cloud-reconstruction step.
All the CO components are narrower than the H{\sc i} components.

\begin{figure*}
\centering
\includegraphics[width=\textwidth]{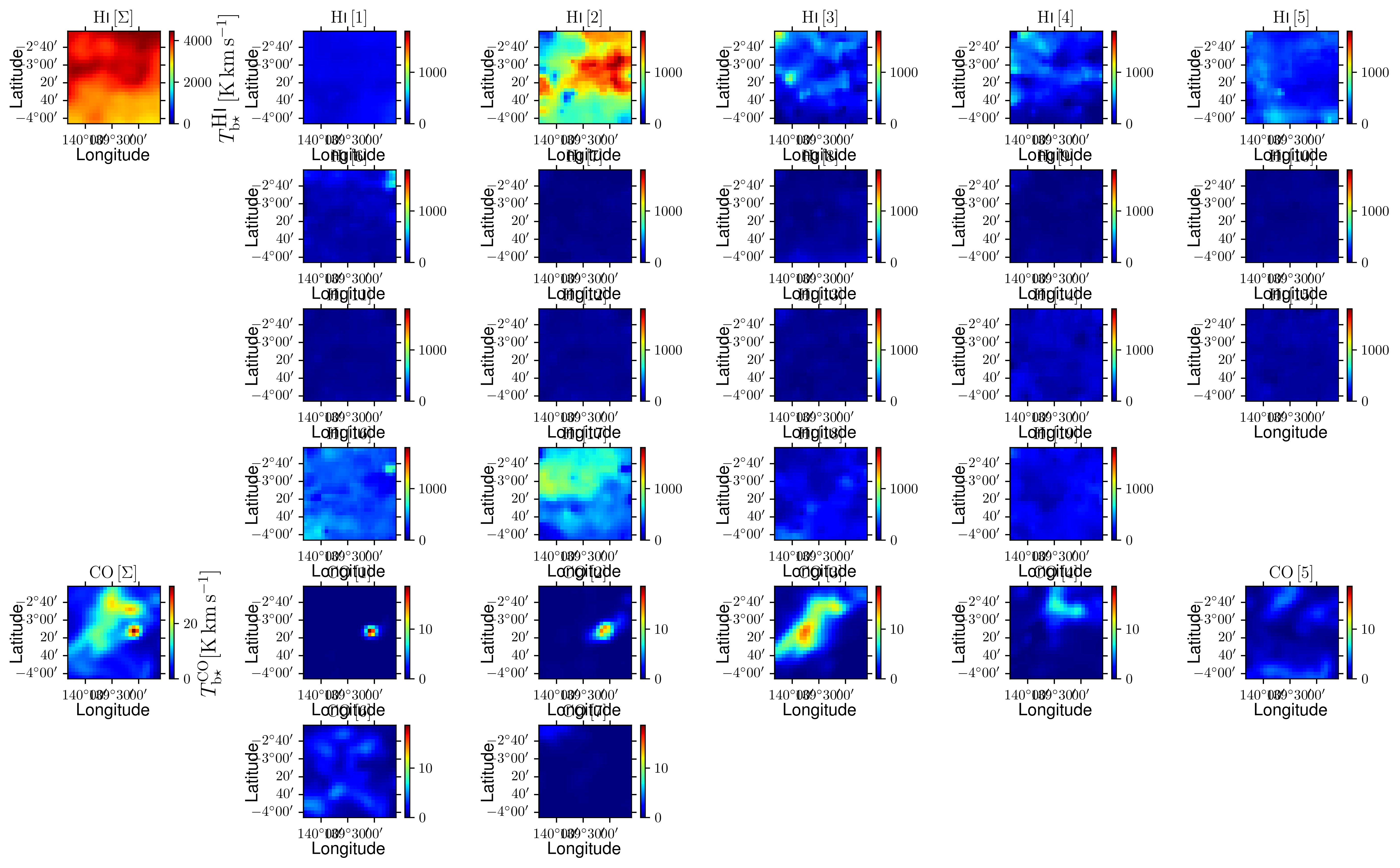}
\caption{
    Reconstructed maps of the opacity-corrected brightness temperatures, $T_{\rm b\star}^{{\rm H}\textsc{i}}$ ({\it top}) and $T_{\rm b\star}^{\rm CO}$ ({\it bottom}), integrated over velocity, toward the G139 region, after application of the Gaussian decomposition algorithm {\tt ROHSA} ({\it leftmost column}), together with the contributions from the individual Gaussian kinematic components ({\it columns~2-6}).
    The maps can be rescaled to dust intensity at 353\,GHz using the best-fit values of the conversion factors derived in Sect.~\ref{sec:application_conversion_factors} (see Fig.~\ref{fig:G139_cornerplot_conversionfactors}).
}
\label{fig:G139_tracers_components_maps}
\end{figure*}

The velocity-integrated maps of all the kinematic components of our two gas tracers are plotted in the top and bottom parts, respectively, of Fig.~\ref{fig:G139_tracers_components_maps},
along with the total reconstructed maps of both tracers (leftmost column), obtained by superposing their respective kinematic components.
The brightest H{\sc i} component is clearly H{\sc i}\,[2], which, we noted earlier, is also prominent in the average H{\sc i} spectrum (top panel of Fig.~\ref{fig:G139_tracers_components_spectra}).
The globally brightest CO component is CO\,[3], which leads to the highest peak in the average CO spectrum (bottom panel of Fig.~\ref{fig:G139_tracers_components_spectra}).
The locally brightest CO feature is the bright core to the right, which, we now see, can be attributed to the CO\,[1]\,-\,CO\,[2] pair at $v \simeq -32\,{\rm km\,s^{-1}}$.

The total reconstructed maps of both gas tracers are also plotted in the fourth row of Fig.~\ref{fig:G139_gasmaps}, where they can be compared to the corresponding re-gridded observational maps in the third row.
The comparison confirms the generally very good quality of the {\tt ROHSA} reconstruction, which we already noted when discussing the average spectra in Fig.~\ref{fig:G139_tracers_components_spectra}.
Furthermore, in the same way as the re-gridded observational maps of both gas tracers were rescaled %%to dust intensity at 353\,GHz 
(see last two sentences in caption of Fig.~\ref{fig:G139_gasmaps}) and combined into a map of the 353\,GHz dust intensity (top-middle panel of Fig.~\ref{fig:G139_dustmaps_tracers}), their reconstructed {\tt ROHSA} maps can be rescaled and combined into a reconstructed {\tt ROHSA} map of the 353\,GHz dust intensity (bottom-middle panel of Fig.~\ref{fig:G139_dustmaps_tracers}).
Here, too, the agreement with the pre-{\tt ROHSA} dust intensity map is very good.

It is interesting to compare our Gaussian decomposition of the CO cube to previous decompositions.
The G139 region is part of the $^{12}$CO cube constructed by \cite{Dame&hdt_2001} from a composite CO survey of the entire Galactic plane with the CfA \& Cerro Tololo 1.2\,m telescopes.
\cite{Straizis&l_2007} divided the G139 region of this cube into three layers with different RVs \citep[as defined earlier by][]{Digel_etal_1996}: the Gould Belt layer, with $v \simeq [-5,+10]\,{\rm km\,s^{-1}}$, the Camelopardalis (Cam) OB1 association layer, with $v \simeq [-20,-5]\,{\rm km\,s^{-1}}$, and the Perseus arm, with $v \simeq [-60,-30]\,{\rm km\,s^{-1}}$.
Using the Galactic rotation curve to convert RVs to distances, they estimated the distances to the clouds of the Gould Belt layer, the Cam OB1 layer, and the Perseus arm at $d \approx [150,300]\,{\rm pc}$, $[800,900]\,{\rm pc}$, and $[2,3]\,{\rm kpc}$, respectively.
Later, \cite{montillaud_etal_2015} identified three kinematic components, with RVs $v \approx -34\,{\rm km\,s^{-1}}$, $-16\,{\rm km\,s^{-1}}$, and $-9\,{\rm km\,s^{-1}}$, in the same $^{12}$CO data from \cite{Dame&hdt_2001} as those studied by \cite{Straizis&l_2007}.
The $v \approx -34\,{\rm km\,s^{-1}}$ component clearly peaks at the position of the bright core in the dust intensity map (top-left panel of Fig.~\ref{fig:G139_planckmaps}),
the $v \approx -16\,{\rm km\,s^{-1}}$ component has a more extended and more diffuse emission, and the $v \approx -9\,{\rm km\,s^{-1}}$ component is fainter.
Referring to the work of \cite{Straizis&l_2007}, they concluded that the $v \approx -34\,{\rm km\,s^{-1}}$ component must be part of the Perseus arm, while the $v \approx -16\,{\rm km\,s^{-1}}$ and $v \approx -9\,{\rm km\,s^{-1}}$ components must belong to the Cam OB1 layer.
In addition, their identification of the $v \approx -34\,{\rm km\,s^{-1}}$ component with the bright core led them to place the latter at a distance $d = 2.5\pm0.5\,{\rm kpc}$.
To make the link with our own study, the $v \approx -34\,{\rm km\,s^{-1}}$, $-16\,{\rm km\,s^{-1}}$, and $-9\,{\rm km\,s^{-1}}$ components of \cite{montillaud_etal_2015} presumably correspond, in our decomposition, to the CO\,[1]\,-\,CO\,[2] pair (centered at $v \simeq -32\,{\rm km\,s^{-1}}$) and to the two peaks formed by the clustered CO\,[3] - CO\,[6] components (at $v \simeq -16$ and $-7\,{\rm km\,s^{-1}}$).

\begin{figure*}
\centering
\includegraphics[width=\textwidth]{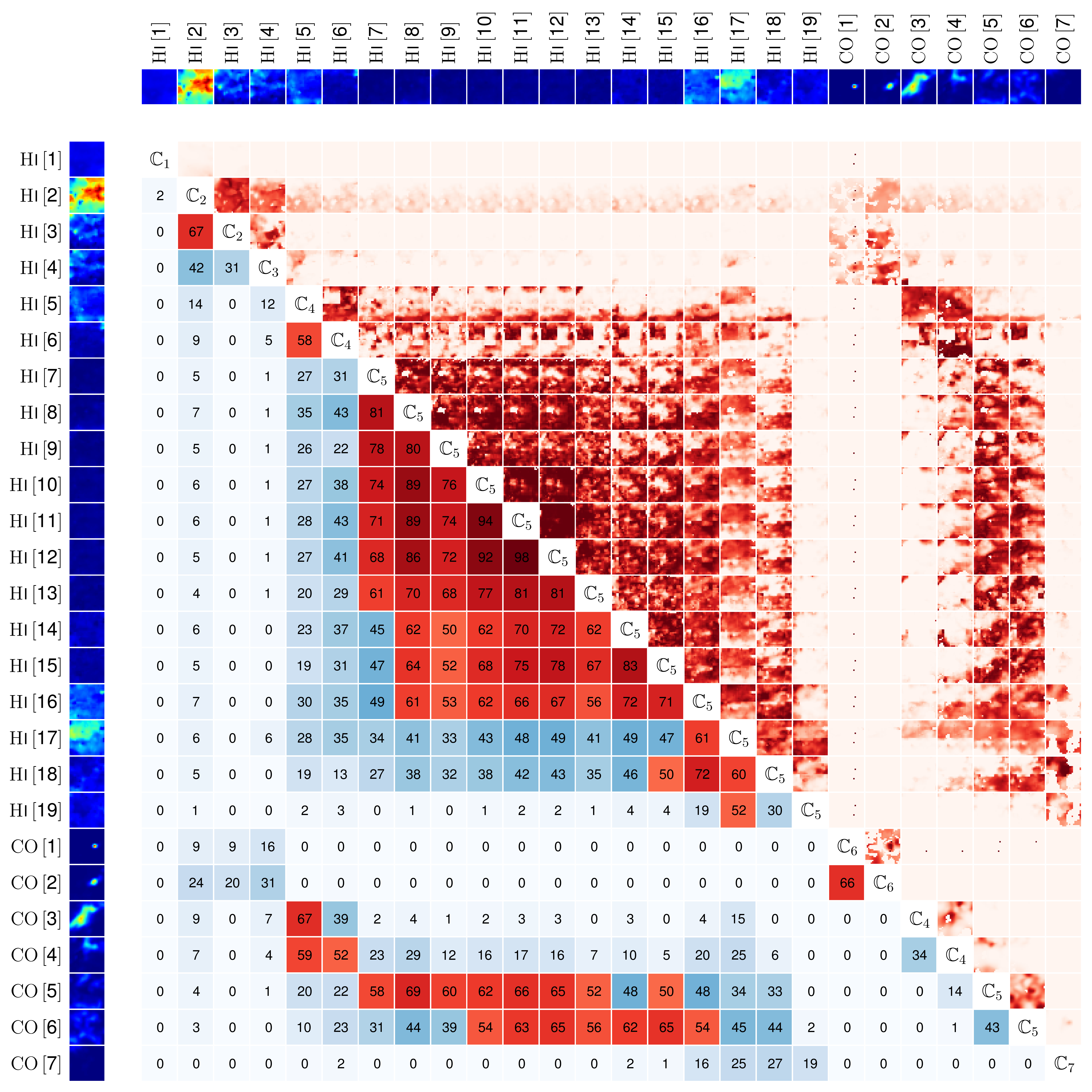}
\caption{
    Correlation coefficient, $\mathcal{C}_{jj'}$ (Eq.~(\ref{eq:correlation})), expressed in percent, for the 325 possible pairs of Gaussian kinematic components $j$ and $j'$ ($j \not= j'$) extracted with {\tt ROHSA} in the H{\sc i} and CO spectral cubes of the G139 region.
    The names of the 26 components, together with the mini-maps of their dust intensities, are shown along the top row and left column.
    Mini-maps of $\mathcal{C}_{jj'}$ are displayed in the upper-right half of the figure. 
    The weighted average values of $\mathcal{C}_{jj'}$ over the $26 \times 26$ pixels, $\langle \mathcal{C}_{jj'} \rangle$ (Eq.~(\ref{eq:correlation_average})), are indicated in the lower-left half, with red or light-blue shading according to whether $\langle \mathcal{C}_{jj'} \rangle \geq 50\,\%$ or $\langle \mathcal{C}_{jj'} \rangle < 50\,\%$ and with a level of shading increasing with increasing $\langle \mathcal{C}_{jj'} \rangle$.
    Pairs for which $\langle \mathcal{C}_{jj'} \rangle \geq 50\,\%$ have their two components $j$ and $j'$ combined and assigned to a same cloud.
    The 26 components are thus grouped into seven different clouds, ${\mathbb{C}}_1$, ..., ${\mathbb{C}}_7$, as indicated along the diagonal.
}
\label{fig:G139_correlations}
\end{figure*}

\subsubsection{Cloud reconstruction}

Moving on to the second step of the procedure described in Sect.~\ref{sec:method_clouds},
we examine the $n_{\rm G}^{\rm tot} = 19 + 7 = 26$ Gaussian kinematic components identified with {\tt ROHSA} and seek to group together, i.e., assign to a same cloud, those with similar velocity profiles.
To that end, we consider every possible pair of components $j$ and $j'$ ($j \not= j'$), compute the correlation coefficient $\mathcal{C}_{jj'}$ (Eq.~(\ref{eq:correlation})) at each of the $26 \times 26$ pixels of our grid, retain the weighted average value of $\mathcal{C}_{jj'}$ over all the pixels, $\langle \mathcal{C}_{jj'} \rangle$ (Eq.~(\ref{eq:correlation_average})), and compare $\langle \mathcal{C}_{jj'} \rangle$ to the velocity-coherence threshold, $\mathcal{C}_{\rm th}$, entering Eq.~(\ref{eq:threshold_correlation}).
Here, for the purpose of illustration, we adopt $\mathcal{C}_{\rm th} = 50\,\%$.
The results obtained for this value of $\mathcal{C}_{\rm th}$ are presented in Sects.~\ref{sec:application_4clouds} -- \ref{sec:application_Bfield}, while
an overview of the results obtained for all the integer values of $\mathcal{C}_{\rm th}$ between 0 and 100\,\% is provided in Sect.~\ref{sec:application_threshold}.
When $\langle \mathcal{C}_{jj'} \rangle \geq \mathcal{C}_{\rm th}$, we combine components $j$ and $j'$ and assign them to a same cloud.
The results of our correlation analysis for the 325 possible pairs of components are reported in Fig.~\ref{fig:G139_correlations}, with mini-maps of $\mathcal{C}_{jj'}$ displayed in the upper-right half and the derived values of $\langle \mathcal{C}_{jj'} \rangle$ indicated in the lower-left half.

It emerges from Fig.~\ref{fig:G139_correlations} that 83 pairs satisfy the condition $\langle \mathcal{C}_{jj'} \rangle \geq \mathcal{C}_{\rm th}$;
for better visibility, the corresponding small squares in the lower-left half of the figure are shaded in red (with increasing level of red as $\langle \mathcal{C}_{jj'} \rangle$ increases), as opposed to light blue for the other pairs.
Amongst the pairs with $\langle \mathcal{C}_{jj'} \rangle \geq \mathcal{C}_{\rm th}$, those having a component in common are further grouped together into a same multicomponent cloud.
The end result is a set of seven clouds, which we name ${\mathbb{C}}_1$, ${\mathbb{C}}_2$, ${\mathbb{C}}_3$, ${\mathbb{C}}_4$, ${\mathbb{C}}_5$, ${\mathbb{C}}_6$, and ${\mathbb{C}}_7$, and which enclose 1, 2, 1, 4, 15, 2, and 1 components, respectively.
The different components of each cloud can be retrieved from the labels with the cloud's name along the diagonal.

\begin{figure*}
\centering
\includegraphics[width=\textwidth]{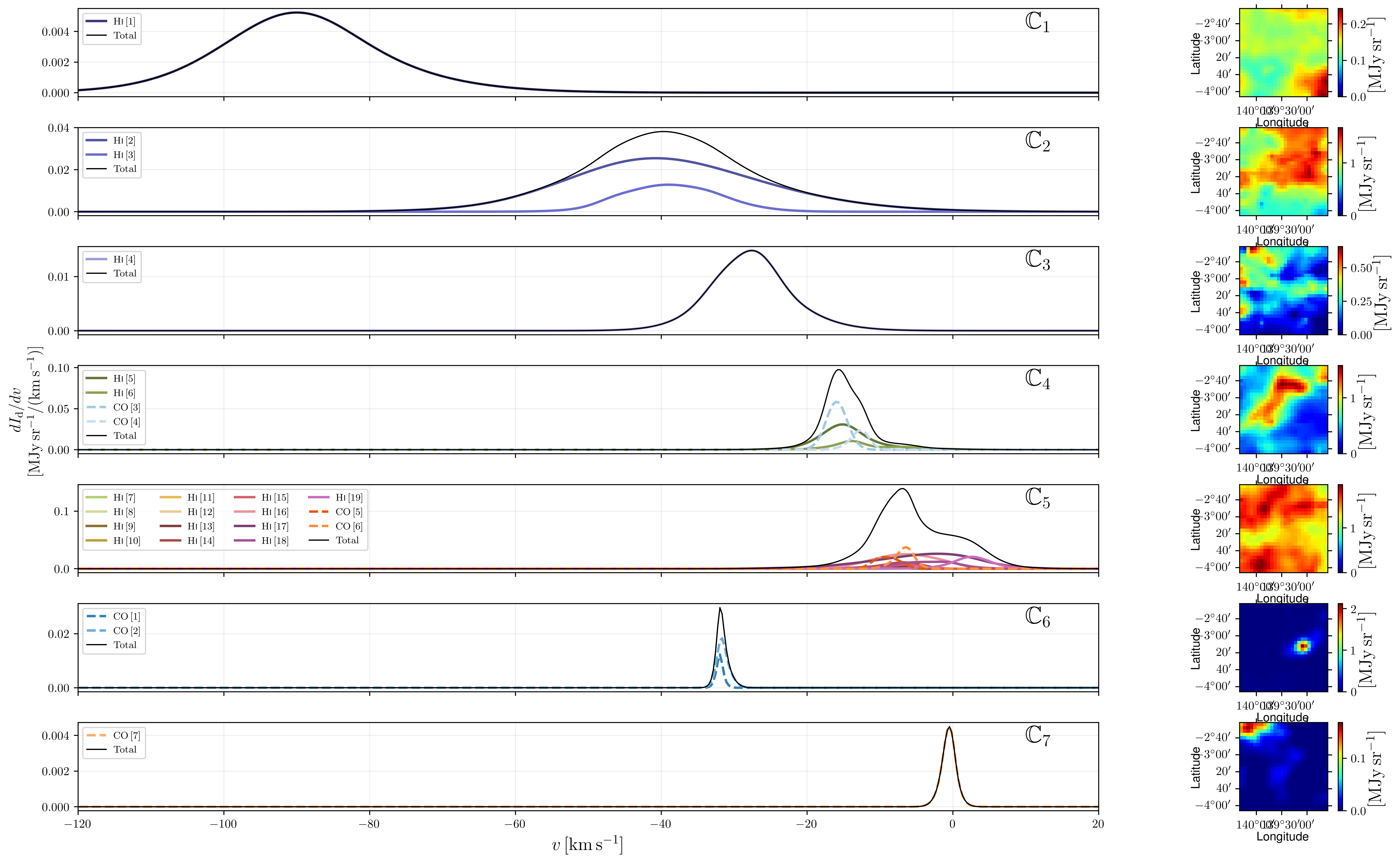}
\caption{
    Reconstructed spectra ({\it left}) and maps ({\it right}) of the 353\,GHz dust intensities, $I_{{\rm d},i}$, of the seven clouds, ${\mathbb{C}}_i$ ($i = 1, ..., 7$), identified in the G139 region.
    The spectra are averaged over the $26 \times 26$ pixels of the $(l,b)$ grid.
    The total spectra are plotted in black solid line, while the spectra of the individual kinematic components are plotted in color, using the same colors as in Fig.~\ref{fig:G139_tracers_components_spectra} and two different linestyles to distinguish the two gas tracers (solid for H{\sc i} and dashed for CO).
}
\label{fig:G139_clouds_components_spectra_maps}
\end{figure*}

Using the best-fit values of the conversion factors derived in Sect.~\ref{sec:application_conversion_factors} (see Fig.~\ref{fig:G139_cornerplot_conversionfactors}), we can now rescale the $T_{\rm b\star}$ spectral cubes of all the kinematic components to dust emission at 353\,GHz.
This common dust scale enables us to combine the different components of each cloud and thus obtain its dust emission spectral cube.
In the left part of Fig.~\ref{fig:G139_clouds_components_spectra_maps}, we plot the dust emission spectra of the seven clouds, averaged over the $26 \times 26$ pixels (black solid lines), together with the spectra of their individual components (color lines). 
The dust intensity maps of the seven clouds are displayed in the right column.

${\mathbb{C}}_1$, ${\mathbb{C}}_2$, and ${\mathbb{C}}_3$ are three purely atomic clouds, containing %%each 3 H{\sc i} components.
1, 2, and 1 H{\sc i} components, respectively.
${\mathbb{C}}_1$ is faint, with an enhancement in the southwest corner;
${\mathbb{C}}_2$ is much brighter, especially in the northwest part;
${\mathbb{C}}_3$ is faint, knotty, and mostly confined to the northeast corner.
${\mathbb{C}}_1$ and ${\mathbb{C}}_2$ both cover a rather broad velocity range, which could perhaps indicate that they are quite extended along the LoS.

${\mathbb{C}}_4$ and ${\mathbb{C}}_5$ are two mixed atomic-molecular clouds.
${\mathbb{C}}_4$ encloses an oblique, elongated CO structure (2 CO components), partially surrounded by an extended H{\sc i} envelope (2 H{\sc i} components).
${\mathbb{C}}_5$ is the richest cloud, with 13 H{\sc i} and 2 CO components, which together fill the entire region.
In both clouds, H{\sc i} appears to surround CO in velocity.

${\mathbb{C}}_6$ and ${\mathbb{C}}_7$ are two purely molecular clouds, with two very close CO components for ${\mathbb{C}}_6$ and a single CO component for ${\mathbb{C}}_7$.
${\mathbb{C}}_6$ is bright and localized both in the sky and in velocity;
it clearly corresponds to the bright core in the observational dust intensity map (top-left panel of Fig.~\ref{fig:G139_planckmaps}).
${\mathbb{C}}_7$ is much fainter, confined to the northeast corner, and also localized in velocity.

Most of the dust emission from the G139 region arises in ${\mathbb{C}}_5$ ($\simeq 42\,\%$), ${\mathbb{C}}_2$ ($\simeq 30\,\%$), and ${\mathbb{C}}_4$ ($\simeq 18\,\%$), with a small, but locally high contribution from ${\mathbb{C}}_6$ ($\simeq 1.4\,\%$).
${\mathbb{C}}_5$ and ${\mathbb{C}}_2$ account for most of the H{\sc i} emission (third panel in the left column of Fig.~\ref{fig:G139_gasmaps}), while ${\mathbb{C}}_4$ and ${\mathbb{C}}_6$ account for most of the CO emission (right column):
${\mathbb{C}}_4$ produces the oblique band north of the southeast-northwest diagonal with the two bright spots at its north end, and ${\mathbb{C}}_6$ is responsible for the high CO peak to the right.

\subsection{Derivation of the magnetic field orientation in each cloud}
\label{sec:application_Bfield}

Our LoS decomposition of the 353\,GHz {\it Planck} dust intensity map of the G139 region in Sect.~\ref{sec:application_4clouds} led to the identification of seven clouds, ${\mathbb{C}}_1$, ..., ${\mathbb{C}}_7$.
For each cloud ${\mathbb{C}}_i$, we obtained a map of the dust intensity, $I_{{\rm d},i}$ (right column of Fig.~\ref{fig:G139_clouds_components_spectra_maps}).
It now remains to determine the magnetic field orientation in each cloud, or, equivalently, its polarization fraction, $\overline{p}_{{\rm d},i}$, and polarization angle, $\overline{\psi}_{{\rm d},i}$.
Following the procedure described in Sect.~\ref{sec:method_Bfield}, we derive the best-fit values of $\overline{p}_{{\rm d},i}$ and $\overline{\psi}_{{\rm d},i}$, together with their uncertainties, $\sigma(\overline{p}_{{\rm d},i})$ and $\sigma(\overline{\psi}_{{\rm d},i})$, by minimizing $\chi_{\rm r}^2$ in Eq.~(\ref{eq:chi_reduced_QU}) through MCMC simulations.
As a prior, we require $\overline{p}_{{\rm d},i} \le 23\,\%$, as suggested by \cite{Planck_XX_2015} (see Appendix~\ref{sec:polfrac}).

\begin{table}
\centering
\caption{
    List of the seven clouds, ${\mathbb{C}}_i$ ($i = 1, ..., 7$), identified toward the G139 region.
}
\begin{threeparttable}
%Tables with captions and notes all the same width. Provides a scheme for tables that have a structured note section, after the caption.
\begin{tabular}{ll|lll|lll}
\toprule\toprule
\!\!\!Cloud & \!\!Kinematic components &
$\overline{p}_{{\rm d},i}$ [\%]\tablefootmark{a} \qquad \qquad & \!\!$\overline{\psi}_{{\rm d},i}$ [deg]\tablefootmark{b} \\
\midrule
${\mathbb{C}}_1$ & \!\!H{\sc i}\,[1] &
$22.9 \pm 0.1$ & \!\!$5.2 \pm 1.3$\!\!\!\! \\
${\mathbb{C}}_2$ & \!\!H{\sc i}\,[2-3] &
$6.4 \pm 0.2$ & \!\!$16.8 \pm 0.8$\!\!\!\! \\
${\mathbb{C}}_3$ & \!\!H{\sc i}\,[4] &
$14.2 \pm 0.4$ & \!\!$89.4 \pm 0.9$\!\!\!\! \\
${\mathbb{C}}_4$ & \!\!H{\sc i}\,[5-6], CO\,[3-4] &
\framebox{$2.2 \pm 0.2$} & \!\!\framebox{$46.3 \pm 1.9$}\!\!\!\! \\
${\mathbb{C}}_5$ & \!\!H{\sc i}\,[7-19], CO\,[5-6] &
$17.9 \pm 0.1$ & \!\!$-3.4 \pm 0.2$\!\!\!\! \\
${\mathbb{C}}_6$ & \!\!CO\,[1-2] &
\framebox{$\mathbf{5.8 \pm 0.3}$} & \!\!\framebox{$\mathbf{66.1 \pm 1.7}$}\!\!\!\! \\
${\mathbb{C}}_7$ & \!\!CO\,[7] &
$22.5 \pm 0.5$ & \!\!$-21.2 \pm 3.4$\!\!\!\! \\
\bottomrule\bottomrule
\end{tabular}
%\begin{tablenotes}
%\end{tablenotes}
\end{threeparttable}
\tablefoot{
\tablefoottext{a}{Best-fit value and standard deviation of the polarization fraction of cloud ${\mathbb{C}}_i$.}
\tablefoottext{b}{Best-fit value and standard deviation of the polarization angle of cloud ${\mathbb{C}}_i$.} \\
The boxed values pertain to the two most interesting molecular clouds: the bright CO core, ${\mathbb{C}}_6$ (bold style), and the oblique, elongated cloud, ${\mathbb{C}}_4$ (normal style).
}
\label{tab:G139_polar}
\end{table}

The best fit has $\chi_{\rm r} = 3.17$. This fairly large value indicates that our model composed of seven clouds with uniform polarization parameters ($\overline{p}_{{\rm d},i}$ and $\overline{\psi}_{{\rm d},i}$) does not reproduce the {\it Planck} polarization maps within the total "observational uncertainties", $\sigma_Q$ and $\sigma_U$, defined below Eq.~(\ref{eq:chi_reduced_QU}).
These uncertainties are dominated by "decomposition errors" in $Q_{\rm d}^{\rm mod}$ and $U_{\rm d}^{\rm mod}$ arising from decomposition errors in the cloud intensities (second term in the r.h.s of Eqs.~(\ref{eq:sigma2_Qd_tot}) -- (\ref{eq:sigma2_Ud_tot})), which generally exceed measurement errors in $Q_{\rm d}^{\rm obs}$ and $U_{\rm d}^{\rm obs}$ (first term) by a factor $\sim 2-3$.
The main reason for the fairly large value of $\chi_{\rm r}$ is that the polarization parameters of our seven clouds are actually not uniform.
Allowing for large-scale variations in these parameters would almost certainly improve the situation.
It could also be that some of our clouds do not form magnetically coherent structures, in the sense that they are actually composed of two or more magnetically distinct regions, with different polarization parameters.

The best-fit values of the free parameters, $\overline{p}_{{\rm d},i}$ and $\overline{\psi}_{{\rm d},i}$, together with their standard deviations, are listed in Table~\ref{tab:G139_polar}.
It appears that the polarization fractions vary widely, from $\overline{p}_{{\rm d},4} \simeq 2\,\%$ to $\overline{p}_{{\rm d},1}$, $\overline{p}_{{\rm d},7}$ hitting our imposed upper limit of $23\,\%$.
Clearly, ${\mathbb{C}}_7$ is trying to make up for the missing matter in the northeast corner (see positive residual in the bottom-right panel of Fig.~\ref{fig:G139_dustmaps_tracers}), while ${\mathbb{C}}_1$ is trying to contribute to the enhanced LoS-averaged polarization fraction in the south (see left panel in the fourth row of Fig.~\ref{fig:G139_dustmaps_tracers_polar} below).
The polarization angles also vary widely, from $|\overline{\psi}_{{\rm d},1}|$, $|\overline{\psi}_{{\rm d},5}|\lesssim 5^\circ$ to $|\overline{\psi}_{{\rm d},3}|\simeq 90^\circ$.
Yet most of the matter is contained within clouds with small polarization angles, consistent with the small values of the LoS-averaged polarization angle measured by {\it Planck} (left panel in the bottom row of Fig.~\ref{fig:G139_dustmaps_tracers_polar}).

\begin{figure*}
\centering
\includegraphics[width=\textwidth]{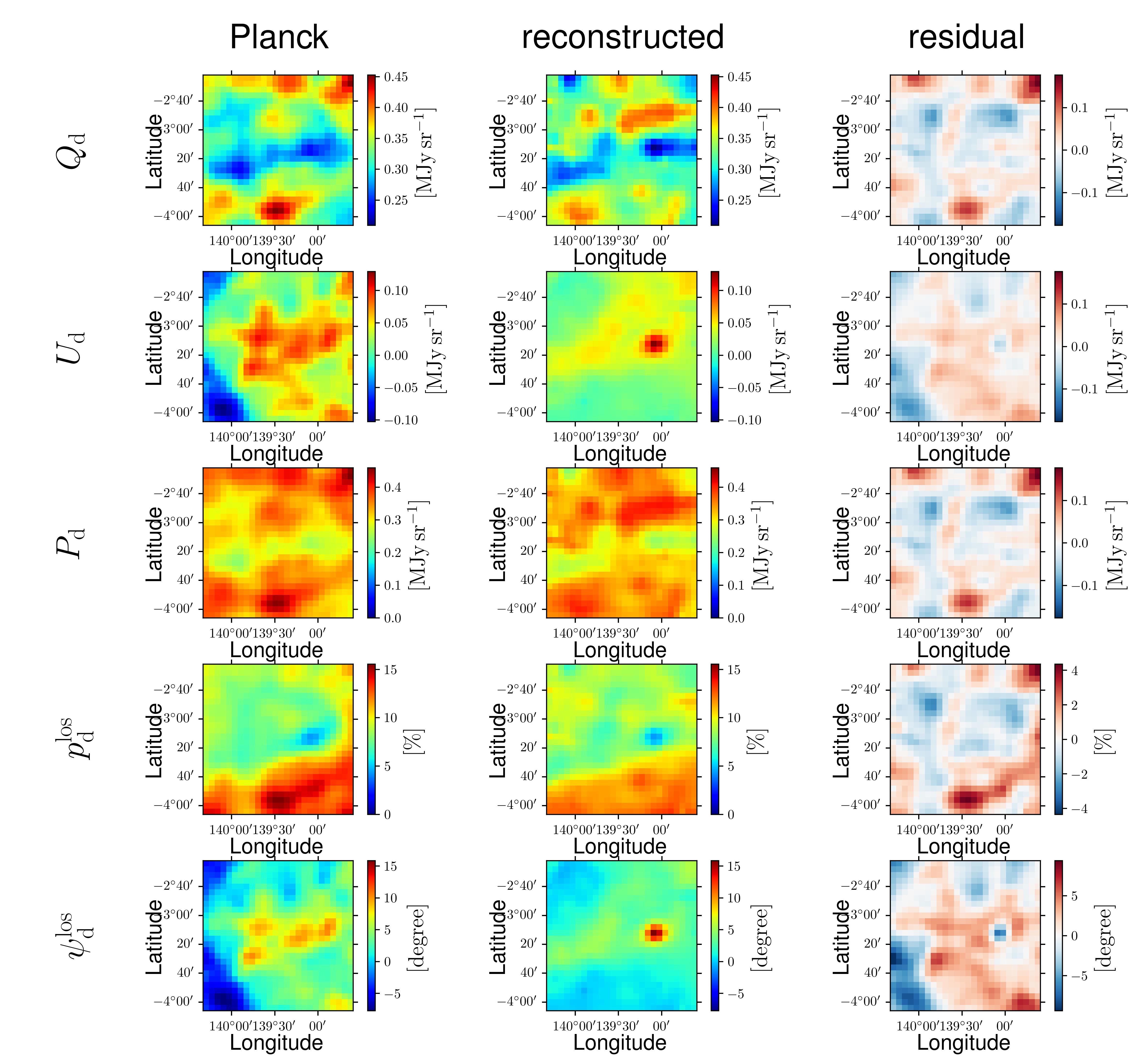}
\caption{
    Maps of the two Stokes parameters for linear polarization, $Q_{\rm d}$ and $U_{\rm d}$, the polarized intensity, $P_{\rm d}$, the LoS-averaged polarization fraction, $p_{\rm d}^{\rm los}$, and the LoS-averaged polarization angle, $\psi_{\rm d}^{\rm los}$, of the dust emission at 353\,GHz toward the G139 region.
    {\it Left}: Observational maps from {\it Planck}. 
    {\it Middle}: Best-fit maps reconstructed with our seven clouds, ${\mathbb{C}}_1$, ..., ${\mathbb{C}}_7$.
    {\it Right}: Maps of the residuals obtained by subtracting the respective reconstructed maps from the observational {\it Planck} maps.
}
\label{fig:G139_dustmaps_tracers_polar}
\end{figure*}

We can now use the best-fit maps of $I_{{\rm d},i}$ (right column of Fig.~\ref{fig:G139_clouds_components_spectra_maps}) in conjunction with the best-fit values of $\overline{p}_{{\rm d},i}$ and $\overline{\psi}_{{\rm d},i}$ (third and fourth columns of Table~\ref{tab:G139_polar}) to compute the Stokes parameters $Q_{{\rm d},i}$ and $U_{{\rm d},i}$ of each cloud ${\mathbb{C}}_i$ (Eqs.~(\ref{eq:stokesQ_cloud}) and (\ref{eq:stokesU_cloud})) as well as the resulting Stokes parameters, $Q_{\rm d}^{\rm mod}$ and $U_{\rm d}^{\rm mod}$ (Eqs.~(\ref{eq:stokesQ_sum}) and (\ref{eq:stokesU_sum})), and the associated polarized intensity, $P_{\rm d}^{\rm mod}$ (Eq.~(\ref{eq:polintensity_real})), LoS-averaged polarization fraction, $p_{\rm d}^{\rm los,mod}$ (Eq.~(\ref{eq:polfrac})), and LoS-averaged polarization angle, $\psi_{\rm d}^{\rm los,mod}$ (Eq.~(\ref{eq:polangle})).
The best-fit maps of $Q_{\rm d}^{\rm mod}$, $U_{\rm d}^{\rm mod}$, $P_{\rm d}^{\rm mod}$, $p_{\rm d}^{\rm los,mod}$, and $\psi_{\rm d}^{\rm los,mod}$, are displayed in the middle column of Fig.~\ref{fig:G139_dustmaps_tracers_polar}, where they can be compared to the observational {\it Planck} maps in the left column.
The observational LoS-averaged polarization fraction, $p_{\rm d}^{\rm los,obs}$, was debiased using the modified asymptotic (MAS) estimator \citep{Planck_XII_2020}. 
The maps of the residuals, i.e., the differences between the observational and reconstructed maps, are shown in the right column.

The map of $P_{\rm d}^{\rm obs}$ (left panel in the third row of Fig.~\ref{fig:G139_dustmaps_tracers_polar}) is globally more uniform than the map of $I_{\rm d}^{\rm obs}$ (left panel of Fig.~\ref{fig:G139_dustmaps_tracers}).
This is because brighter [fainter] regions are generally less [more] polarized, as can be seen by comparing the maps of $I_{\rm d}^{\rm obs}$ (left panel of Fig.~\ref{fig:G139_dustmaps_tracers}) and $p_{\rm d}^{\rm los,obs}$ (left panel in the fourth row of Fig.~\ref{fig:G139_dustmaps_tracers_polar}).
The map of $P_{\rm d}^{\rm obs}$ actually exhibits two faint spots along a central horizontal band.
The east faint spot and the westernmost part of the west faint spot correspond to regions that are both fainter and less polarized, whereas the easternmost part of the west faint spot is bright and weakly polarized.
Our model does a good job at reproducing the map of $P_{\rm d}^{\rm obs}$, including the two faint spots (see middle panel in the third row of Fig.~\ref{fig:G139_dustmaps_tracers_polar}).
The similarity between the maps of $P_{\rm d}^{\rm mod}$ and $I_{{\rm d},5}$ (fifth panel in the right column of Fig.~\ref{fig:G139_clouds_components_spectra_maps}) suggest that the bright and strongly polarized ($\overline{p}_{{\rm d},5} \simeq 18\,\%$) cloud ${\mathbb{C}}_5$ provides a dominant contribution to $P_{\rm d}^{\rm mod}$.

The map of $p_{\rm d}^{\rm los,obs}$ (left panel in the fourth row of Fig.~\ref{fig:G139_dustmaps_tracers_polar}) is roughly divided into a strongly polarized region south and a more weakly polarized region north, with a distinct low-polarization spot (blue region).
All these features are nicely reproduced in our model (middle panel), which also provides a natural explanation for the low-polarization spot.
This spot arises precisely at the location of ${\mathbb{C}}_6$, which is very bright and has a polarization angle very different from those of the other bright clouds, ${\mathbb{C}}_5$ and ${\mathbb{C}}_2$, detected along its LoS (see fourth column of Table~\ref{tab:G139_polar}).
As a result, ${\mathbb{C}}_6$ contributes constructively to $I_{\rm d}^{\rm mod}$, leading to a bright spot in the map of $I_{\rm d}^{\rm mod}$ (Fig.~\ref{fig:G139_dustmaps_tracers}), and contributes destructively to $P_{\rm d}^{\rm mod}$, leading to a dip in the map of $P_{\rm d}^{\rm mod}$ and a low-polarization spot in the map of $p_{\rm d}^{\rm los,mod}$ (Fig.~\ref{fig:G139_dustmaps_tracers_polar}).

The map of $\psi_{\rm d}^{\rm los,obs}$ (left panel in the fifth row of Fig.~\ref{fig:G139_dustmaps_tracers_polar}) is fairly homogeneous, with $|\psi_{\rm d}^{\rm los,obs}| \lesssim 10^\circ$ everywhere.
The four large-$\psi_{\rm d}^{\rm los,obs}$ spots (orange regions) near the middle are associated with low $p_{\rm d}^{\rm los,obs}$, which most likely indicates the presence of depolarizing clouds along the LoS.
The reason why the low-polarization spot in the map of $p_{\rm d}^{\rm los,obs}$ is not associated with a particularly large $\psi_{\rm d}^{\rm los,obs}$ (it actually sits between two large-$\psi_{\rm d}^{\rm los,obs}$ spots) is probably because the putative depolarizing cloud has a polarization angle relatively close to $90^\circ$.\footnote{A cloud with parameters $I_{{\rm d},1}$, $p_{{\rm d},1}$, $\psi_{{\rm d},1}$ against a brighter emission background with parameters $I_{{\rm d},0}$, $p_{{\rm d},0}$, $\psi_{{\rm d},0} = 0^\circ$ causes depolarization and rotation of the polarization orientation.
The resulting LoS-averaged polarization fraction and angle are given by
\begin{equation}
p_{\rm d}^{\rm los} = 
\frac
{\sqrt{
P_{{\rm d},0}^2 + P_{{\rm d},1}^2 + 2 \ P_{{\rm d},0} \ P_{{\rm d},1} \ \cos (2\psi_{{\rm d},1})
}}
{I_{{\rm d},0} + I_{{\rm d},1}}
\end{equation}
and
\begin{equation}
\tan (2\psi_{\rm d}^{\rm los}) = 
\frac
{P_{{\rm d},1} \ \sin (2\psi_{{\rm d},1})}
{P_{{\rm d},0} + P_{{\rm d},1} \ \cos (2\psi_{{\rm d},1})} \ ,
\end{equation}
with $P_{{\rm d},0} = (I_{{\rm d},0} \, p_{{\rm d},0})$ and $P_{{\rm d},1} = (I_{{\rm d},1} \, p_{{\rm d},1})$.
The maximum rotation of the polarization orientation occurs for
\begin{equation}
\cos (2\psi_{{\rm d},1}) = - \frac{P_{{\rm d},1}}{P_{{\rm d},0}} \ ,
\end{equation}
which, in the case $P_{{\rm d},1} \ll P_{{\rm d},0}$, reduces to $|\psi_{{\rm d},1}| \simeq 45^\circ$.
}
The model map of $\psi_{\rm d}^{\rm los,mod}$ (middle panel) is even more homogeneous than its observational counterpart, except toward ${\mathbb{C}}_6$, where $\psi_{\rm d}^{\rm los,mod}$ reaches $\simeq 16^\circ$.
The homogeneous, low-$|\psi_{\rm d}^{\rm los,mod}|$ background can be mostly attributed to the widespread, bright, and strongly polarized cloud ${\mathbb{C}}_5$, which dominates the polarized emission and has a small polarization angle ($\overline{\psi}_{{\rm d},5} \simeq -3^\circ$).
The jump in $\psi_{\rm d}^{\rm los,mod}$ toward ${\mathbb{C}}_6$ can be explained by this very bright and moderately polarized cloud having a much larger polarization angle ($\overline{\psi}_{{\rm d},6} \simeq 66^\circ$).
The discrepancies between $\psi_{\rm d}^{\rm los,obs}$ and $\psi_{\rm d}^{\rm los,mod}$ are mostly due to our assumption that each cloud has a uniform polarization angle.
In particular, the behavior of $\psi_{\rm d}^{\rm los,obs}$ in the vicinity of ${\mathbb{C}}_6$ suggests that the true polarization angle of ${\mathbb{C}}_6$ has values closer to $90^\circ$ in the bright center of the cloud (hence a small impact on $\psi_{\rm d}^{\rm los,obs}$) and values closer to $45^\circ$ on either side (hence a more significant increase of $\psi_{\rm d}^{\rm los,obs}$).

Once we have derived the best-fit values of the polarization fraction, $\overline{p}_{{\rm d},i}$, and polarization angle, $\overline{\psi}_{{\rm d},i}$, of every cloud ${\mathbb{C}}_i$, we can, in principle, obtain the best-fit orientation of its internal magnetic field, $\overline{\boldvec{B}}_i$, defined by the orientation angle of $\overline{\boldvec{B}}_i$ in the PoS, $\overline{\psi}_{B,i}$ (Eq.~(\ref{eq:polangle_cloud_inverted})), and the inclination angle of $\overline{\boldvec{B}}_i$ to the PoS, $\overline{\gamma}_{B,i}$ (Eq.~(\ref{eq:polfrac_cloud_inverted})).

The small values of $|\overline{\psi}_{{\rm d},1}|$ and $|\overline{\psi}_{{\rm d},5}|$ indicate that the faint ${\mathbb{C}}_1$ and the dominant ${\mathbb{C}}_5$ both have nearly horizontal $\overline{\boldvec{B}}_\perp$.
At the other extreme, $|\overline{\psi}_{{\rm d},3}|\simeq 90^\circ$ corresponds to a vertical $\overline{\boldvec{B}}_\perp$;
however, the faint ${\mathbb{C}}_3$, which appears to be mostly confined to the northeast corner, is probably part of a larger cloud that extends beyond the G139 region, so its derived polarization parameters might not be representative of this larger cloud.
In between, the bright ${\mathbb{C}}_2$ and the faint ${\mathbb{C}}_7$ have slightly tilted $\overline{\boldvec{B}}_\perp$, whereas the two most interesting molecular clouds, ${\mathbb{C}}_6$ and ${\mathbb{C}}_4$, have significantly tilted $\overline{\boldvec{B}}_\perp$.
Our previous argument regarding $\psi_{\rm d}^{\rm los,obs}$ suggests that $\overline{\boldvec{B}}_{\perp,6}$ is more strongly tilted in the bright center of ${\mathbb{C}}_6$ than around it.
Also noteworthy is that $\overline{\boldvec{B}}_{\perp,4}$ appears to be nearly aligned with the elongated main segment of ${\mathbb{C}}_4$.

The difficulty with $\overline{\gamma}_{B,i}$ is that the maximum polarization fraction, $(\overline{p}_{{\rm d},i})_{\rm max}$, entering Eq.~(\ref{eq:polfrac_cloud_inverted}) is unknown.
Below Eq.~(\ref{eq:polfrac_cloud_inverted}), we suggested adopting $(\overline{p}_{{\rm d},i})_{\rm max} = 23\,\%$.
Evidently, this is not a reasonable choice for ${\mathbb{C}}_1$ and ${\mathbb{C}}_7$, whose polarization fractions would have exceeded $23\,\%$ had they not been constrained by our imposed upper limit of $23\,\%$ (see third column of Table~\ref{tab:G139_polar}).
All we can reasonably conclude regarding the magnetic field inclinations to the PoS is that $\overline{\boldvec{B}}_1$ and $\overline{\boldvec{B}}_7$ are probably close to the PoS; $\overline{\boldvec{B}}_2$, $\overline{\boldvec{B}}_3$, $\overline{\boldvec{B}}_5$, and $\overline{\boldvec{B}}_6$ are probably moderately inclined to the PoS; and $\overline{\boldvec{B}}_4$ is probably close to the LoS. 
An additional conclusion can potentially be drawn regarding ${\mathbb{C}}_4$, for which we just noted that $\overline{\boldvec{B}}_{\perp,4}$ is nearly aligned with the elongated main segment of the cloud.
If this alignment also exists in 3D, the main segment of ${\mathbb{C}}_4$ must also be close to the LoS, with the implication that it is actually much more elongated in 3D than in the PoS.

\subsection{Results obtained for other values of the velocity-coherence threshold, $\mathcal{C}_{\rm th}$}
\label{sec:application_threshold}

The 26 Gaussian kinematic components extracted with {\tt ROHSA} each have a well-determined dust intensity, $\tilde{I}_{{\rm d},j}$,
and the total dust intensity, $I_{\rm d}^{\rm mod}$ (Eqs.~(\ref{eq:totintensity_sum}) and (\ref{eq:totintensity_cloud})), is just the sum of the 26 $\tilde{I}_{{\rm d},j}$,
independent of how the 26 components are subsequently grouped into clouds.
In contrast, the polarization fraction and angle of a given component are not determined by {\tt ROHSA}, but are those of the cloud that this component is assigned to.
As a result, the Stokes parameters, $Q_{\rm d}^{\rm mod}$ and $U_{\rm d}^{\rm mod}$ (Eqs.~(\ref{eq:stokesQ_sum_cloud}) and (\ref{eq:stokesU_sum_cloud})), as well as the associated polarized intensity, $P_{\rm d}^{\rm mod}$ (Eq.~(\ref{eq:polintensity_real})), LoS-averaged polarization fraction, $p_{\rm d}^{\rm los,mod}$ (Eq.~(\ref{eq:polfrac})), and LoS-averaged polarization angle, $\psi_{\rm d}^{\rm los,mod}$ (Eq.~(\ref{eq:polangle})), are sensitive to the exact distribution of the 26 components between different clouds.
This distribution, in turn, is governed by the value adopted for the velocity-coherence threshold, $\mathcal{C}_{\rm th}$, entering Eq.~(\ref{eq:threshold_correlation}). 

\begin{table*}
\centering
\caption{
    Results obtained for the first eight cloud configurations identified in the G139 region.
}
\begin{threeparttable}
%Tables with captions and notes all the same width. Provides a scheme for tables that have a structured note section, after the caption.
\begin{tabular}{llcllll}
\toprule\toprule
Configuration & Range of $\mathcal{C}_{\rm th}$ [\%]\tablefootmark{a} & Clouds & Kinematic components & 
$\overline{p}_{{\rm d},i}$ [\%]\tablefootmark{b} \qquad \qquad & $\overline{\psi}_{{\rm d},i}$ [deg]\tablefootmark{c} \qquad \qquad & $\chi_{\rm r}$\tablefootmark{d} \\
\midrule
1 cloud & $[0,1]$ & 
${\mathbb{C}}_1$ &
H{\sc i}\,[1-19], CO\,[1-7] &
$8.7 \pm 0.1$ & $2.7 \pm 0.1$ & 4.66  \\
\noalign{\medskip}
2 clouds & $[2,14]$ & 
${\mathbb{C}}_1$ & 
H{\sc i}\,[1] & 
$22.9 \pm 0.1$ & $-1.5 \pm 0.8$ & 4.53 \\
& & ${\mathbb{C}}_2$ & 
H{\sc i}\,[2-19], CO\,[1-7] &
$8.2 \pm 0.1$ & $3.2 \pm 0.1$ \\
\noalign{\medskip}
3 clouds & $[15,27]$ & 
${\mathbb{C}}_1$ & 
H{\sc i}\,[1] & 
$22.9 \pm 0.1$ & $-1.5 \pm 0.8$ & 4.30 \\
& & ${\mathbb{C}}_2$ & 
H{\sc i}\,[2-4], CO\,[1-2] &
$4.1 \pm 0.1$ & $25.2 \pm 0.8$ \\
& & ${\mathbb{C}}_3$ & 
H{\sc i}\,[5-19], CO\,[3-7] &
$11.5 \pm 0.1$ & $-1.1 \pm 0.2$ \\
\noalign{\medskip}
4 clouds & $[28,31]$ & 
${\mathbb{C}}_1$ & 
H{\sc i}\,[1] & 
$22.9 \pm 0.1$ & $-1.6 \pm 0.8$ & 4.29 \\
& & ${\mathbb{C}}_2$ & 
H{\sc i}\,[2-4], CO\,[1-2] &
$4.1 \pm 0.1$ & $25.2 \pm 0.8$ \\
& & ${\mathbb{C}}_3$ & 
H{\sc i}\,[5-19], CO\,[3-6] &
$11.6 \pm 0.1$ & $-1.0 \pm 0.2$ \\
& & ${\mathbb{C}}_4$ & 
CO\,[7] &
$22.5 \pm 0.5$ & $-63.7 \pm 3.6$ \\
\noalign{\medskip}
5 clouds & $[32,41]$ & 
${\mathbb{C}}_1$ & 
H{\sc i}\,[1] & 
$22.9 \pm 0.1$ & $-1.6 \pm 0.8$ & 4.29 \\
& & ${\mathbb{C}}_2$ & 
H{\sc i}\,[2-4] &
$4.5 \pm 0.2$ & $21.5 \pm 1.0$ \\
& & ${\mathbb{C}}_3$ & 
H{\sc i}\,[5-19], CO\,[3-6] &
$11.2 \pm 0.1$ & $-0.9 \pm 0.2$ \\
& & ${\mathbb{C}}_4$ & 
CO\,[1-2] &
$3.5 \pm 0.3$ & $40.8 \pm 2.8$ \\
& & ${\mathbb{C}}_5$ & 
CO\,[7] &
$22.5 \pm 0.5$ & $-64.4 \pm 3.6$ \\
\noalign{\medskip}
6 clouds & $[42,43]$ & 
${\mathbb{C}}_1$ & 
H{\sc i}\,[1] & 
$22.9 \pm 0.1$ & $-1.8 \pm 0.8$ & 4.14 \\
& & ${\mathbb{C}}_2$ & 
H{\sc i}\,[2-3] &
$7.2 \pm 0.2$ & $14.9 \pm 0.7$ \\
& & ${\mathbb{C}}_3$ & 
H{\sc i}\,[4] &
$13.0 \pm 0.4$ & $89.3 \pm 1.0$ \\
& & ${\mathbb{C}}_4$ & 
H{\sc i}\,[5-19], CO\,[3-6] &
$11.2 \pm 0.1$ & $-0.9 \pm 0.2$ \\
& & ${\mathbb{C}}_5$ & 
CO\,[1-2] &
$3.4 \pm 0.3$ & $43.5 \pm 2.9$ \\
& & ${\mathbb{C}}_6$ & 
CO\,[7] &
$22.2 \pm 0.8$ & $-26.4 \pm 4.2$ \\
\noalign{\medskip}
7 clouds & $[44,51]$ & 
${\mathbb{C}}_1$ & 
H{\sc i}\,[1] & 
$22.9 \pm 0.1$ & $5.2 \pm 1.3$ & 3.17 \\
& & ${\mathbb{C}}_2$ & 
H{\sc i}\,[2-3] &
$6.4 \pm 0.2$ & $16.8 \pm 0.8$ \\
& & ${\mathbb{C}}_3$ & 
H{\sc i}\,[4] &
$14.2 \pm 0.4$ & $89.4 \pm 0.9$ \\
& & ${\mathbb{C}}_4$ & 
H{\sc i}\,[5-6], CO\,[3-4] &
\framebox{$2.2 \pm 0.2$} & \framebox{$46.3 \pm 1.9$} \\
& & ${\mathbb{C}}_5$ & 
H{\sc i}\,[7-19], CO\,[5-6] &
$17.9 \pm 0.1$ & $-3.4 \pm 0.2$ \\
& & ${\mathbb{C}}_6$ & 
CO\,[1-2] &
\framebox{$\mathbf{5.8 \pm 0.3}$} & \framebox{$\mathbf{66.1 \pm 1.7}$} \\
& & ${\mathbb{C}}_7$ & 
CO\,[7] &
$22.5 \pm 0.5$ & $-21.2 \pm 3.4$ \\
\noalign{\medskip}
8 clouds & $[52,58]$ & 
${\mathbb{C}}_1$ & 
H{\sc i}\,[1] & 
$22.9 \pm 0.1$ & $4.8 \pm 1.3$ & 3.14 \\
& & ${\mathbb{C}}_2$ & 
H{\sc i}\,[2-3] &
$6.4 \pm 0.2$ & $17.2 \pm 0.8$ \\
& & ${\mathbb{C}}_3$ & 
H{\sc i}\,[4] &
$13.8 \pm 0.4$ & $89.5 \pm 1.1$ \\
& & ${\mathbb{C}}_4$ & 
H{\sc i}\,[5-6], CO\,[3-4] &
\framebox{$2.1 \pm 0.2$} & \framebox{$42.5 \pm 2.0$} \\
& & ${\mathbb{C}}_5$ & 
H{\sc i}\,[7-18], CO\,[5-6] &
$17.2 \pm 0.1$ & $-3.1 \pm 0.2$ \\
& & ${\mathbb{C}}_6$ & 
H{\sc i}\,[19] &
$22.9 \pm 0.1$ & $-5.4 \pm 1.0$ \\
& & ${\mathbb{C}}_7$ & 
CO\,[1-2] &
\framebox{$\mathbf{6.0 \pm 0.3}$} & \framebox{$\mathbf{66.2 \pm 1.7}$} \\
& & ${\mathbb{C}}_8$ & 
CO\,[7] &
$22.5 \pm 0.5$ & $-21.1 \pm 3.8$ \\
\bottomrule\bottomrule
\end{tabular}
%\begin{tablenotes}
%\end{tablenotes}
\end{threeparttable}
\tablefoot{
\tablefoottext{a}{$\mathcal{C}_{\rm th}$ is the threshold imposed on the weighted-averaged correlation coefficient, $\langle \mathcal{C}_{jj'} \rangle$ (Eq.~(\ref{eq:correlation_average})), as a condition for components $j$ and $j'$ to be assigned to a same cloud (Eq.~(\ref{eq:threshold_correlation})).}
\tablefoottext{b}{Best-fit values and standard deviations of the polarization fractions of the different clouds ${\mathbb{C}}_i$ found in every configuration.}
\tablefoottext{c}{Best-fit values and standard deviations of the polarization angles of the different clouds ${\mathbb{C}}_i$ found in every configuration.}
\tablefoottext{d}{Best-fit reduced $\chi$.} \\
The boxed values pertain to the two most interesting molecular clouds: the bright CO core (bold style) and the oblique, elongated cloud (normal style).
}
\label{tab:G139_dependence_Cth}
\end{table*}

\begin{sidewaysfigure*}
\centering
\includegraphics[width=\textwidth]{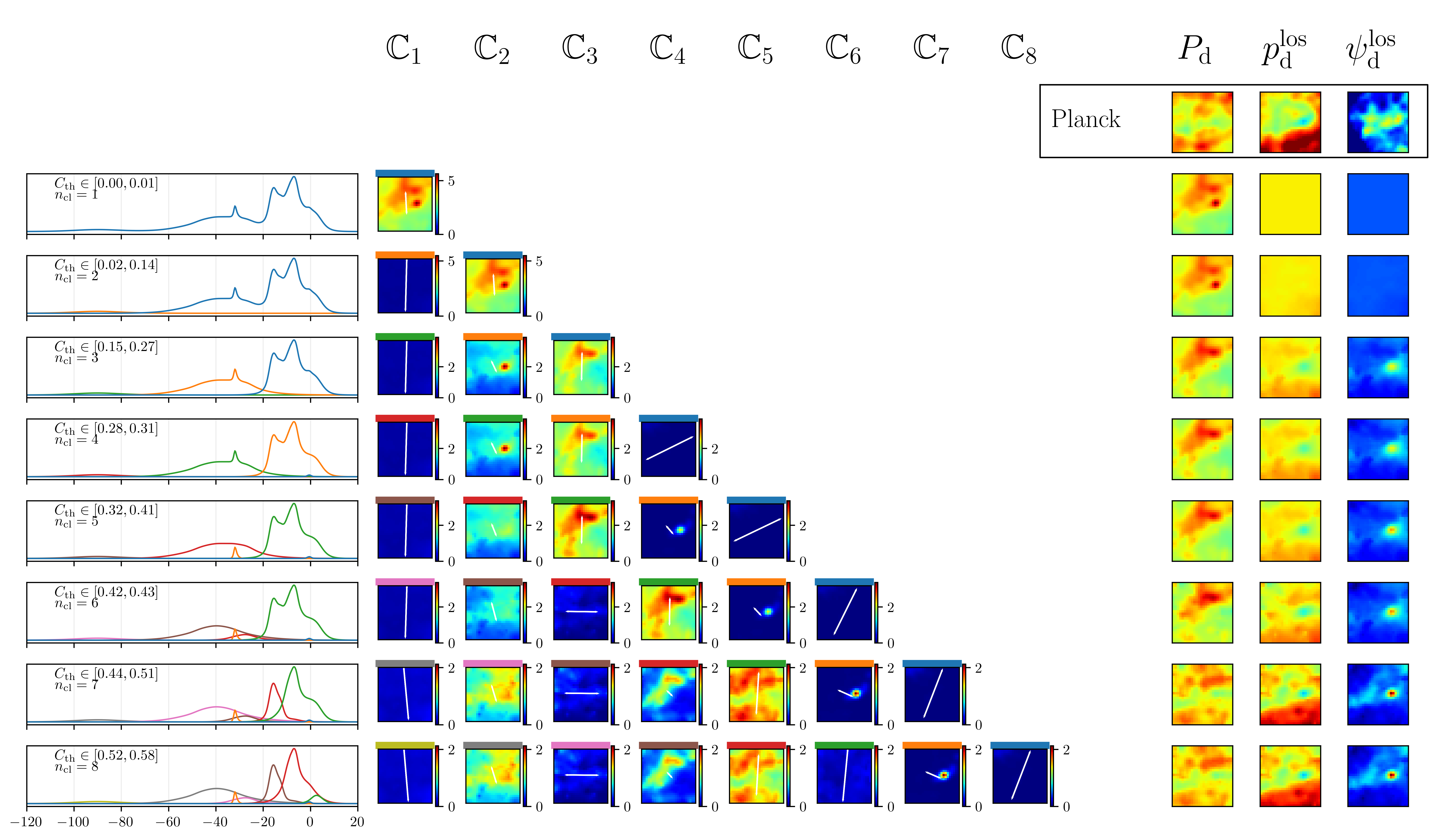}
\caption{
    Results obtained for the eight cloud configurations described in Table~\ref{tab:G139_dependence_Cth}.
    {\it Left}: Spatially-averaged spectra of the 353\,GHz dust intensity of the $n_{\rm cl}$ clouds, ${\mathbb{C}}_i$ ($i = 1, \ ... \ , \ n_{\rm cl}$), all in the same (arbitrary) units. %%identified in every range of $\mathcal{C}_{\rm th}$.
    {\it Center}: Maps of the 353\,GHz dust intensity (in ${\rm MJy\,sr^{-1}}$) of the $n_{\rm cl}$ clouds.
    The polarization fractions, $\overline{p}_{{\rm d},i}$, and polarization angles, $\overline{\psi}_{{\rm d},i}$, of the clouds are indicated by the length and orientation of the overplotted white headless vectors.
    To help make the connection between spectra and maps, each map is topped with a thick band having the same color as the associated spectrum.
    {\it Right}: Maps of the resulting polarized intensity, $P_{\rm d}^{\rm mod}$, LoS-averaged polarization fraction, $p_{\rm d}^{\rm los,mod}$, and LoS-averaged polarization angle, $\psi_{\rm d}^{\rm los,mod}$, all in the same units as in Fig.~\ref{fig:G139_dustmaps_tracers_polar}.
    For reference, the corresponding observational {\it Planck} maps are shown in the top row.
}
\label{fig:G139_dependence_Cth}
\end{sidewaysfigure*}

The purpose of this section is to examine the impact of $\mathcal{C}_{\rm th}$ on the polarization results.
By considering all the integer values of $\mathcal{C}_{\rm th}$ in the range $[0,100\,\%]$, we are led to identify different cloud configurations with increasing numbers of clouds, from $n_{\rm cl} = 1$ to 26.
In Table~\ref{tab:G139_dependence_Cth} and Fig.~\ref{fig:G139_dependence_Cth}, we present the results obtained for the first eight configurations, with $n_{\rm cl} = 1$ to 8, corresponding to $\mathcal{C}_{\rm th} = 0$ to $58\,\%$.
For every configuration, Table~\ref{tab:G139_dependence_Cth} lists the relevant range of $\mathcal{C}_{\rm th}$, the {\tt ROHSA} kinematic components of all the clouds, the best-fit values and standard deviations of their polarization parameters, and the best-fit $\chi_{\rm r}$, while Fig.~\ref{fig:G139_dependence_Cth} displays the spatially-averaged spectra of all the clouds, their dust intensity maps together with their polarization half-vectors, and the reconstructed maps of $P_{\rm d}^{\rm mod}$, $p_{\rm d}^{\rm los,mod}$, and $\psi_{\rm d}^{\rm los,mod}$.

The single cloud formed at small values of $\mathcal{C}_{\rm th}$ loses its weak, high-velocity H{\sc i}\,[1] component as soon as $\mathcal{C}_{\rm th}$ reaches $2\,\%$.
It then breaks up into two mixed atomic-molecular clouds when $\mathcal{C}_{\rm th}$ reaches $15\,\%$: the localized cloud ${\mathbb{C}}_2$, which includes the bright CO core, and the pervasive cloud ${\mathbb{C}}_3$, which contains the rest of the CO gas.
${\mathbb{C}}_3$ loses its very weak, low-velocity CO\,[7] component at $\mathcal{C}_{\rm th} = 28\,\%$, while ${\mathbb{C}}_2$ successively loses the bright CO core (CO\,[1-2]) at $\mathcal{C}_{\rm th} = 32\,\%$ and the knotty H{\sc i}\,[4] component at $\mathcal{C}_{\rm th} = 42\,\%$.
At $\mathcal{C}_{\rm th} = 44\,\%$, an oblique, elongated chunk (H{\sc i}\,[5-6], CO\,[3-4]) splits off from the dominant cloud, leaving behind a mostly atomic cloud (H{\sc i}\,[7-19], CO\,[5-6]).
Further fissions successively occur above $\mathcal{C}_{\rm th} = 51\,\%$ until every cloud reduces to a single kinematic component.

The general quality of the fit, measured by the value of $\chi_{\rm r}$, gradually improves as $\mathcal{C}_{\rm th}$ increases from 0 to $42\,\%$ -- and, accordingly, $n_{\rm cl}$ increases from 1 to 6.
This is because at each division the two new clouds are allowed to take on more representative polarization parameters, presumably closer to reality. 
When $\mathcal{C}_{\rm th}$ reaches $44\,\%$, $\chi_{\rm r}$ drops by $23\,\%$ and the fit suddenly looks much better.
What happens is that the oblique, elongated cloud (H{\sc i}\,[5-6], CO\,[3-4]) breaks away with a polarization angle very different from that of its parent cloud.
As a result, it acts as a depolarizing cloud, which manages to reproduce a large portion of the region of lower polarization in the map of $p_{\rm d}^{\rm los,obs}$ (second-to-last map in the top row of Fig.~\ref{fig:G139_dependence_Cth}) -- in the same way as the bright CO core was argued in Sect.~\ref{sec:application_Bfield} to explain the localized low-polarization spot in the map of $p_{\rm d}^{\rm los,obs}$.
In the rest of this subsection, these two depolarizing clouds are referred to as ${\mathbb{C}}_4^{[7]}$ and ${\mathbb{C}}_6^{[7]}$, respectively, based on their names in the seven-cloud configuration.

As $\mathcal{C}_{\rm th}$ keeps increasing above $51\,\%$, the emergence of new clouds only leads to marginal improvements in the reconstructed polarization maps, with no significant decrease of $\chi_{\rm r}$.
Most importantly, the polarization parameters of ${\mathbb{C}}_4^{[7]}$ and ${\mathbb{C}}_6^{[7]}$ remain stable until these clouds themselves break apart, which we take as evidence that they have nearly reached their true values.
Physically, when ${\mathbb{C}}_4^{[7]}$ becomes a separate cloud, it directly takes on the polarization parameters that make it possible to reproduce not only the observed Stokes parameters in its own direction,
but also the general polarized background against which ${\mathbb{C}}_6^{[7]}$ acts as a small depolarizing cloud.
${\mathbb{C}}_6^{[7]}$ then automatically adjusts its polarization parameters to reproduce the low-polarization spot observed at its location.

To sum up, the status of ${\mathbb{C}}_4^{[7]}$ and ${\mathbb{C}}_6^{[7]}$ as depolarizing clouds appears to be robust, and the best-fit values of their polarization parameters can be considered to be trustworthy.

\section{Discussion and conclusions}
\label{sec:conclusion}

In this paper, we present a new method designed (1) to identify along the LoS the different clouds that contribute to the observed dust emission and (2) to estimate the orientations of their internal magnetic fields.
The cloud identification is performed %%in two steps, 
with the help of three kinematic gas tracers: the H{\sc i} 21\,cm, $^{12}$CO 2.6\,mm, and $^{13}$CO 2.7\,mm emission lines.
The 3D spectral cubes of these three tracers are corrected for opacity saturation, the corrected $^{12}$CO and $^{13}$CO cubes are combined into a single CO cube, and the H{\sc i} and combined CO cubes are each decomposed with the algorithm {\tt ROHSA} into several spatially coherent Gaussian kinematic components.
All the kinematic components from both tracers are rescaled to dust intensity, and those (from either tracer) with similar velocity profiles are grouped together into clouds.
The result is a set of $n_{\rm cl}$ clouds, ${\mathbb{C}}_i$ ($i = 1, \ ... \ , \ n_{\rm cl}$), with given dust intensities, $I_{{\rm d},i}(l,b)$.
The estimation of their magnetic field orientations rests on the linear polarization of the observed dust emission, described by the two Stokes parameters, $Q_{\rm d}$ and $U_{\rm d}$. 
By decomposing the latter in the basis formed by the 2D dust intensities of the $n_{\rm cl}$ clouds (Eqs.~(\ref{eq:stokesQ_sum_cloud}) and (\ref{eq:stokesU_sum_cloud})), we can obtain the polarization parameters (polarization fraction, $\overline{p}_{{\rm d},i}$, and polarization angle, $\overline{\psi}_{{\rm d},i}$) of each cloud, ${\mathbb{C}}_i$, which, in turn, lead to its magnetic field orientation.

As an illustration of our method, we proposed a first application to the G139 region, for which we had access to 353\,GHz {\it Planck} maps of the polarized dust emission as well as spectral cubes of the H{\sc i} 21\,cm, $^{12}$CO 2.6\,mm, and $^{13}$CO 2.7\,mm emission lines.
The exact number of clouds present along the LoS is poorly constrained, such that too much reality should not be ascribed to the exact values of their individual polarization parameters (listed in Table~\ref{tab:G139_dependence_Cth}).
These parameters show great variability:
the polarization fractions span a broad range from $\simeq 2\,\%$ up to our imposed upper limit of 23\,\% and the polarization angles span most of the range $[-90^{\circ}, +90^{\circ}]$.
The dominant cloud (${\mathbb{C}}_5$ in the seven-cloud configuration) consistently has a very small polarization angle, which accounts for most of the observed low-$\psi_{\rm d}^{\rm los,obs}$ background.
The two molecular clouds that stand out against this background in the seven-cloud configuration (${\mathbb{C}}_6$ and ${\mathbb{C}}_4$) have much larger polarization angles ($\simeq 65^\circ$ and $\simeq 45^\circ$, respectively), such that they act as depolarizing clouds and produce the localized low-polarization spot and a large portion of the more extended low-polarization region, respectively, in the map of $p_{\rm d}^{\rm los,obs}$. 

The broad ranges obtained for the clouds' polarization parameters imply that their magnetic fields are variously inclined to the PoS, from nearly perpendicular to nearly parallel, and variously tilted to the Galactic plane, from nearly horizontal to nearly vertical.
The magnetic field of the dominant cloud is consistently found to be nearly horizontal, in accordance with the general orientation of the interstellar magnetic field in the Galactic disk.
The magnetic fields of the two depolarizing molecular clouds are significantly tilted to the Galactic plane, which again is not surprising given that the formation of molecular clouds is often accompanied by rotation and shearing of magnetic field lines.
${\mathbb{C}}_6$ shows evidence that its bright center has rotated more than its periphery, while ${\mathbb{C}}_4$ appears to have its magnetic field nearly aligned with its elongated main segment (at least in the PoS).

Our study highlights the need to incorporate the LoS dimension and to understand how different clouds along the LoS contribute to the observed polarized dust emission.
Had we only relied on 2D sky maps of the polarized dust emission and tried to read polarization angles directly off the map of $\psi_{\rm d}^{\rm los,obs}$,
we would have mistakenly concluded that the bright CO core causes no more than a small distortion in a nearly horizontal magnetic field, and we would have completely missed the contribution from the oblique, elongated molecular cloud near the center of the G139 region.
Despite the degeneracies in the LoS decomposition, the values derived here for the polarization parameters of these two clouds appear to be robust, as they remain stable from the seven-cloud configuration where the oblique, elongated cloud becomes a separate entity to the higher-$n_{\rm cl}$ configuration where each cloud breaks apart.
These values also admit a plausible, simple physical interpretation. In principle, our method can be applied to any region of the sky for which polarization and kinematic data are available. 

Below, we discuss the main assumptions underlying our method and comment on its limitations.

First, our study relies on two kinds of tracers: polarized thermal emission from dust and spectral lines of H{\sc i} and CO (including $^{12}$CO and $^{13}$CO).
Our first important assumption is that these two kinds of tracers are linearly related through Eq.~(\ref{eq:finalexpr_totintensity}).
This, in turn, assumes that 
(1) although dust accounts for only $\sim 1\,\%$ in mass of the interstellar matter, it is well mixed in with the gas, and its column density is proportional to that of the gas;
(2) the emissivity of dust grains is uniform across each of the atomic and molecular media;
(3) H{\sc i} and CO are complementary tracers of the gas, which together detect all the gas along the LoS, with no omission and no overlap;
and (4) the brightness temperatures of the considered spectral lines are properly corrected for opacity saturation.
The validity of this multiple assumption can be assessed with the help of Fig.~\ref{fig:G139_dustmaps_tracers}, which compares the dust intensity map reconstructed with the two gas tracers (top-middle panel) to the observational {\it Planck} map (left panel).
The map of the residuals (top-right panel) indicates that the reconstruction is generally good to within a few percent, although three regions show residuals of up to $\sim 20\,\%$.
Residuals around the bright CO core suggest that our H{\sc i} and CO tracers combined miss a small fraction of the gas along the LoS (the so-called dark gas).
Other residuals are probably linked to our opacity corrections, which were shown to be quite sensitive to the excitation temperatures and their uncertainties.
 The adopted values of $T^{{\rm H}\textsc{i}}$ and $\sigma(T^{{\rm H}\textsc{i}}) $ are particularly critical,  affecting most of the G139 region.
The values of $T^{{\rm CO}}$ and $\sigma(T^{{\rm CO}})$ could potentially have an even greater impact in saturated $^{12}$CO regions, but this impact is considerably reduced by our using a combination of $^{12}$CO and $^{13}$CO data.
Another, more minor source of residuals could be that dust emission does not perfectly trace the gas distribution (nonuniform conversion factors).

Second, the matter associated with the observed dust emission is decomposed, on purely kinematic grounds into several clouds along the LoS.
This LoS decomposition is by no means unique:
both the Gaussian kinematic decomposition of each tracer with {\tt ROHSA} and the grouping of all kinematic components into separate clouds involve free parameters whose values are chosen a bit arbitrarily. 
This is unavoidable, as the very definition of an interstellar cloud, the criteria used to locate its physical boundaries, and the extent to which adjacent structures can be grouped together into a single cloud are all somewhat subjective.
The {\tt ROHSA} decomposition of each tracer turns out to be very good, as the spatially-averaged spectrum and the velocity-integrated map reconstructed with all the kinematic components of each tracer are very close to their observational counterparts.
The component grouping into clouds has no impact on the reconstructed dust intensity, but it does affect the reconstructed Stokes parameters for linear polarization.
This is why we looked into different cloud configurations having from $n_{\rm cl} = 1$ to 26 clouds. 
We found that the reconstruction of the Stokes parameters improves as $n_{\rm cl}$ increases from 1 to 7, then remains quite stable as $n_{\rm cl}$ increases even further.

Our method has a wide range of applications. 
One of them is to derive the relative orientation angles between filamentary structures and the ambient magnetic field in star-forming regions, with the aim of gaining insight into the role played by magnetic fields in the processes of filament formation and star formation.
With that perspective in mind, we intend to apply our recently developed method for \textbf{Fil}ament \textbf{D}etection and \textbf{Re}construction \textbf{a}t \textbf{M}ultiple \textbf{S}cales \citep[FilDReaMS;][]{carriere_etal_2022a, carriere_etal_2022b} to a large sample of star-forming regions.
This kind of application would benefit from higher angular-resolution data for both the gas distribution and dust polarization.

Our proposed new method requires a lot more testing and validation than we have provided in the present paper.
Systematic tests should be conducted to examine the degeneracies in the LoS decomposition and their impact on the derived cloud parameters, $I_{{\rm d},i}$, $\overline{p}_{{\rm d},i}$, and $\overline{\psi}_{{\rm d},i}$.
More generally, it would be very useful to identify the conditions under which the method produces reliable results and to assess the degree of accuracy of the results.
We could, for instance, start with simple toy models, involving an increasing number of partly overlapping homogeneous clouds.
The method could then be applied to mock data and to simulations, with an increasing number of more realistic clouds.
In each case, it would be possible to quantify the accuracy with which the cloud parameters are recovered and to determine how the validity of the reconstruction depends on various factors, such as the relative (total and polarized) intensities of the clouds, their spatial distribution, and the presence of depolarizing clouds.

Along the same lines, it would be very interesting to compare the results of our LoS decomposition to the results of tomographic decomposition with stellar polarization data \citep[see, e.g.,][]{pelgrims_etal_24}.
Both methods have their intrinsic limitations, and comparing them would make it possible not only to improve each of them, but also to connect the RVs of the dust-emitting clouds to their distances.

Our results could also be confronted with complementary magnetic field observations, such as Zeeman measurements.
This could help to lift (at least part of) the degeneracies and improve the robustness of our results.
Moreover, this would lead to a more complete view of the magnetic fields of the dominant clouds.

\begin{acknowledgements}
We wish to extend our deepest thanks to Isabelle Ristorcelli, with whom we had many interesting and lively discussions related to our study. 
We also thank Mika Juvela, Dana Alina, Julien Montillaud, and Tie Liu, as well as the anonymous referee. 
This work was supported by the Programme National “Physique et Chimie du Milieu Interstellaire” (PCMI) of CNRS/INSU with INC/INP co-funded by CEA and CNES.
We made use of the data from the Milky Way Imaging Scroll Painting (MWISP) project, which is a multiline survey in $^{12}$CO/$^{13}$CO/C$^{18}$O along the northern galactic plane with the PMO-13.7m telescope. We are grateful to all the members of the MWISP working group, particularly the staff members at th PMO-13.7m telescope, for their long-term support. MWISP was sponsored by National Key R\&D Program of China with grants 2023YFA1608000 \& 2017YFA0402701 and by CAS Key Research Program of Frontier Sciences with grant QYZDJ-SSW-SLH047.
\end{acknowledgements}

\bibliographystyle{aa}
\bibliography{biblio}

\begin{appendix}

\section{Opacity corrections}
\label{sec:opacity}

\subsection{Opacity corrections for H{\sc i}}
\label{sec:opacity_HI}

For notational simplicity, superscript H{\sc i} is dropped throughout this subsection.

In a homogeneous medium in thermal equilibrium, the brightness temperature of the H{\sc i} 21\,cm line at LoS velocity $v$, $T_{\rm b}(v)$, can be written in terms of the kinetic temperature, $T$, and the optical depth, $\tau_v$, as
\begin{equation}
T_{\rm b}(v) = T \ \left( 1 - e^{-\tau_v} \right) \ \cdot
\label{eq:Tb}
\end{equation}
In the optically thin regime ($\tau_v \ll 1$), $T_{\rm b}(v) = T\,\tau_v$.
In the general case, we can correct $T_{\rm b}(v)$ for opacity saturation 
and define an opacity-corrected or "de-saturated" brightness temperature, 
\begin{equation}
T_{\rm b\star}(v) = T \ \tau_v \ \cdot
\label{eq:Tb*}
\end{equation}
If the medium is out of thermal equilibrium, 
$T$ should be understood as an excitation temperature (the spin temperature in this case).
By combining Eqs.~(\ref{eq:Tb}) and (\ref{eq:Tb*}), we can easily relate $T_{\rm b\star}$ to $T_{\rm b}$, either in terms of $\tau_v$,
\begin{equation}
T_{\rm b\star} = T_{\rm b} \ \frac{\tau_v}{1 - e^{-\tau_v}} \ ,
\label{eq:Tb*_Tb_tau}
\end{equation}
or 
in terms of $T$,
\begin{equation}
T_{\rm b} = T \ \left(
1 - \exp{\Big( - \frac{T_{\rm b\star}}{T} \Big)}
\right) \ ,
\label{eq:Tb_Tb*_T}
\end{equation}
which is equivalent to
\begin{equation}
\boxed{
T_{\rm b\star} = - T \ \ln \Big( 1 - \frac{T_{\rm b}}{T} \Big)
} \ \cdot
\label{eq:Tb*_Tb_T}
\end{equation}
The optical depth, $\tau_v$, %%at distance $s$ from the observer 
is directly related to the hydrogen column density, $N_{\rm H}$, and the temperature, $T$, \begin{equation}
\tau_v = {\cal C} \ \frac{N_{\rm H}}{T} \ \phi(v) \ ,
\label{eq:tau}
\end{equation}
where ${\cal C}$ is a constant for the considered spectral line and $\phi(v)$ is the so-called line profile function \citep[e.g.,][]{rybicki&l_79}.
Substituting Eq.~(\ref{eq:tau}) into Eq.~(\ref{eq:Tb*}) and integrating over $v$ leads to a relation between the velocity-integrated opacity-corrected brightness temperature and the hydrogen column density,
\begin{equation}
\boxed{
\int T_{\rm b\star}(v) \ dv = {\cal C} \ N_{\rm H}
} \ ,
\label{eq:coldensity}
\end{equation}
when use is made of the normalization condition $\int \phi(v) \ dv = 1$. 

In an inhomogeneous medium, Eq.~(\ref{eq:coldensity}) remains valid, but Eq.~(\ref{eq:Tb*_Tb_T}) strictly makes sense only if the spin temperature is uniform.
If this is not the case, it might still be possible to use Eq.~(\ref{eq:Tb*_Tb_T}), with $T$ replaced by an appropriate effective temperature, $T_{\rm eff}$.
Remember that H{\sc i} gas in the ISM can basically be found in two different phases: a cold phase with $T_{\rm c} \approx 80~{\rm K}$ and a warm phase with $T_{\rm w} \approx 8\,000~{\rm K}$ \citep[e.g.,][]{ferriere_2020}, where subscripts c and w denote the cold and warm phases, respectively.
The total optical depth can then be written as
\begin{equation}
\tau_v = \tau_{v,{\rm c}} + \tau_{v,{\rm w}} \ ,
\label{eq:tau_2phases}
\end{equation}
with, in view of Eq.~(\ref{eq:tau}),
\begin{equation}
\tau_{v,{\rm i}} 
\ = \ {\cal C} \ \frac{N_{\rm H,i}}{T_{\rm i}} \ \phi_{\rm i}(v) 
\ = \ {\cal C} \ \frac{f_{\rm i}(v)}{T_{\rm i}} \ N_{\rm H} \ \phi(v) \ ,
\label{eq:tau_1phase}
\end{equation} 
%%for ${\rm i =}$ c and w, 
where $f_{\rm i}(v)$ is the mass fraction of phase i at velocity $v$, for i = c, w.
Similarly, Eq.~(\ref{eq:Tb*}) becomes
\begin{equation}
T_{\rm b\star}(v) = T_{\rm c} \ \tau_{v,{\rm c}} + T_{\rm w} \ \tau_{v,{\rm w}} \ ,
\label{eq:Tb*_2phases}
\end{equation} 
which can also be written as 
\begin{equation}
T_{\rm b\star}(v) = T_{\rm eff}(v) \ \tau_v \ ,
\label{eq:Tb*_2phases_Teff}
\end{equation} 
where the effective temperature is the weighted harmonic mean of $T_{\rm c}$ and $T_{\rm w}$:
\begin{equation}
T_{\rm eff}(v) = 
\left( \frac{f_{\rm c}(v)}{T_{\rm c}} 
+ \frac{f_{\rm w}(v)}{T_{\rm w}} \right)^{-1} \ \cdot
\label{eq:Teff}
\end{equation}
Clearly, unless the mass fraction of the cold phase at velocity $v$ is much smaller than that of the warm phase, $T_{\rm eff} \simeq T_{\rm c} / f_{\rm c}$.

The similarity between Eq.~(\ref{eq:Tb*_2phases_Teff}) and Eq.~(\ref{eq:Tb*}) suggests that the most reasonable approach is to adopt $T = T_{\rm eff}$ in Eq.~(\ref{eq:Tb*_Tb_T}).
Incidentally, this approach would be exact if the cold and warm phases were well mixed along the LoS, with a uniform mass ratio, because then 
$T_{\rm eff}$ would also be uniform, and we would formally be back to the case of a homogeneous medium.
The problem is that $T_{\rm eff}$ varies across the H{\sc i} spectrum.
However, aside from the requirement $T_{\rm eff} \ge T_{\rm c}$, 
$T_{\rm eff}$ does not need to be determined with accuracy across the entire spectrum.
The only spectral regions where the exact value of $T_{\rm eff}$ is critical are regions where $T_{\rm b}$ approaches (or exceeds) $T_{\rm c}$.
But these regions are almost always dominated by cold gas, so they can be described by $T_{\rm eff} = T_{\rm c}$.
Spectral regions not dominated by cold gas have $T_{\rm b}$ well below $T_{\rm c}$, and hence well below $T_{\rm eff}$, so their $T_{\rm b\star}$ inferred from Eq.~(\ref{eq:Tb*_Tb_T}) with $T = T_{\rm eff}$ is little sensitive to the adopted value of $T_{\rm eff}$, %%in Eq.~(\ref{eq:Tb*_Tb_T}), 
which can again be set to $T_{\rm c}$.
Altogether, $T_{\rm b\star}$ can be approximated by Eq.~(\ref{eq:Tb*_Tb_T}) with $T = T_{\rm eff} = T_{\rm c} = 80~{\rm K}$ throughout the spectrum.
This choice for $T$ results in a small overestimation of $N_{\rm H}$.

The uncertainty in $T_{\rm b\star}$, $\sigma(T_{\rm b\star})$, is the quadratic sum of the uncertainties arising from $T_{\rm b}$ and from $T$:
\begin{equation}
\sigma^2(T_{\rm b\star}) = 
\left[ \sigma^2(T_{\rm b\star}) \right]_{T_{\rm b}} 
+ \left[ \sigma^2(T_{\rm b\star}) \right]_{T} \ \cdot
\label{eq:sigma2_Tb*}
\end{equation}
The uncertainty arising from $T_{\rm b}$, $\left[ \sigma(T_{\rm b\star}) \right]_{T_{\rm b}}$, directly follows from the measurement error in $T_{\rm b}$, $\sigma(T_{\rm b})$:
\begin{equation}
\left[ \sigma(T_{\rm b\star}) \right]_{T_{\rm b}} 
\ = \ \frac{\partial T_{\rm b\star}}{\partial T_{\rm b}} \ \sigma(T_{\rm b})
\ = \ \frac{T}{T - T_{\rm b}} \ \sigma(T_{\rm b}) \ ,
\label{eq:sigma_Tb*_Tb}
\end{equation}
where the second identity is obtained with the help of Eq.~(\ref{eq:Tb*_Tb_T}).
Similarly, the uncertainty arising from $T$, $\left[ \sigma(T_{\rm b\star}) \right]_{T}$, reads
\begin{equation}
\left[ \sigma(T_{\rm b\star}) \right]_{T} 
\ = \ \frac{\partial T_{\rm b\star}}{\partial T} \ \sigma(T)
\ = \ - \left[ \ln \Big( 1 - \frac{T_{\rm b}}{T} \Big)
+ \frac{T_{\rm b}}{T - T_{\rm b}} \right]
\ \sigma(T) \ \cdot
\label{eq:sigma_Tb*_T}
\end{equation}
Here, we adopt $\sigma(T) = 20~{\rm K}$ \citep[e.g.,][]{ferriere_2020}.

\subsection{Opacity corrections for $^{12}$CO}
\label{sec:opacity_CO}

In this subsection, superscript $^{12}$CO is dropped unless there is a possible ambiguity with $^{13}$CO, in which case superscripts $^{12}$CO and $^{13}$CO are reduced to $^{12}$ and $^{13}$, respectively.

In principle, we can derive the opacity-corrected brightness temperature of the $^{12}$CO 2.6\,mm line, $T_{\rm b\star}^{^{12}}$, in the same way as explained in Sect.~\ref{sec:opacity_HI} for the H{\sc i} 21\,cm line.
There are, however, a few important differences.

First, Eq.~(\ref{eq:Tb}) is strictly valid only in the Rayleigh-Jeans regime ($h \nu \ll k T$) and in the absence of a significant background radiation, both of which are satisfied in the case of H{\sc i}, but not in the case of $^{12}$CO.
Indeed, the higher frequency of the $^{12}$CO 2.6\,mm line and the lower temperature of molecular gas cause small deviations from the Rayleigh-Jeans law and require taking the $2.73~{\rm K}$ background radiation into account.
Moreover, the observed $^{12}$CO signal may be subject to beam dilution.
Under these conditions, the correct expression of the brightness temperature reads \citep[e.g.,][]{ho&mr_82},
\begin{equation}
T_{\rm b}(v) = \eta_{\rm b} \ \big( J(T_{\rm ex}) - J(T_{\rm bg}) \big) \ \left( 1 - e^{-\tau_v} \right), \ 
\label{eq:Tb_mol}
\end{equation}
where $\eta_{\rm b}$ is the beam
dilution factor,
\begin{equation}
J(T) = \frac{T_0}{\exp{\Big( \frac{T_0}{T} \Big)} - 1}
\label{eq:planck_law}
\end{equation}
is the intensity in temperature units,
$T_0 = h \nu / k$ the intrinsic temperature, $T_{\rm ex}$ the excitation temperature, and $T_{\rm bg} = 2.73~{\rm K}$ the blackbody background temperature.
In the case of the $^{12}$CO 2.6\,mm line, $T_0 = 5.53~{\rm K}$, $J(T_{\rm ex}\!=\!20~{\rm K}) = 17.4~{\rm K}$, and $J(T_{\rm bg}) = 0.84~{\rm K}$.
Although the correct expression is now Eq.~(\ref{eq:Tb_mol}), we are still allowed to use Eq.~(\ref{eq:Tb}) provided $T$ refers to the corrected excitation temperature, $\eta_{\rm b} \ \big[ J(T_{\rm ex}) - J(T_{\rm bg}) \big]$.
Eq.~(\ref{eq:Tb*_Tb_T}) then provides a first estimate of $T_{\rm b\star}^{^{12}}$,
\begin{equation}
%%T_{\rm b\star}^{(1)} = - T \ \ln \Big( 1 - \frac{T_{\rm b}^{12}}{T} \Big) \ \cdot
(T_{\rm b\star}^{^{12}})_{1} = - T \ \ln \Big( 1 - \frac{T_{\rm b}^{^{12}}}{T} \Big) \ \cdot
\label{eq:Tb*_Tb_T_1}
\end{equation}

Second, in contrast to the H{\sc i} 21\,cm line, the $^{12}$CO 2.6\,mm line is often strongly saturated ($T_{\rm b}^{^{12}} \to T$), with the implication that $(T_{\rm b\star}^{^{12}})_{1}$ is highly uncertain near the spectral peaks.
This leads us to consider a second estimate of $T_{\rm b\star}^{^{12}}$, based on the optically much thinner $^{13}$CO 2.7\,mm line,
\begin{equation}
(T_{\rm b\star}^{^{12}})_{2} 
\ = \ {\mathtt r} \ T_{\rm b\star}^{^{13}}
\ = \ {\mathtt r} \ \left[ - T \ \ln \Big( 1 - \frac{T_{\rm b}^{^{13}}}{T} \Big) \right], \ 
\label{eq:Tb*_Tb_T_2}
\end{equation}
where ${\mathtt r}$ is the $^{12}$CO/$^{13}$CO abundance ratio.
In writing Eq.~(\ref{eq:Tb*_Tb_T_2}), we implicitly assume that $^{13}$CO has the same excitation temperature as $^{12}$CO, and we neglect the slight differences in $J(T_{\rm ex})$ and $J(T_{\rm bg})$ (Eq.~(\ref{eq:Tb_mol})) and in ${\cal C}$ (Eq.~(\ref{eq:coldensity})) arising from the slight frequency difference between the $^{12}$CO and $^{13}$CO lines.
We further assume that the $^{12}$CO/$^{13}$CO abundance ratio is equal to the $^{12}$C/$^{13}$C isotopic ratio, for which we adopt the value ${\mathtt r} = 68 \pm 15$ \citep{milam_etal_2005}.
The advantage of considering both estimates of $T_{\rm b\star}^{^{12}}$ together is that they nicely complement each other: as the CO column density increases, $T_{\rm b}^{^{12}}$ approaches $T$, and thus $(T_{\rm b\star}^{^{12}})_{1}$ becomes more uncertain, but at the same time $T_{\rm b}^{^{13}}$ increasingly rises above the noise level, and thus $(T_{\rm b\star}^{^{12}})_{2}$ becomes more reliable.

The temperature, $T$, is a priori unknown, while its exact value is critical for the derivation of $(T_{\rm b\star}^{^{12}})_{1}$ (Eq.~(\ref{eq:Tb*_Tb_T_1})) near the $^{12}$CO spectral peaks.
Here, we consider that the $^{12}$CO line is fully saturated ($T_{\rm b}^{^{12}} \simeq T$) at the 3D positions of the $^{13}$CO spectral peaks exceeding a certain threshold, $(T_{\rm b}^{^{13}})_{\rm thr}$,\footnote{
For the application presented in Sect.~\ref{sec:application}, we adopt $(T_{\rm b}^{^{13}})_{\rm thr} = 4.25~{\rm K}$, 
to include the $n$ highest peaks that are clearly fully saturated in $^{12}$CO, as indicated by their $T_{\rm b}^{^{12}} / T_{\rm b}^{^{13}}$ ratio being very close to that of the strongest peak.
The corresponding uncertainty in $T$, rounded to the nearest whole number, is $\sigma(T) = 5~{\rm K}$.
}
and we retain the small subset of fully saturated $T_{\rm b}^{^{12}}$.
We might be tempted to assign their average value to $T$, but this would automatically lead to a number of unphysical voxels with $T_{\rm b}^{^{12}} > T$.
Instead, we assign the largest fully saturated $T_{\rm b}^{^{12}}$, $(T_{\rm b}^{^{12}})_{\rm max}$, to $T$ and the rms deviation of the fully saturated $ T_{\rm b}^{^{12}}$from $(T_{\rm b}^{^{12}})_{\rm max}$ to the uncertainty in $T$, $\sigma(T)$.
Our estimates for $T$ and $\sigma(T)$ are obviously biased toward the brightest $^{13}$CO-emitting regions, which are also the regions where the $^{12}$CO line is the most saturated, and hence where the opacity correction is the most critical.
We note, however, that our particular choice $T = (T_{\rm b}^{^{12}})_{\rm max}$ leads to a systematic underestimation of the opacity correction in these regions.
We also note that our subset of fully saturated $T_{\rm b}^{^{12}}$ might miss unresolved bright sources, for which the measured $T_{\rm b}^{^{13}}$ is artificially reduced by beam dilution (factor $\eta_{\rm b}$ in Eq.~(\ref{eq:Tb_mol})).

The uncertainties in $(T_{\rm b\star}^{^{12}})_{1}$ and $(T_{\rm b\star}^{^{12}})_{2}$ are given by equations similar to Eq.~(\ref{eq:sigma2_Tb*}):
\begin{equation}
\sigma^2 \big( (T_{\rm b\star}^{^{12}})_{1} \big) = 
\left[ \sigma^2 \big( (T_{\rm b\star}^{^{12}})_{1} \big) \right]_{T_{\rm b}^{^{12}}} 
+ \left[ \sigma^2 \big( (T_{\rm b\star}^{^{12}})_{1} \big) \right]_{T}
\label{eq:sigma2_Tb*_1}
\end{equation}
and
\begin{equation}
\sigma^2 \big( (T_{\rm b\star}^{^{12}})_{2} \big) = 
\left[ \sigma^2 \big( (T_{\rm b\star}^{^{12}})_{2} \big) \right]_{T_{\rm b}^{^{13}}} 
+ \left[ \sigma^2 \big( (T_{\rm b\star}^{^{12}})_{2} \big) \right]_{T} 
+ \left[ \sigma^2 \big( (T_{\rm b\star}^{^{12}})_{2} \big) \right]_{\mathtt r} \ ,
\label{eq:sigma2_Tb*_2}
\end{equation}
where (see Eqs.~(\ref{eq:sigma_Tb*_Tb}) and \ref{eq:sigma_Tb*_T}))
\begin{equation}
\left[ \sigma \big( (T_{\rm b\star}^{^{12}})_{1} \big) \right]_{T_{\rm b}^{^{12}}} 
%%\ = \ \frac{\partial T_{\rm b\star}}{\partial T_{\rm b}} \ \sigma(T_{\rm b})
\ = \ \frac{T}{T - T_{\rm b}^{^{12}}} \ \sigma(T_{\rm b}^{^{12}}) \ ,
\label{eq:sigma_Tb*_Tb_1}
\end{equation}
\begin{equation}
\left[ \sigma \big( (T_{\rm b\star}^{^{12}})_{1} \big) \right]_{T} 
%%\ = \ \frac{\partial (T_{\rm b\star}^{^{12}})_{1}}{\partial T} \ \sigma(T)
\ = \ - \left[ \ln \Big( 1 - \frac{T_{\rm b}^{^{12}}}{T} \Big)
+ \frac{T_{\rm b}^{^{12}}}{T - T_{\rm b}^{^{12}}} \right]
\ \sigma(T) \ ,
\label{eq:sigma_Tb*_T_1}
\end{equation}
\begin{equation}
\left[ \sigma \big( (T_{\rm b\star}^{^{12}})_{2} \big) \right]_{T_{\rm b}^{^{13}}} 
\ = \ {\mathtt r} \ \frac{T}{T - T_{\rm b}^{^{13}}} \ \sigma(T_{\rm b}^{^{13}}) \ ,
\label{eq:sigma_Tb*_Tb_2}
\end{equation}
\begin{equation}
\left[ \sigma \big( (T_{\rm b\star}^{^{12}})_{2} \big) \right]_{T} 
\ = \ - {\mathtt r} \ \left[ \ln \Big( 1 - \frac{T_{\rm b}^{^{13}}}{T} \Big)
+ \frac{T_{\rm b}^{^{13}}}{T - T_{\rm b}^{^{13}}} \right]
\ \sigma(T) \ ,
\label{eq:sigma_Tb*_T_2}
\end{equation}
and
\begin{equation}
\left[ \sigma \big( (T_{\rm b\star}^{^{12}})_{2} \big) \right]_{{\mathtt r}} 
%%\ = \ \frac{\partial (T_{\rm b\star}^{^{12}})_{2}}{\partial {{\mathtt r}}} \ \sigma({\mathtt r})
\ = \ - T \ \ln \Big( 1 - \frac{T_{\rm b}^{^{13}}}{T} \Big) \ \sigma({\mathtt r}) \ \cdot
\label{eq:sigma_Tb*_r_2}
\end{equation}
In the above equations, $\sigma(T_{\rm b}^{^{12}})$ and $\sigma(T_{\rm b}^{^{13}})$ are the measurement errors in $T_{\rm b}^{^{12}}$ and $T_{\rm b}^{^{13}}$, respectively, $\sigma(T)$ is the uncertainty in $T$ (discussed in the paragraph preceding Eq.~(\ref{eq:sigma2_Tb*_1})), and $\sigma({\mathtt r})$ is the uncertainty in ${\mathtt r}$, for which we adopt $\sigma({\mathtt r}) = 15$ \citep[][see paragraph following Eq.~(\ref{eq:Tb*_Tb_T_2})]{milam_etal_2005}.
%%(see paragraph following Eq.~(\ref{eq:Tb*_Tb_T_2})).

It is important to realize that Eqs.~(\ref{eq:sigma_Tb*_Tb_1}) and (\ref{eq:sigma_Tb*_T_1}) are strictly valid only in the limit of small errors.
When $T_{\rm b}^{12}$ approaches $T$, the small-error approximation fails, and Eqs.~(\ref{eq:sigma_Tb*_Tb_1}) and (\ref{eq:sigma_Tb*_T_1}) should be replaced by 
\begin{equation}
\left[ \sigma \big( (T_{\rm b\star}^{^{12}})_{1} \big) \right]_{T_{\rm b}^{^{12}}} 
\ = \ \big( (T_{\rm b\star}^{^{12}})_{1} \big) \big( T_{\rm b}^{^{12}} + \sigma(T_{\rm b}^{^{12}}) \big)
- \big( (T_{\rm b\star}^{^{12}})_{1} \big) \big( T_{\rm b}^{^{12}} \big)
\label{eq:sigma_Tb*_Tb_1_bis}
\end{equation}
and
\begin{equation}
\left[ \sigma \big( (T_{\rm b\star}^{^{12}})_{1} \big) \right]_{T} 
\ = \ \big( (T_{\rm b\star}^{^{12}})_{1} \big) ( T - \sigma(T) )
- \big( (T_{\rm b\star}^{^{12}})_{1} \big) (T) \ ,
\label{eq:sigma_Tb*_T_1_bis}
\end{equation}
respectively.
In case $T_{\rm b}^{^{12}} + \sigma(T_{\rm b}^{^{12}}) \ge T$, the r.h.s. of Eq.~(\ref{eq:sigma_Tb*_Tb_1_bis}) is undetermined (see Eq.~(\ref{eq:Tb*_Tb_T_1})), and we simply set $\left[ \sigma \big( (T_{\rm b\star}^{^{12}})_{1} \big) \right]_{T_{\rm b}^{^{12}}}$ to an arbitrarily large value.
Similarly for $\left[ \sigma \big( (T_{\rm b\star}^{^{12}})_{1} \big) \right]_{T}$ in case $T - \sigma(T) \le T_{\rm b}^{^{12}}$.

The best combined estimate of %%$T_{\rm b\star}^{^{12}}$ 
the opacity-corrected brightness temperature of the $^{12}$CO line, $T_{\rm b\star}^{^{12}}$, is the value that minimizes the sum of the weighted (by the inverse-variance) deviations squared from our two estimates, $(T_{\rm b\star}^{^{12}})_{1}$ and $(T_{\rm b\star}^{^{12}})_{2}$ -- or, equivalently, the weighted (by the inverse-variance) mean of $(T_{\rm b\star}^{^{12}})_{1}$ and $(T_{\rm b\star}^{^{12}})_{2}$,
\begin{equation}
\boxed{
(T_{\rm b\star}^{^{12}{\rm CO}})_{\rm best}
\ = \ \frac{\displaystyle
\frac{(T_{\rm b\star}^{^{12}})_{1}}{\sigma^2 \big( (T_{\rm b\star}^{^{12}})_{1} \big)} + \frac{(T_{\rm b\star}^{^{12}})_{2}}{\sigma^2 \big((T_{\rm b\star}^{^{12}})_{2} \big)}
}{\displaystyle
\frac{1}{\sigma^2 \big((T_{\rm b\star}^{^{12}})_{1} \big)} + \frac{1}{\sigma^2 \big((T_{\rm b\star}^{^{12}})_{2} \big)}
}
} \ \cdot
\label{eq:Tb*_best}
\end{equation}
As expected, $(T_{\rm b\star}^{^{12}{\rm CO}})_{\rm best}$ is closer to the better-constrained of our two estimates, $(T_{\rm b\star}^{^{12}})_{1}$ and $(T_{\rm b\star}^{^{12}})_{2}$.
The uncertainty in $(T_{\rm b\star}^{^{12}{\rm CO}})_{\rm best}$, $\sigma \big((T_{\rm b\star}^{^{12}{\rm CO}})_{\rm best} \big)$, is given by
\begin{equation}
\sigma^2 \big( (T_{\rm b\star}^{^{12}{\rm CO}})_{\rm best} \big) = \left( 
\frac{1}{\sigma^2 \big( (T_{\rm b\star}^{^{12}})_{1} \big)} 
+ \frac{1}{\sigma^2 \big( (T_{\rm b\star}^{^{12}})_{2} \big)} 
\right)^{-1} \,,
\label{eq:sigma2_Tb*_best}
\end{equation}
as is easily verified in the case of Gaussian distributions.
Thus, the uncertainty in $(T_{\rm b\star}^{^{12}{\rm CO}})_{\rm best}$ depends only on the uncertainties in $(T_{\rm b\star}^{^{12}})_{1}$ and $(T_{\rm b\star}^{^{12}})_{2}$, not on the difference between $(T_{\rm b\star}^{^{12}})_{1}$ and $(T_{\rm b\star}^{^{12}})_{2}$.
Moreover, $\sigma \big( (T_{\rm b\star}^{^{12}{\rm CO}})_{\rm best} \big)$ is smaller than each of $\sigma \big( (T_{\rm b\star}^{^{12}})_{1} \big)$ and $\sigma \big( (T_{\rm b\star}^{^{12}})_{2} \big)$, which is in line with our expectation that combining two different estimates of $T_{\rm b\star}^{^{12}}$ reduces the final uncertainty.

\section{Observational values of the LoS-averaged polarization fraction}
\label{sec:polfrac}

The all-sky map of the 353\,GHz polarized emission from Galactic dust presented by \cite{Planck_XIX_2015} revealed a large scatter in $p_{\rm d}^{\rm los}$, with measured values ranging  
from the noise limit up to a maximum, $(p_{\rm d}^{\rm los})_{\rm max}$, which depends on the hydrogen column density, $N_{\rm H}$:
for $N_{\rm H} \lesssim 10^{21}\,{\rm cm}^{-2}$, $(p_{\rm d}^{\rm los})_{\rm max} \simeq 20\,\%$;
for $10^{21}\,{\rm cm}^{-2} \lesssim N_{\rm H} \lesssim 10^{22}\,{\rm cm}^{-2}$, $(p_{\rm d}^{\rm los})_{\rm max}$ decreases steadily from $\simeq 20\,\%$ to $\simeq 12\,\%$;
then $(p_{\rm d}^{\rm los})_{\rm max}$ drops more steeply, down to $\lesssim 5\,\%$ for $N_{\rm H} \gtrsim 2 \times 10^{22}\,{\rm cm}^{-2}$.
Toward nearby dense cores, with $N_{\rm H} \gtrsim 10^{22}\,{\rm cm}^{-2}$, $p_{\rm d}^{\rm los}$ also systematically decreases with increasing $N_{\rm H}$.\footnote{
The derived values of $(p_{\rm d}^{\rm los})_{\rm max}$ depend on the exact criterion used to define $(p_{\rm d}^{\rm los})_{\rm max}$ and, more sensitively, on the angular resolution of the map.
The values of $(p_{\rm d}^{\rm los})_{\rm max}$ quoted here refer to the 99.99\,\% percentile of $(p_{\rm d}^{\rm los})$ in the considered $N_{\rm H}$ bin, at a resolution of $1^\circ$ (see upper solid red curve in Fig.~2 of \cite{Planck_XX_2015}).
}
The observed scatter in $p_{\rm d}^{\rm los}$ arises from a combination of factors, including (1) spatial variations in the intrinsic polarization fraction and/or the efficiency of grain alignment, (2) spatial variations in the magnetic field inclination to the PoS, and (3) depolarization due to fluctuations in the PoS magnetic field orientation both along the LoS and across the telescope beam.
The decrease of $(p_{\rm d}^{\rm los})_{\rm max}$ with increasing $N_{\rm H}$ can be attributed to either a gradual loss of dust grain alignment in dense regions shielded from the ambient UV radiation field 
(assuming radiative torques are largely responsible for dust grain alignment; see, e.g., \cite{draine&w_1997, hoang&l_2008})
or a gradual rise in the LoS + beam depolarization due to fluctuations in the PoS magnetic field orientation over long path lengths.
In any case, the LoS-averaged $(p_{\rm d}^{\rm los})_{\rm max}$ provides a lower limit to the local $(p_{\rm d}^{\rm loc})_{\rm max}$.

In a companion paper, \cite{Planck_XX_2015} analyzed the 353\,GHz polarized dust emission from ten nearby molecular clouds located away from the Galactic plane and displaying a variety of physical conditions and polarization properties.
They detected the largest $p_{\rm d}^{\rm los}$ in the most diffuse regions, and they confirmed the decrease of $(p_{\rm d}^{\rm los})_{\rm max}$ with increasing $N_{\rm H}$ above a certain threshold, the value of which actually depends on the considered cloud and is generally $\simeq (1-3) \times 10^{21}\,{\rm cm}^{-2}$.
The 2D distribution of ($p_{\rm d}^{\rm los}$,$N_{\rm H}$) in each cloud falls below the upper envelope, $(p_{\rm d}^{\rm los})_{\rm max}$ vs. $N_{\rm H}$, of the all-sky distribution obtained by \cite{Planck_XIX_2015}.
\cite{Planck_XX_2015} further compared their observed polarization maps to synthetic maps obtained with anisotropic, turbulent MHD simulations, in which they assumed a uniform polarization fraction parameter, $p_0 = 20\,\%$, corresponding to $(p_{\rm d}^{\rm loc})_{\rm max} = 23\,\%$ (see their Eq.~(8), where their intrinsic polarization fraction, $p_{\rm i}$, corresponds to our $(p_{\rm d}^{\rm loc})_{\rm max}$).
The assumption of uniform $(p_{\rm d}^{\rm loc})_{\rm max}$ is probably reasonable insofar as the dense cores appearing in the simulations are only weakly shielded from the ambient UV radiation.
The simulations were able to reproduce the main statistical trends of the observed polarization maps (excluding dense cores), with the only source of depolarization arising from the magnetic field (both its inclination to the PoS and the spatial fluctuations in its PoS orientation), i.e., without invoking any loss of dust grain alignment. 
They also showed that a good estimate of $(p_{\rm d}^{\rm loc})_{\rm max}$ could be inferred from the measured $(p_{\rm d}^{\rm los})_{\rm max}$ in diffuse regions where $\boldvec{B}$ lies in the PoS and is ordered on large scales.

The statistical analysis of \cite{Planck_XIX_2015} was later refined by \cite{Planck_XII_2020},
who relied on the third public release of {\it Planck} data and debiased the estimates of the polarization fraction with the help of the modified asymptotic (MAS) estimator.
The new results were found to be consistent with those of \cite{Planck_XIX_2015}.
In particular, the maximum value of the measured LoS-averaged polarization fraction was found to be $(p_{\rm d}^{\rm los})_{\rm max} = 22.0^{+3.5}_{-1.4}\,\%$.

In dense molecular cores, $p_{\rm d}^{\rm los}$ appears to be much lower than in the large-scale diffuse medium.
A number of studies reported observations of a drop in the polarization fraction toward dense cores (see, e.g., \cite{ward-Thompson_2000, alves_2014}; and more references in the review by \cite{pattle_2019}). These so-called polarization holes could result from a loss of grain alignment efficiency or from the tangling of magnetic field lines, within the beam or along the LoS. 
In a statistical analysis of {\it Planck} cold clumps \citep{PGCC_2016}, Ristorcelli et al. (in prep.) found a significant decrease of $p_{\rm d}^{\rm los}$ toward the clump centers, and they argued that this decrease is due to both grain misalignment and the magnetic field complex geometry.
Typical values near clump centers are $p_{\rm d}^{\rm los} \sim 2 - 4\,\%$.

\section{Expressions of the best-fit polarization parameters}
\label{sec:bestfit_polar_parameters}

The procedure used to derive the polarization parameters of each cloud ${\mathbb{C}}_i$ is to minimize the $\chi_{\rm r}^2$ associated with the Stokes parameters for linear polarization (Eq.~(\ref{eq:chi_reduced_QU}), with $Q_{\rm d}^{\rm mod}$ and $U_{\rm d}^{\rm mod}$ given by Eqs.~(\ref{eq:stokesQ_sum_cloud}) and (\ref{eq:stokesU_sum_cloud}), respectively).
Here, for simplicity, we assume $\sigma_Q = \sigma_U = \sigma$.
Minimizing Eq.~(\ref{eq:chi_reduced_QU}) with respect to $\overline{p}_{{\rm d},j}$, then with respect to $\overline{\psi}_{{\rm d},j}$, successively yields
\begin{equation}
\begin{aligned}
& \mathlarger{\sum}_{n_{\rm pix}} \
\frac{1}{\sigma^2} \ p_{\rm d}^{\rm los,obs} \ I_{\rm d}^{\rm obs} \ I_{{\rm d},j} \
\cos{\Big( 2 \,\big( \psi_{\rm d}^{\rm los,obs} - \overline{\psi}_{{\rm d},j} \big) \Big)} \\
& = 
\mathlarger{\sum}_{n_{\rm pix}} \
\frac{1}{\sigma^2} \ \sum_{i=1}^{n_{\rm cl}} \ \overline{p}_{{\rm d},i} \ I_{{\rm d},i} \ I_{{\rm d},j} \
\cos{\Big( 2 \,\big( \overline{\psi}_{{\rm d},i} - \overline{\psi}_{{\rm d},j} \big) \Big)}
\end{aligned}
\label{eq:bestfit_fromQ}
\end{equation}
and
\begin{equation}
\begin{aligned}
& \mathlarger{\sum}_{n_{\rm pix}} \
\frac{1}{\sigma^2} \ p_{\rm d}^{\rm los,obs} \ I_{\rm d}^{\rm obs} \ I_{{\rm d},j} \
\sin{\Big( 2 \,\big( \psi_{\rm d}^{\rm los,obs} - \overline{\psi}_{{\rm d},j} \big) \Big)} \\
& = 
\mathlarger{\sum}_{n_{\rm pix}} \
\frac{1}{\sigma^2} \ \sum_{i=1}^{n_{\rm cl}} \ \overline{p}_{{\rm d},i} \ I_{{\rm d},i} \ I_{{\rm d},j} \
\sin{\Big( 2 \,\big( \overline{\psi}_{{\rm d},i} - \overline{\psi}_{{\rm d},j} \big) \Big)} \ \cdot
\end{aligned}
\label{eq:bestfit_fromU}
\end{equation}

These equations become physically more transparent 
in the case of a single cloud ${\mathbb{C}}_i$, where they reduce to
\begin{equation}
\overline{p}_{{\rm d},i} \ = \ 
\frac{\displaystyle 
\mathlarger{\sum}_{n_{\rm pix}} \
p_{\rm d}^{\rm los,obs} \ \frac{I_{{\rm d},i}^2}{\sigma^2} \ 
\cos{\Big( 2 \,\big( \psi_{\rm d}^{\rm los,obs} - \overline{\psi}_{{\rm d},i} \big) \Big)}
}
{\displaystyle 
\mathlarger{\sum}_{n_{\rm pix}} \
\frac{I_{{\rm d},i}^2}{\sigma^2}
}
\label{eq:bestfit_polfrac}
\end{equation}
and
\begin{equation}
\mathlarger{\sum}_{n_{\rm pix}} \
p_{\rm d}^{\rm los,obs} \ \frac{I_{{\rm d},i}^2}{\sigma^2} \
\sin{\Big( 2 \,\big( \psi_{\rm d}^{\rm los,obs} - \overline{\psi}_{{\rm d},i} \big) \Big)}
\ = \ 0 \ ,
\label{eq:bestfit_polangle}
\end{equation}
when use is made of the identity $I_{\rm d}^{\rm obs} = I_{{\rm d},i}$.
If $\psi_{\rm d}^{\rm los,obs}$ does not fluctuate too much across the map, such that $\big( \psi_{\rm d}^{\rm los,obs} - \overline{\psi}_{{\rm d},i} \big)$ remains small, Eq.~(\ref{eq:bestfit_polangle}) can further be approximated by
\begin{equation}
\overline{\psi}_{{\rm d},i} \ = \ 
\frac{\displaystyle 
\mathlarger{\sum}_{n_{\rm pix}} \
\psi_{\rm d}^{\rm los,obs} \ \frac{p_{\rm d}^{\rm los,obs} \ I_{{\rm d},i}^2}{\sigma^2}
}
{\displaystyle 
\mathlarger{\sum}_{n_{\rm pix}} \
\frac{p_{\rm d}^{\rm los,obs} \ I_{{\rm d},i}^2}{\sigma^2}
} \ \cdot
\label{eq:bestfit_polangle_approx}
\end{equation}

Equation~(\ref{eq:bestfit_polfrac}) shows that the best-fit polarization fraction of cloud ${\mathbb{C}}_i$, $\overline{p}_{{\rm d},i}$, is a weighted average over all the pixels of the LoS-averaged polarization fraction, $p_{\rm d}^{\rm los,obs}$, 
reduced by a depolarization factor, $\cos{\Big( 2 \,\big( \psi_{\rm d}^{\rm los,obs} - \overline{\psi}_{{\rm d},i} \big) \Big)}$, due to fluctuations in the LoS-averaged polarization angle, $\psi_{\rm d}^{\rm los,obs}$, across the map.
Eq.~(\ref{eq:bestfit_polangle_approx}), for its part, shows that the best-fit polarization angle of cloud ${\mathbb{C}}_i$, $\overline{\psi}_{{\rm d},i}$, is a weighted average over all the pixels of the LoS-averaged polarization angle, $\psi_{\rm d}^{\rm los,obs}$.

\end{appendix}

\end{document}